\documentclass[journal]{IEEEtran}

\usepackage{algorithm}
\usepackage{algpseudocode}
\usepackage[T1]{fontenc}
\usepackage{multirow}
\usepackage[bookmarks=false]{hyperref}
\usepackage{cite}
\usepackage[algo2e]{algorithm2e} 
\usepackage[utf8x]{inputenc}
\usepackage{ulem,xpatch}
\usepackage{longtable}
\usepackage{colortbl}
\usepackage{tikz}
\usepackage{subcaption}
\usepackage{xcolor}
\usetikzlibrary{arrows,shapes,positioning,shadows,trees}
\usepackage{rotating, adjustbox}
\usepackage{pgfplots}
\usepgfplotslibrary{groupplots}
\pgfplotsset{compat=1.12}
\usepackage{tkz-euclide}
\usetikzlibrary{patterns}
\usetikzlibrary{knots,calc}
\usepackage{algpseudocode}
\usepackage{supertabular} 
\usepackage{graphicx}
\usepackage{lineno,hyperref}
\usepackage{amsmath}
\usepackage{amsfonts}
\usepackage{amssymb}
\usepackage{booktabs}
\usepackage{flushend}
\usepackage{smartdiagram}
\def\BibTeX{{\rm B\kern-.05em{\sc i\kern-.025em b}\kern-.08em
    T\kern-.1667em\lower.7ex\hbox{E}\kern-.125emX}}

\usepackage{pifont}
\usepackage{multirow}
\usepackage{booktabs}
\usepackage{tabularx}
\usepackage[inline]{enumitem}  
\usepackage{float}
\usepackage[vlines]{tabularht}
\usepackage{rotating}
\usepackage{array,multirow,makecell}
\modulolinenumbers[5]
\usepackage{lscape}
\usepackage[cmintegrals]{newtxmath}

\usetikzlibrary{shadows}
\usepackage{forest}
\usepackage{fontawesome}
\usepackage{comment}
\usepackage{diagbox}
\usepackage{caption}

\definecolor{morange}{rgb}{0.8,0.2,0}
 
\IEEEoverridecommandlockouts

\definecolor{Gray}{gray}{0.9}
\newcolumntype{g}{>{\columncolor{Gray}}p}

\usepackage{etoolbox}

\newcommand{\PreserveBackslash}[1]{\let\temp=\\#1\let\\=\temp}
\newcolumntype{C}[1]{>{\PreserveBackslash\centering}p{#1}}

\newcolumntype{P}[1]{>{\PreserveBackslash\centering}p{#1}}

\definecolor{caribbeangreen}{rgb}{0.0, 0.8, 0.6}
\newcommand{\noteothmane}[1]{\textcolor{black}{{ #1}}} 

\let\mybibitem\bibitem
\renewcommand{\bibitem}[1]{%
\ifstrequal{#1}{rathee2021design}{\color{black}\mybibitem{#1}}
{\ifstrequal{#1}{barka2021sthm}{\color{black}\mybibitem{#1}}
{\color{black}\mybibitem{#1}}}%
}

\begin{document}

\title{Edge Learning for 6G-enabled Internet of Things: A Comprehensive Survey of Vulnerabilities, Datasets, and Defenses}

\author{Mohamed~Amine~Ferrag, \IEEEmembership{Senior Member, IEEE}, Othmane Friha, Burak Kantarci, \IEEEmembership{Senior Member, IEEE}, Norbert Tihanyi, \IEEEmembership{Member, IEEE}, Lucas Cordeiro,  Merouane Debbah,  \IEEEmembership{Fellow, IEEE}, Djallel~Hamouda, Muna Al-Hawawreh, Kim-Kwang Raymond Choo, \IEEEmembership{Senior Member, IEEE} 

\thanks{M. A. Ferrag is the corresponding author.}
\thanks{M. A. Ferrag is with Technology Innovation Institute, 9639 Masdar City, Abu Dhabi, UAE email: mohamed.ferrag@tii.ae}
\thanks{O. Friha is with Networks and Systems Laboratory (LRS), Badji Mokhtar-Annaba University, B.P.12, Annaba 23000, Algeria, email: othmane.friha@univ-annaba.org}
\thanks{B. Kantarci is with School of Electrical Engineering and Computer Science, University of Ottawa, Ottawa, ON, Canada email: Burak.Kantarci@uOttawa.ca
}
\thanks{N. Tihanyi is with Technology Innovation Institute, 9639 Masdar City, Abu Dhabi, United Arab Emirates email: norbert.tihanyi@tii.ae}
\thanks{L. Cordeiro is with the University of Manchester, Manchester, UK.  email: lucas.cordeiro@manchester.ac.uk}
\thanks{M. Debbah is with Khalifa University of Science and Technology, P O Box 127788, Abu Dhabi, UAE email: merouane.debbah@ku.ac.ae}
\thanks{D. Hamouda is with Labstic Laboratory, Department of Computer Science, Guelma University, B.P. 401, 24000, Algeria. e-mail: hamouda.djallel@univ-guelma.dz}
\thanks{Muna Al-Hawawreh is with School of Information Technology, Deakin University, Australia email: muna.alhawawreh@deakin.edu.au}
\thanks{K.-K. R. Choo is with the Department of Information Systems and Cyber Security, University of Texas at San Antonio, San Antonio, TX 78249-0631, USA. email: raymond.choo@fulbrightmail.org}
}

\maketitle

\begin{abstract}
The deployment of the fifth-generation (5G) wireless networks in the Internet of Everything (IoE) applications and future networks (e.g., sixth-generation (6G) networks) has raised several operational challenges and limitations, for example in terms of security and privacy. Edge learning is an emerging approach to training models across distributed clients while ensuring data privacy. Such an approach when integrated in future network infrastructures (e.g., 6G) can potentially solve challenging problems such as resource management and behavior prediction. However, edge learning (including distributed deep learning) is known to be susceptible to tampering and manipulation. This survey article provides a holistic review of the extant literature focusing on edge learning-related vulnerabilities and defenses for 6G-enabled Internet of Things (IoT) systems. Existing machine learning approaches for 6G–IoT security and machine learning-associated threats are broadly categorized based on learning modes, namely: centralized, federated, and distributed. Then, we provide an overview of enabling emerging technologies for 6G–IoT intelligence. We also provide a holistic survey of existing research on attacks against machine learning and classify threat models into eight categories, namely: backdoor attacks, adversarial examples, combined attacks, poisoning attacks, Sybil attacks, byzantine attacks, inference attacks, and dropping attacks. In addition, we provide a comprehensive and detailed taxonomy and a comparative summary of the state-of-the-art defense methods against edge learning-related vulnerabilities. Finally, as new attacks and defense technologies are realized, new research and future overall prospects for 6G-enabled IoT are discussed \footnote{This paper has been accepted for publication in  IEEE Communications Surveys \& Tutorials: https://ieeexplore.ieee.org/document/10255264}.
\end{abstract}

\begin{IEEEkeywords}
\textcolor{black}{Edge Learning}, 6G, IoT, Federated Learning, AI vulnerabilities, Security.
\end{IEEEkeywords}

\section*{List of Abbreviations}
\begin{tabular}{ll}
5G  & Fifth-Generation\\
6G & Sixth-Generation\\
AEA & Auto-Encoder with Attention\\
AI & Artificial Intelligence\\
APT & Advanced Persistent Threat\\
BCD  & Bayesian Compromise Detection\\
CNN  & Convolutional Neural Network\\
CoAP & Constrained Application Protocol\\
CPMS  & Control Plane Micro Services\\
CVAE  & Conditional Variational Autoencoder\\
\textcolor{black}{CNDF} & \textcolor{black}{Core Network Data Analytics Function}\\
DL  & Deep Learning\\
DLT  & Distributed Ledger Technologies\\
DP  & Differential Privacy\\
DPI  & Deep Packet Inspection\\
DQN  & Deep Q-network\\
DNN & Deep Neural Network\\
DRL  & Deep Reinforcement Learning\\
eMBB &  Enhanced Mobile Broadband\\
FD  & Federated Distillation\\
FDD  & Frequency Division Duplexing\\
FGSM &  Fast Gradient Sign Method\\
FL  & Federated Learning\\
GAN  & Generative Adversarial Network\\
GRUs &  Gated Recurrent Units \\
HAR  & Human Activity Recognition\\
HE  & Homomorphic Encryption\\
HFL  & Horizontal Federated Learning\\
\textcolor{black}{HAI} & \textcolor{black}{Human-Centered Artificial Intelligence}\\
IID  & Independent and Identically Distributed\\
IoE &  Internet of Everything\\
IoMT &  Internet of Medical Things\\
IoT  & Internet of Things\\
IIoT & Industrial Internet of Things\\
JSCC  & Joint Source-Channel Coding\\
KPCA  & Kernel Principal Component Analysis\\
KPI  & Key Performance Indicators\\
\end{tabular} 
    
    \begin{tabular}{ll}
      \\
\textcolor{black}{LSTM} & \textcolor{black}{Long Short Term Memory} \\
MAC  & Medium Access Control\\
MDP  & Markov Decision Process\\
MQTT & Message Queuing Telemetry Transport\\
\textcolor{black}{MEC}  & \textcolor{black}{Multi-Access Edge Computing} \\
MIA  & Membership Inference Attack\\
MIMO &  Multiple-Input Multiple-Output\\
\textcolor{black}{MTD} & \textcolor{black}{Moving Target Defense} \\
mMIMO &  massive MIMO\\
MTD  & Moving Target Defense\\
NFV  & Network Functions Virtualization\\
\textcolor{black}{NOMA} & \textcolor{black}{Non-Orthogonal Multi-Access}\\
NI  & Network Intelligence\\
NIDS  & Network Intrusion Detection Systems\\
NLP  & Natural Language Processing\\
NOMA &  Non-Orthogonal Multi-Access \\
Non-IID  & Non-Independent and Identically Distributed\\
ODT &  Opportunistic Data Transfer\\
P2P &  Peer-to-Peer\\
 \textcolor{black}{PoF} &  \textcolor{black}{Proof of Federation}\\
RL  & Reinforcement Learning\\
 \textcolor{black}{RNN} &  \textcolor{black}{Recurrent Neural Network} \\
\textcolor{black}{RIS} & \textcolor{black}{Reconfigurable Intelligent Surface}\\
SDN  & Software-Defined Networking\\
SGD  & Stochastic Gradient Descent\\
SNN &  Self-Sustaining Network\\
\textcolor{black}{SHA} & \textcolor{black}{Secure Hash Algorithm}\\
\textcolor{black}{SVM} & \textcolor{black}{Support Vector Machine}\\
SONs &  Self-Organizing Networks\\
TFL  & Transfer Federated Learning\\
UE  & User Equipment\\
UPMS  & User Plane Micro Services\\
URLLC &  Ultra-Reliable, Low-Latency Communications\\
VFL  & Vertical Federated Learning\\
\textcolor{black}{VRF} & \textcolor{black}{Verifiable Random Function}\\
VLC  & Visible Light Communication\\
VM & Virtual Machine\\
XAI &  Explainable AI\\
XR  & Extended Reality\\
xURLLC &  eXtreme URLLC\\
    \end{tabular}

\section{Introduction}\label{sec:1}

We are experiencing accelerated development, increasing adoption, and innovative combinations of information and communication technologies, such as cloud and edge computing, the Internet of Things (IoT), massive data analytics, and Artificial Intelligence (AI). This widespread endorsement is driven by many factors, including the widespread availability of broadband communications, which are expected to move the world into an all-connected zone. One instance of such a combination is the integration of AI into the fifth-generation (5G) wireless networks. However, it is only intended to operate in specific areas under specific conditions (massive data and robust computing)~\cite{tariq2020speculative,batista2023smart}. This alliance is expected to be much tighter in future generations, starting with the upcoming sixth-generation (6G) wireless networks. AI is expected to be a core component in it \cite{dang2020should,osorio2022towards}. In addition, Multi-Access Edge Computing (MEC) provides the possibility of processing large volumes of data by edge devices and distributed learning paradigms such as Federated Learning (FL) which enable multiple parties to collaborate in building shared Machine Learning (ML) models without sharing their data. There is a lot of interest in making the edge intelligent, an emerging research area known as \textit{distributed edge learning}~\cite{zhu2020toward,naeem2023security}.

\begin{figure*}
\centering
\includegraphics[width=1\textwidth]{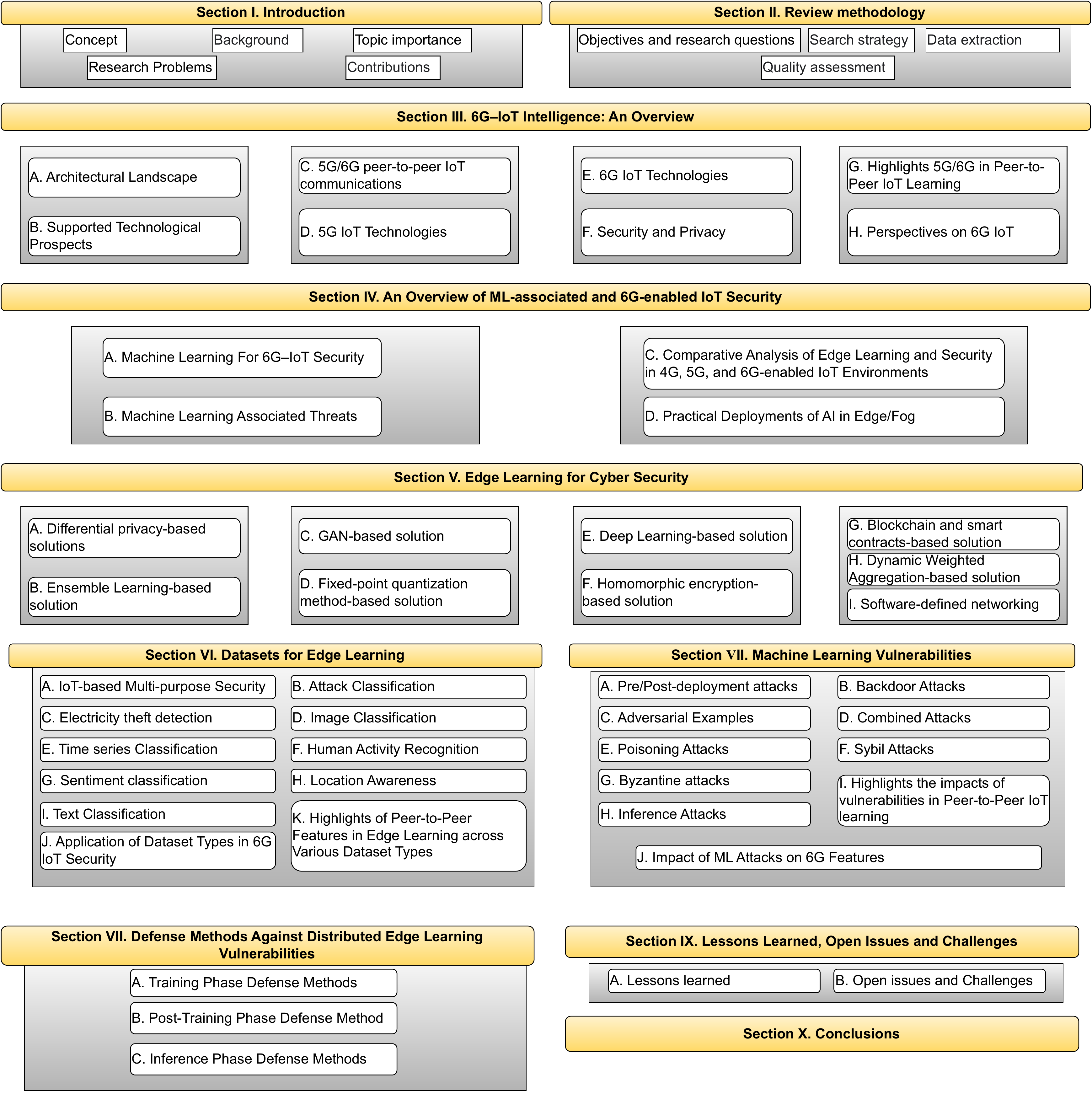}
\caption{\textcolor{black}{Structure of this survey paper.}}
\label{fig:figall}
\end{figure*}

The key driver for the advancement of wireless networks has been the requirement for higher data rates, which necessitated a continuous boost in network capacity. The ongoing rise of the Internet of Everything (IoE), which Cisco defines as the "\textit{networked connection of people, process, data, and things}"\footnote{https://www.cisco.com/c/dam/en\_us/about/business\-insights/docs/ioe\-value\-index\-faq.pdf}, in which billions of devices are plugged in and exchanging large amounts of data continuously, has led to a fundamental upgrade from enhanced mobile broadband (eMBB) services to ultra-reliable, low-latency communications (URLLC) \cite{saad2019vision}. While currently commercialized 5G in the ground will comfortably handle core IoE and URLLC services, as well as potentially going to support fixed access to mmWave frequencies, early 5G deployments are expected to utilize sub-6 GHz frequencies to support mobility, making it questionable whether they can provide the IoE applications of future intelligent cities \cite{saad2019vision}. To address these issues, 6G networks are envisioned to be the solution with AI as an indispensable component \cite{siriwardhana2021ai}.

\begin{figure*}
\centering
\includegraphics[width=1\textwidth]{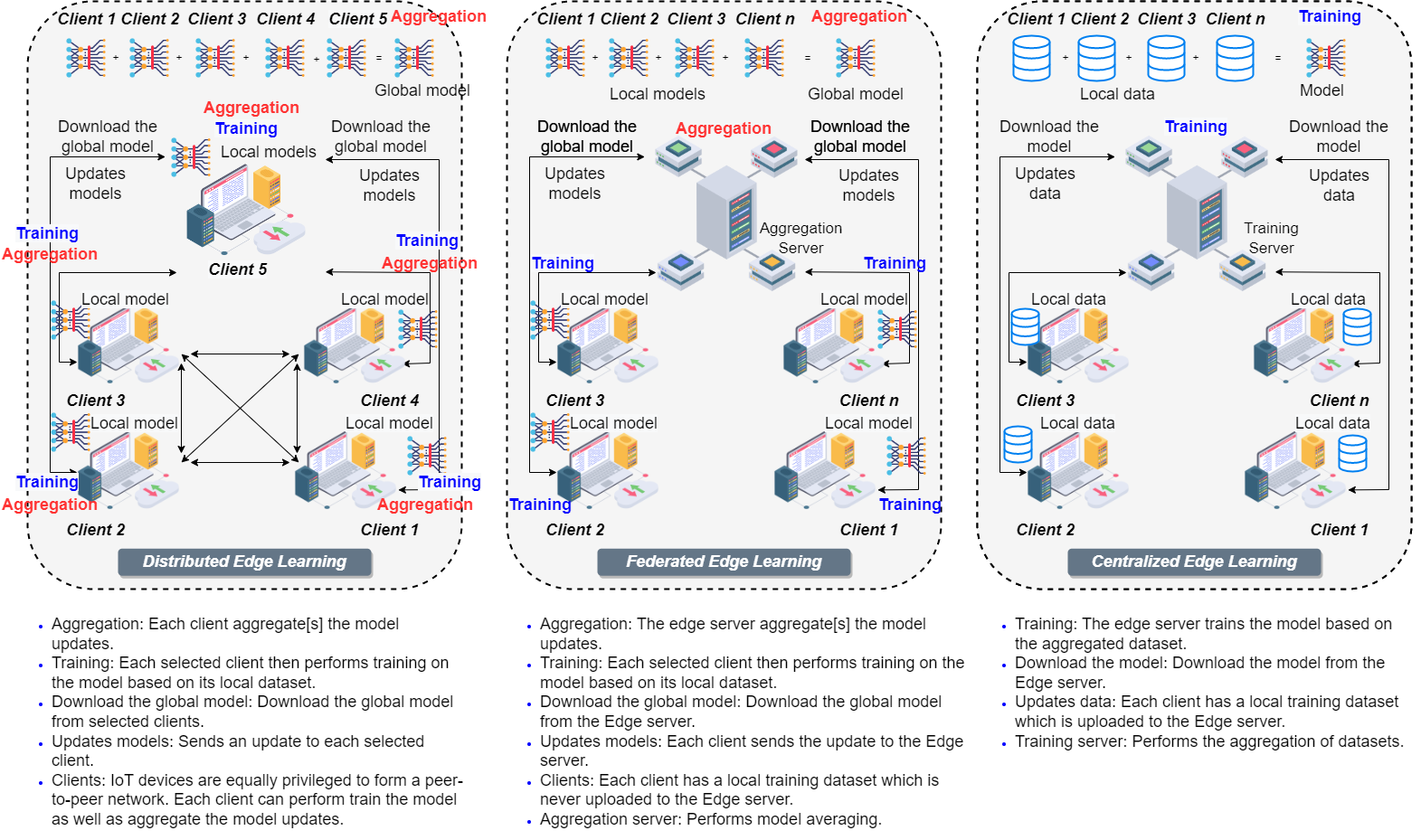}
\caption{\noteothmane{Distributed Edge Learning vs. Federated Edge Learning vs. Centralized Edge Learning.}}
\label{fig:fignew}
\end{figure*}

Currently, not only have AI and IoT demonstrated their potential benefits in various spheres, but their synergistic effect is seen as a \textcolor{black}{critical} factor in transforming the future, including Industry 4.0, Agriculture 4.0, and 6G communication networks. \textcolor{black}{An excellent} practical, real-world example of such cooperation is the development of research into autonomous vehicles. Furthermore, from an economic standpoint, both fields are emerging. In its 2022 AI Index report \footnote{https://aiindex.stanford.edu/wp-content/uploads/2022/03/2022-AI-Index-Report\_Master.pdf}, the Stanford Institute for Human-Centered Artificial Intelligence (HAI) states that investment in AI surpassed \$46 billion to reach \$93.5 billion between 2020 and 2021, with the \textcolor{black}{most significant} growth in investment coming from the global private sector. In addition, in the same context, the worldwide cellular IoT market size is projected to expand to \$61 billion by 2026 (from \$31 billion in 2022) \footnote{https://www.juniperresearch.com/press/cellular-iot-market-value-to-exceed-61b-globally}. The preceding statistics point out the success and breakthrough achieved through AI and IoT, which are virtually employed across all areas of daily lives, ranging from military, industry, healthcare, education, and entertainment, to name a few. 

\begin{figure*}
\centering
\includegraphics[width=0.8\textwidth]{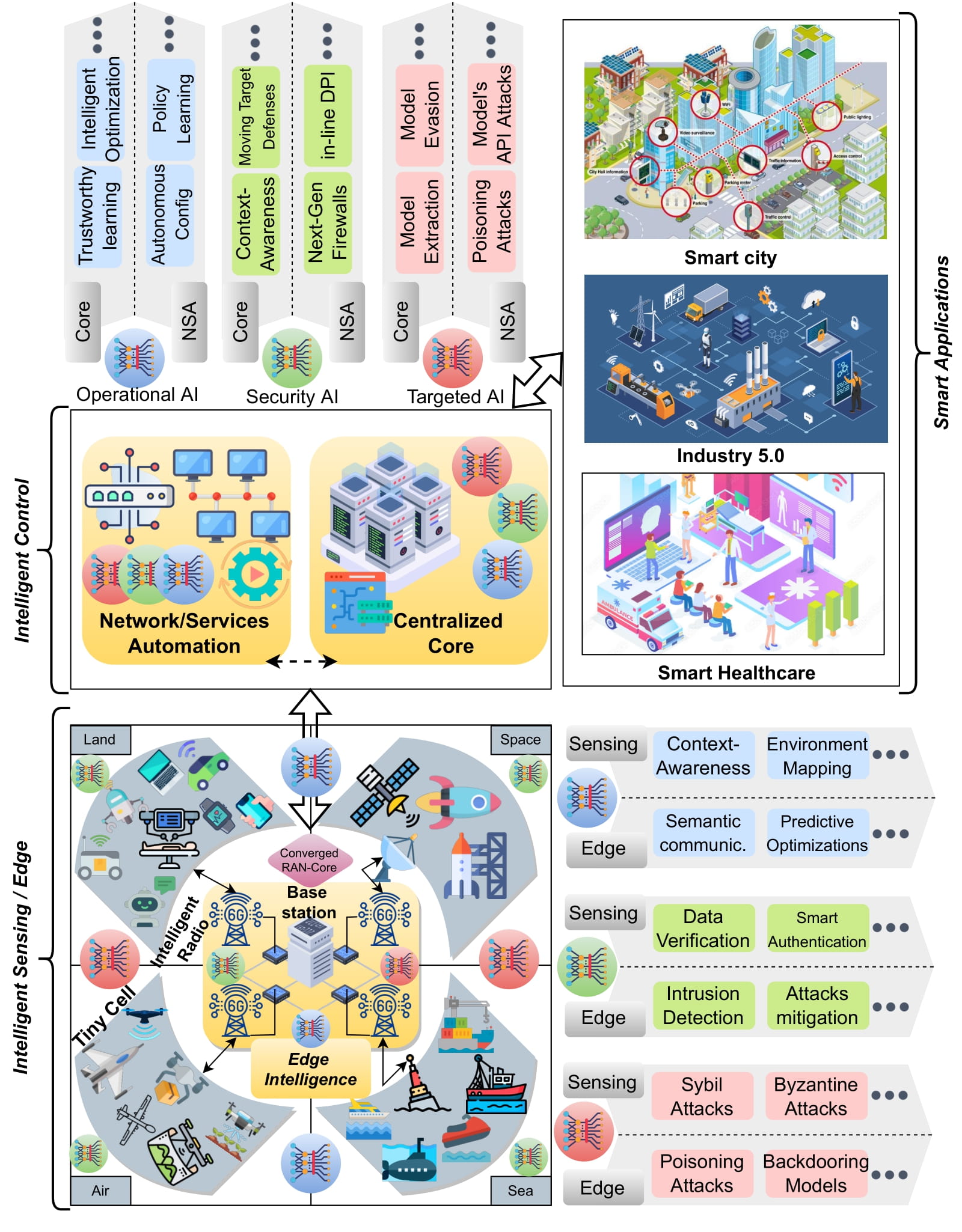}
\caption{\noteothmane{AI as a native component of 6G concerning the operational, defensive, and targeted perspectives: 1) Intelligent Sensing/Edge: This comprises two primary components: the first is the data generation aspect, which includes devices, systems, and processes from which data originates; the second is the edge layer, where certain cloud processing tasks are executed at the network's periphery; 2) Intelligent Control: This pertains to smart network management, primarily at the network core (e.g., Core Network Data Analytics Function (CNDF) \cite{chouman2022towards}); 3) Smart Applications: These encompass present and future intelligent applications that utilize the network; Additionally, we consider various AI perspectives within the network, such as operational, defensive, and targeted.}}
\label{fig:fig1}
\end{figure*}

Although both AI and IoT are considered somewhat mature from a variety of perspectives, including efficiency and operability \cite{ferrag2021federated}, as far as their security is concerned, they are viewed as being in the early developmental stages, and further progress is required to strengthen their robustness against cyber-attacks \cite{oseni2021security, kuzlu2021role, shejwalkar2022back}. To illustrate, in 2021, a widely used neural networks framework was found to have more than 300 classical security vulnerabilities, including overflows, memory corruption, bypasses, information leaks, and code executions \footnote{https://www.cvedetails.com/product/53738/Google-Tensorflow.html}. A riskier pathway is the adversarial example, where core-engineered manipulations or more refined AI methods are employed to produce synthetic input data to trigger malfunctions in the targeted AI systems. As an example, a group of Skylight researchers identified a specific bias toward a specific pattern in the Cylance AI-based antivirus product, which enabled the development of a workaround by adding a selected list of strings to a malware package, altering identification scores significantly, and avoiding malware detection, with a 100\% success rate for the top 10 malware (2019), and near 90\% for a broader sample of 380+ malware \footnote{https://skylightcyber.com/2019/07/18/cylance-i-kill-you/}. Not to mention the widespread attacks on Industrial IoT-based infrastructures in recent years, especially with sophisticated reconnaissance tools and search engines such as Shodan becoming publicly available on the Internet. We aim to shed light on these important issues within this paper (Figure \ref{fig:figall}).

\textcolor{black}{As we delve into cybersecurity and ML, we must recognize the interconnectedness and interdependence of two primary concepts: ``security for ML'' and ``ML for security''. The former focuses on utilizing ML techniques to address traditional security attacks, such as denial of service (DoS), man-in-the-middle, malware, and intrusion detection. In contrast, the latter deals with the inherent vulnerabilities and potential attacks targeting ML systems, such as data poisoning. This paper explores both concepts concurrently rather than presenting them as separate entities. This approach may initially appear to create confusion; however, it is a deliberate decision driven by our belief that the two domains cannot be studied in isolation. By examining their relationship holistically, we aim to provide a comprehensive understanding of the challenges and opportunities that arise when using ML in security applications. Throughout the paper, we will demonstrate how advancements in 'security for ML' can contribute to more robust and resilient 'ML for security solutions and vice versa. We believe that emphasizing the interplay between these areas is crucial for encouraging researchers and practitioners to develop innovative, integrated strategies that ensure the security and reliability of ML-based systems.}

Figure \ref{fig:fignew} illustrates the differentiation between Distributed Edge Learning, Federated Edge Learning, and Centralized Edge Learning. \textcolor{black}{Centralized Edge Learning refers to the traditional approach to machine learning where data is collected and stored in a central location, such as an edge server or data center. In this approach, a single machine-learning model is trained using all the data available in the central location. The model is then used to make predictions on new data. On the other hand, Federated Edge Learning is a decentralized approach to machine learning where data is stored on multiple devices or servers. In this approach, the model is trained collaboratively using data from all the devices or servers without sharing the raw data. The devices or servers send model updates to a central edge server, aggregating them to form a new model. This process is repeated iteratively until the model converges to a desired level of accuracy. Distributed edge learning refers to a machine learning paradigm where a large-scale model is trained across a network of distributed devices, such as IoT devices. In this approach, each device or client is responsible for training the model using its local dataset and then sending the updated model parameters to aggregators, which combines these updates to produce a new \textcolor{black}{model version}. Each device or client is equally privileged and forms a peer-to-peer network. Each device can communicate directly with its neighbors without a central coordinating authority. This decentralized approach is highly scalable and fault-tolerant, as it can continue to operate even if some devices fail or leave the network.}

The increased reliability and the enhanced utility of AI and IoT are \textcolor{black}{apparent} and potentially \textcolor{black}{helpful}. \textcolor{black}{Still, they} also lead to a novel and unique attack surface with cyber vulnerabilities that resemble the traditional vulnerabilities through primitive tampering and probing or a fresh category of vulnerabilities like adversarial AI. Figure \ref{fig:fig1} illustrates the effectiveness of AI and IoT as enablers of future 6G systems while focusing on the three \textcolor{black}{leading} roles of AI: 1) operative, 2) defender, and 3) target. The main objective of this paper is to provide a comprehensive and in-depth review of threats and challenges faced by AI-based IoT systems and infrastructures. While doing so, the article focuses on the Machine Learning (ML) subset and its different learning paradigms, namely centralized, federated, and decentralized approaches. In addition, we discuss possible effective countermeasures that can be employed to protect these systems. The contributions of this study are summarized below:

\begin{itemize}
    \item An overview of enabling emerging technologies for 6G–IoT intelligence.
    \item  A detailed report on the datasets used by the scientific community for experimenting and evaluating \textcolor{black}{Edge Learning} on cyber attacks.
    \item Presentation of the threat model of attacks against machine learning; and classification into eight categories:  backdoor attacks, adversarial examples, combined attacks, poisoning attacks, Sybil attacks, byzantine attacks, inference attacks, and drop attacks.
    \item A comprehensive taxonomy and a side-by-side comparison of the state-of-the-art defense methods against federated machine learning vulnerabilities.
    \item Presentation of the security and privacy challenges and opportunities for federated machine learning in 6G-enabled IoTs. 
\end{itemize}

The structure of this article is organized as shown in Figure \ref{fig:figall}. \textcolor{black}{Section \ref{sec:new1} provides the review methodology.} Section \ref{sec:2} sheds light on state of the art in ML in 6G-enabled IoT security and ML-associated threats. Section \ref{sec:3} presents an overview of enabling emerging technologies for 6G–IoT intelligence. \textcolor{black}{ Section \ref{sec:edge} discusses the use of edge learning and its application in cybersecurity.} \textcolor{black}{Section \ref{sec:44} provides a detailed report on the datasets used by the scientific community.} Section \ref{sec:4} reviews the threat models attacks against machine learning and provides a classification into eight categories. Section \ref{sec:5} provides a taxonomy and a side-by-side comparison of the state-of-the-art defense methods. Then, we discuss new research directions and future overall prospects for 6G-enabled IoTs in Section \ref{sec:7}. Lastly, Section \ref{sec:8} presents concluding remarks.

\begin{table*}[]
\centering
\setlength{\tabcolsep}{2.5pt}
\renewcommand{\arraystretch}{1}
\caption{{\color{black} Related studies on edge learning for 6G-enabled IoT applications.}}
\label{tab:tabrelatedwork}
\scriptsize
\centering
{\color{black} \begin{tabular}{|p{1.2in}|p{0.2in}|p{0.7in}|p{0.7in}|p{0.7in}|p{0.7in}|p{0.7in}|p{0.4in}|p{0.7in}|}
\hline
\textbf{Reference} & \textbf{Year} & \textbf{ML for 6G–IoT security} & \textbf{Vision for 6G networks security
} & \textbf{ML-associated threat (Centralized)} & \textbf{ML-associated threat (Federated)} & \textbf{ML-associated threat (Distributed)} & \textbf{Datasets}  & \textbf{Defenses methods against ML vulnerabilities} \\ \hline
\textcolor{black}{Yang \textit{et al.} \cite{yang2019federated}} & \textcolor{black}{2019} & \textcolor{black}{No} &\textcolor{black}{No} & \textcolor{black}{No}  & \textcolor{black}{Partial}  &  \textcolor{black}{No} & \textcolor{black}{No}  & \textcolor{black}{No}\\ \hline
Sun \textit{et al.} \cite{sun2020machine} & 2020 & Partial  & No & Partial & Partial  & No & No &No \\ \hline
Hussain \textit{et al.} \cite{hussain2020machine} & 2020 & Yes & No & Yes & No & No & No &  Partial \\ \hline
Sodhro \textit{et al.} \cite{sodhro2020toward} & 2020 & No & No & No & No & No & No &  No \\ \hline
Mohanta \textit{et al.} \cite{mohanta2020survey} &  2020& Yes & No & Yes & No & No & No & Partial\\ \hline
Lyu \textit{et al.} \cite{lyu2020threats} & 2020 & No & No & Partial & Yes & Partial & No & Partial\\ \hline
Fang \textit{et al.} \cite{fang20215g} & 2021 & No & No & No & No & No & No &  No \\ \hline

\textcolor{black}{Wahab \textit{et al.} \cite{wahab2021federated} }& \textcolor{black}{2021} & \textcolor{black}{No} &\textcolor{black}{No} & \textcolor{black}{No}  & \textcolor{black}{Partial}  &  \textcolor{black}{No} & \textcolor{black}{No}  & \textcolor{black}{Partial}\\ \hline

\textcolor{black}{Guo \textit{et al.} \cite{guo2021enabling} }& \textcolor{black}{2021} & \textcolor{black}{No} &\textcolor{black}{No} & \textcolor{black}{No}  & \textcolor{black}{No}  &  \textcolor{black}{No} & \textcolor{black}{No}  & \textcolor{black}{No}\\ \hline

\textcolor{black}{Chen \textit{et al.} \cite{chen2022reconfigurable} }& \textcolor{black}{2021} & \textcolor{black}{No} &\textcolor{black}{No} & \textcolor{black}{No}  & \textcolor{black}{No}  &  \textcolor{black}{No} & \textcolor{black}{No}  & \textcolor{black}{No}\\ \hline

\textcolor{black}{Nguyen \textit{et al.} \cite{nguyen20216g} }& \textcolor{black}{2021} & \textcolor{black}{Partial} &\textcolor{black}{Partial} & \textcolor{black}{No}  & \textcolor{black}{No}  &  \textcolor{black}{No} & \textcolor{black}{No}  & \textcolor{black}{No}\\ \hline

\textcolor{black}{Tang \textit{et al.} \cite{tang2021survey} }& \textcolor{black}{2021} & \textcolor{black}{Partial} &\textcolor{black}{Partial} & \textcolor{black}{No}  & \textcolor{black}{No}  &  \textcolor{black}{No} & \textcolor{black}{No}  & \textcolor{black}{No}\\ \hline

\textcolor{black}{Nguyen \textit{et al.} \cite{nguyen2021federated}} & \textcolor{black}{2021} & \textcolor{black}{No} &\textcolor{black}{No} & \textcolor{black}{No}  & \textcolor{black}{Partial}  &  \textcolor{black}{No} & \textcolor{black}{No}  & \textcolor{black}{No}\\ \hline

\textcolor{black}{Alazab \textit{et al.} \cite{alazab2021federated}} & \textcolor{black}{2021} & \textcolor{black}{Partial} &\textcolor{black}{No} & \textcolor{black}{No}  & \textcolor{black}{Partial}  &  \textcolor{black}{No} & \textcolor{black}{No}  & \textcolor{black}{No}\\ \hline
Zaman \textit{et al.} \cite{zaman2021security} & 2021 & Yes & No & Yes & No & No & No & Partial\\ \hline
Mothukuri \textit{et al.} \cite{mothukuri2021survey}  & 2021 & No & No & Partial  & Yes & Partial & No & Partial\\ \hline
\textcolor{black}{Ghimire and Rawat \cite{ghimire2022recent} }& \textcolor{black}{2022} & \textcolor{black}{No} &\textcolor{black}{No} & \textcolor{black}{No}  & \textcolor{black}{Partial}  &  \textcolor{black}{No} & \textcolor{black}{No}  & \textcolor{black}{No}\\ \hline
\textcolor{black}{Ma \textit{et al.} \cite{ma2022state}} &  \textcolor{black}{2022}& \textcolor{black}{No} &\textcolor{black}{No} & \textcolor{black}{No}  & \textcolor{black}{No}  &  \textcolor{black}{No} & \textcolor{black}{No}  & \textcolor{black}{No}\\ \hline

\textcolor{black}{Lu \textit{et al.} \cite{lu2022reinforcement}} &  \textcolor{black}{2022}& \textcolor{black}{Partial} &\textcolor{black}{No} & \textcolor{black}{Partial}  & \textcolor{black}{No}  &  \textcolor{black}{No} & \textcolor{black}{No}  & \textcolor{black}{No}\\ \hline

Vaezi \textcolor{black}{ \textit{et al.} \cite{vaezi2022cellular}} &  \textcolor{black}{2022}& \textcolor{black}{No} &\textcolor{black}{No} & \textcolor{black}{No}  & \textcolor{black}{No}  &  \textcolor{black}{No} & \textcolor{black}{No}  & \textcolor{black}{No}\\ \hline
Fadlullah \textcolor{black}{ \textit{et al.} \cite{fadlullah2022balancing}} &  \textcolor{black}{2022}& \textcolor{black}{Partial} &\textcolor{black}{Partial} & \textcolor{black}{No}  & \textcolor{black}{No}  &  \textcolor{black}{No} & \textcolor{black}{No}  & \textcolor{black}{Partial}\\ \hline

\textcolor{black}{Serghiou \textit{et al.} \cite{serghiou2022terahertz}} & \textcolor{black}{2022} & \textcolor{black}{No} &\textcolor{black}{No} & \textcolor{black}{No}  & \textcolor{black}{No}  &  \textcolor{black}{No} & \textcolor{black}{No}  & \textcolor{black}{No}\\ \hline

\textcolor{black}{Liu \textit{et al.} \cite{liu2022privacy}} &  \textcolor{black}{2022}& \textcolor{black}{No} &\textcolor{black}{No} & \textcolor{black}{No}  & \textcolor{black}{No}  &  \textcolor{black}{No} & \textcolor{black}{No}  & \textcolor{black}{No}\\ \hline
\textcolor{black}{Boobalan \textit{et al.} \cite{boobalan2022fusion}} & \textcolor{black}{2022} & \textcolor{black}{No} &\textcolor{black}{No} & \textcolor{black}{No}  & \textcolor{black}{Partial}  &  \textcolor{black}{No} & \textcolor{black}{No}  & \textcolor{black}{No}\\ \hline
\textcolor{black}{Yang \textit{et al.} \cite{yang2022federated}} & \textcolor{black}{2022} & \textcolor{black}{Partial} &\textcolor{black}{No} & \textcolor{black}{No}  & \textcolor{black}{No}  &  \textcolor{black}{No} & \textcolor{black}{No}  & \textcolor{black}{No}\\ \hline
Sarker \textit{et al.} \cite{sarker2022internet} &  2022 & Yes & No & Yes & No & No & No & Partial\\ \hline
\textcolor{black}{Azari \textit{et al.} \cite{azari2022evolution}} & \textcolor{black}{2022} & \textcolor{black}{No} &\textcolor{black}{No} & \textcolor{black}{No}  & \textcolor{black}{No}  &  \textcolor{black}{No} & \textcolor{black}{No}  & \textcolor{black}{No}\\ \hline
Qian \textit{et al.} \cite{qian2022distributed} &  2022 & No & No & Partial & Partial & Yes & No & Partial\\ \hline
Ma \textit{et al.} \cite{ma2022trusted} & 2022 & No & No & Partial  & Partial & Yes & No & Partial\\ \hline
\textcolor{black}{Zhang \textit{et al.} \cite{zhang2022federated}} & \textcolor{black}{2022} & \textcolor{black}{No} &\textcolor{black}{No} & \textcolor{black}{No}  & \textcolor{black}{No}  &  \textcolor{black}{No} & \textcolor{black}{No}  & \textcolor{black}{No}\\ \hline
Veith \textit{et al.} \cite{veith2023road} & 2023  & No & Partial  & No & No & No & No & No \\ \hline
Mao \textit{et al.} \cite{mao2023security} & 2023  & Partial & Partial & No & Partial & No & No &  Partial \\ \hline
Alotaibi \textit{et al.} \cite{alotaibi2023securing} & 2023 & Yes & Partial & Partial & Partial & No & No & Partial \\ \hline
Hua \textit{et al.} \cite{hua2023edge }& 2023 & No & Partial & No & Partial & No & No & No \\ \hline
Al-Quraan \textit{et al.} \cite{al2023edge} & 2023 & No  & No & No & No& No & No &No \\ \hline
Xia \textit{et al.} \cite{xia2023poisoning} & 2023 & No  & No & No & Partial  & No & No &No \\ \hline
\textcolor{black}{Issa \textit{et al.} \cite{issa2023blockchain}} & \textcolor{black}{2023} & \textcolor{black}{No} &\textcolor{black}{No} & \textcolor{black}{No}  & \textcolor{black}{No}  &  \textcolor{black}{No} & \textcolor{black}{No}  & \textcolor{black}{No}\\ \hline
\textcolor{black}{Zhu \textit{et al.} \cite{zhu2023blockchain}} & \textcolor{black}{2023} & \textcolor{black}{No} &\textcolor{black}{No} & \textcolor{black}{No}  & \textcolor{black}{No}  &  \textcolor{black}{No} & \textcolor{black}{No}  & \textcolor{black}{No}\\ \hline
\textcolor{black}{Chafii \textit{et al.} \cite{chafii2023twelve}} & \textcolor{black}{2023} & \textcolor{black}{No} &\textcolor{black}{No} & \textcolor{black}{No}  & \textcolor{black}{No}  &  \textcolor{black}{No} & \textcolor{black}{No}  & \textcolor{black}{No}\\ \hline
\textcolor{black}{Wang \textit{et al.} \cite{wang2023road}} & \textcolor{black}{2023} & \textcolor{black}{No} &\textcolor{black}{No} & \textcolor{black}{No}  & \textcolor{black}{No}  &  \textcolor{black}{No} & \textcolor{black}{No}  & \textcolor{black}{No}\\ \hline
\textcolor{black}{Khalid \textit{et al.} \cite{khalid2023reconfigurable}} & \textcolor{black}{2023} & \textcolor{black}{No} &\textcolor{black}{No} & \textcolor{black}{No}  & \textcolor{black}{No}  &  \textcolor{black}{No} & \textcolor{black}{No}  & \textcolor{black}{No}\\ \hline
Our study & 2023 & Yes & Yes & Yes & Yes & Yes & Yes & Yes \\ \hline
\end{tabular}}
\end{table*}

\begin{table*}[]
\centering
\setlength{\tabcolsep}{2.5pt}
\renewcommand{\arraystretch}{1}
\caption{{\color{black}Research questions.}}
\label{tab:tabnew1}
\scriptsize
\centering
{\color{black}\begin{tabular}{|p{0.2in}|p{3in}|p{3in}|p{0.5in}|}
\hline
\textbf{} & \textbf{Question} & \textbf{Objective} & \textbf{Section} \\ \hline
  RQ1       & What are the current state-of-the-art reviews on ML-related and 6G-enabled IoT security, and how can they be classified into ML-based security for 6G-IoT systems and ML-associated threats? &  To review and categorize the studies into ML-based security for 6G-IoT systems and ML-associated threats. The research can enhance the security of 6G-IoT systems and identify potential threats facing ML-associated paradigms.  &  Section \ref{sec:2}   \\ \hline
  RQ2     & What are the potential benefits, challenges, and implications of using AI-based Network Intelligence (NI) as the backbone for network management in 6G and beyond, particularly in the context of the IoE?  &   To provide insights into the architectural landscape, technological prospects, and security and privacy concerns associated with using NI in future sophisticated networks and applications.   &  Section \ref{sec:3}    \\ \hline
  RQ3  & What are the main characteristics and purposes of the datasets used in the scientific community to experiment and evaluate machine learning techniques for cyber attacks, and how can they be classified into different categories based on their content?  &   To provide a detailed report on the datasets used by the scientific community for experimenting and evaluating machine learning techniques on cyber attacks  &    Section \ref{sec:44}              \\ \hline
  RQ4     & What are the types of attacks and vulnerabilities are machine learning systems susceptible to, particularly in 6G-IoT Networks, and how can they be classified based on the attacker's knowledge, the type of attack employed, and the final objective?  &   To identify and classify the most common types of attacks and vulnerabilities that machine learning systems face in 6G-IoT Networks, and to provide a comprehensive understanding of these threats based on the attacker's knowledge, the type of attack employed, and the final objective.    &  Section \ref{sec:4}     \\ \hline
  RQ5    & What are the current state-of-the-art methods for securing machine learning systems in 6G-IoT systems where AI is a key player?  &  To identify and analyze the current defense mechanisms against ML attacks and provide insights into the most effective ways to protect machine learning systems against potential security risks.               &     Section \ref{sec:5}  \\ \hline
  RQ6   &  What are the major challenges and open issues in enhancing cyber security in IoT and AI, and how can the scientific community ensure a fully secure cyber environment for future networks (i.e., 6G and beyond)?  &    To identify and explore the challenges associated with creating reliable and trustworthy learning environments for 6G-IoT intelligence and to propose potential solutions to these challenges.     &   Section \ref{sec:7}   \\ \hline
\end{tabular}}
\end{table*}

\begin{table*}[]
\centering
\setlength{\tabcolsep}{2.5pt}
\renewcommand{\arraystretch}{1}
\caption{{\color{black}Quality assessment questionnaire.}}
\label{tab:tabnew3}
\scriptsize
\centering
{\color{black}
\begin{tabular}{|p{0.2in}|p{2in}|p{2in}|p{0.8in}|}
\hline
\textbf{No.} & \textbf{Question} & \textbf{Description} & \textbf{Relevant to the research question} \\ \hline
Q1 & Were the research questions clearly stated and appropriate? & Evaluate the clarity and appropriateness of the research questions & RQ1, RQ2, RQ3, RQ4, RQ5, RQ6\\ \hline
Q2 & Did the authors provide a comprehensive review of the state-of-the-art in the relevant areas? & Evaluate the comprehensiveness of the literature review & RQ1, RQ2, RQ4, RQ5, RQ6 \\ \hline
Q3 & Were the datasets used in the study appropriate and relevant? & Evaluate the suitability and significance  of the datasets used in the study & RQ3 \\ \hline
Q4 & Were the methods used to classify the studies, datasets, attacks, and vulnerabilities appropriate and effective? & Review the appropriateness and effectiveness of the classification methods used & RQ1 \\ \hline
Q5 & Were the proposed solutions' potential benefits, challenges, and implications discussed in detail? & Evaluate the depth and comprehensiveness of the proposed solutions & RQ1, RQ2, RQ4, RQ5, RQ6 \\ \hline
Q6 & Were the limitations of the proposed solutions discussed? & Evaluate the limits of the proposed solutions & RQ2, RQ6 \\ \hline
Q7 & Were the potential applications of the proposed solutions discussed in detail? & Review the thoroughness of the discussions on the potential applications of the proposed solutions & RQ2, RQ5\\ \hline
Q8 & Were the conclusions and recommendations based on the results and analysis presented in the study? & Evaluate the degree to which the conclusions and recommendations are supported by the results and analysis & RQ1, RQ2, RQ3, RQ4, RQ5, RQ6 \\ \hline
\end{tabular}}
\end{table*}

\begin{table*}[]
\centering
\setlength{\tabcolsep}{2.5pt}
\renewcommand{\arraystretch}{1}
\caption{{\color{black}Usage of Machine Learning Paradigms in 6G-enabled IoT Applications.}}
\label{tab:tabpara}
\scriptsize
\centering
{\color{black}
\begin{tabular}{|p{1.1in}|p{2.2in}|p{2.2in}|}
\hline
\multicolumn{1}{|c|}{\textbf{Paradigm}} & \multicolumn{1}{c|}{\textbf{Characteristics}}                                                                                                                                                                                                                                      & \textbf{Usage in 6G-enabled IoT}                                                                                                                                                                                                                               \\ \hline
Supervised Learning                     & - Models are trained on labeled data \newline - The goal is to make predictions on new, unseen data                                                              & Build predictive models that can detect anomalies in large datasets, such as predicting device failures or detecting anomalies in IoT network traffic                                                                                                          \\ \hline
Unsupervised Learning                   & - Models are trained on unlabeled data.\newline - The goal is to identify patterns or structures within the data                                          & Identifying patterns and insights from unstructured data in 6G-enabled IoT systems, such as clustering devices based on their behavior or detecting anomalies in sensor data                                                                                \\ \hline
Reinforcement Learning                  & - Models take actions based on feedback from the environment.\newline - The goal is to learn the optimal policy that maximizes reward over time                      & Optimizing the performance of devices and networks in 6G-enabled IoT, such as optimizing the energy consumption of IoT devices by learning from the device's environment and adjusting its behavior accordingly                                                   \\ \hline
Semi-Supervised Learning                & - Models are trained on a combination of labeled and unlabeled data \newline - The goal is to improve performance by leveraging the unlabeled data           & Improving the performance of models in 6G-enabled IoT by leveraging both labeled and unlabeled data, such as classifying devices in the network by using both labeled and unlabeled data                                                                      \\ \hline
Transfer Learning                       & - Models use knowledge learned from one task to improve performance on another task.\newline - The goal is to reduce the amount of data needed to train the model  & Improving the performance of models in 6G-enabled IoT by leveraging knowledge learned from one task to improve performance on another task, such as improving the accuracy of intrusion detection in IoT devices by using pre-trained models on similar datasets \\ \hline
\end{tabular}}
\end{table*}

\section{\color{black}Review methodology}\label{sec:new1}


\textcolor{black}{In this section, we provide the review methodology that we employed to investigate the vulnerabilities, datasets, and defenses of \textcolor{black}{Edge Learning} for 6G-enabled IoT systems. We begin by outlining our objectives and research questions, followed by our search strategy, data extraction process, and quality assessment criteria.}

\subsection{\color{black}Objectives and research questions}
\textcolor{black}{In this research study, we aim to investigate the vulnerabilities, datasets, and defenses of \textcolor{black}{edge learning} for 6G-enabled IoT systems. \textcolor{black}{We aim} to understand the security threats posed to these systems and the methods researchers propose to protect them. We will analyze the vulnerabilities of machine learning and advanced deep learning techniques used to differentiate between benign and malicious traffic patterns. Our research will also explore the growing interest in defense solutions and their potential effectiveness in securing IoT systems against machine learning large-scale attacks. Table \ref{tab:tabnew1} provides the particular identified research questions to address the goals outlined above.}

\subsection{\color{black}Search strategy}
\textcolor{black}{To identify the literature for analysis in this paper, a keyword search was conducted using terms such as "Federated Learning for 6G-enabled IoT", "Edge Learning for 6G-enabled IoT", "Distributed Learning for 6G-enabled IoT", "Federated Learning for Cyber Security", and "ML Attacks" in academic databases such as SCOPUS, Web of Science, ACM Digital Library, IEEE, Wiley, and Springer. This search produced substantial results, but some relevant primary sources may have been missed. Only proposed ML attacks and Defenses methods for 6G-enabled IoT were collected, and each source was evaluated based on criteria such as reputation, relevance, originality, publication date (between $2015$ and $2023$), and influence in the field. Reputation was evaluated based on the author's previous work and expertise in the field. The Reputation was evaluated based on the author's previous work and expertise in the field. The Originality was evaluated based on the novelty of the ideas presented in the source. The Influence was evaluated based on the impact of the source in the field. The Sources cited more frequently were considered to have a higher influence and were given a higher score. The evaluation criteria used in this study ensured that the sources selected for analysis were of high quality, relevant to the research topic, and contained innovative and influential ideas. After evaluation, the sources were ranked based on their overall score, and the top sources were selected for analysis.}

\begin{table}[]
\centering
\setlength{\tabcolsep}{2.5pt}
\renewcommand{\arraystretch}{1}
\caption{{\color{black}Data extraction relevant fields.}}
\label{tab:tabnew2}
\scriptsize
\centering
{\color{black}\begin{tabular}{|p{0.7in}|p{2.3in}|}
\hline
\textbf{Field} & \textbf{Description} \\ \hline
Reference & Provide the title of the research paper and its citation details  \\ \hline
Year &  Year of publication \\ \hline
Dataset & Provides the datasets used as a benchmark experiment in the evaluation of ML vulnerabilities  \\ \hline
Attack category &  The different types of attacks against ML techniques, such as Sybil attack, Backdoor attack, Poisoning attack, and Adversarial attack, are classified under attack categories \\ \hline
 Attack type &  This describes various types of attacks against ML techniques, such as miscellaneous attacks, RNN backdoor attacks, Sybil-based poisoning attacks, Federated poisoning attacks, imperceptible backdoor patterns, Clean-label poisoning attacks, Poisonous label attacks, and more \\ \hline
Machine learning  &   Provides the machine learning techniques used in the \textcolor{black}{Edge Learning} \\ \hline
Learning mode &  This describes various types of learning modes, including centralized learning, distributed learning, federated learning, and transfer learning \\ \hline
Targeted ML &   Provides the machine learning techniques that are affected by attacks and vulnerabilities \\ \hline
ML phases &   This describes various types of ML model development lifecycles, including, Post-deployment and pre-deployment\\ \hline
Attack Goal &  Provides the goal of an attack against a machine learning system, including Data Poisoning, Adversarial Examples, Model Stealing, Model Evasion, ...etc. \\ \hline
Attack Description &  Describes an attack against a machine learning system \\ \hline
The attacker's knowledge  & Provides the attacker's knowledge (e.g., the trained parameters, the learning algorithm, the feature values, and the training set) that can be used to launch sophisticated attacks, evade detection by traditional security systems, and create targeted attacks  \\ \hline
 Attack mode &   This describes various types of attack modes, including centralized learning, distributed learning, federated learning \\ \hline
The attacker's goal - Vulnerabilities & Provides the goal of an attacker targeting machine learning systems that can vary depending on the specific context and motivations of the attacker \\ \hline
Mitigation solution & Provides the mitigation solutions that can be implemented to address machine learning vulnerabilities  \\ \hline
Attacks examples & Provides the threat models of adversarial examples generation designed to cause a machine learning model to make incorrect predictions or decisions  \\ \hline
Attacks for the specific tasks  & Provides the different types of machine learning tasks (e.g., Classification, Recognition, Image Segmentation, ... etc.)  \\ \hline
White/blue/Grey box & The techniques used to test and exploit the vulnerability of ML models in different levels of access and knowledge about the model's internal workings  \\ \hline
Adversarial Example Generation &  Describes the adversarial example generation  \\ \hline
Defense framework &  Provides the defense frameworks in mitigating machine learning vulnerabilities and safeguarding the integrity and security of AI-powered systems \\ \hline
 Threat model & The threat model of machine learning vulnerabilities posed by adversarial attacks, data poisoning, model inversion, and other malicious activities that exploit weaknesses in the machine learning pipeline  \\ \hline
 Classifiers &   \\ \hline
Pros (+)  Open Issues (-)  &  Provides the Pros (+) and Open Issues (-) of the defense framework against ML vulnerabilities\\ \hline
Defense methods  &  This includes the classification of defense methods against \textcolor{black}{Edge Learning} vulnerabilities (e.g.,  Training phase defense methods, Post-training phase defense methods, Inference phase defense methods) \\ \hline
Defense mechanisms & This includes the classification of defense mechanisms used in defense Defense mechanisms against \textcolor{black}{Edge Learning} vulnerabilities (e.g.,  Privacy leakage defense mechanisms, Sybil attacks defense mechanisms, ... etc.)  \\ \hline
Defense strategy &  Provides the defense strategy adopted by defense mechanisms against \textcolor{black}{Edge Learning} vulnerabilities (e.g.,  Bio-inspired, Reputation-Awareness, Federated Filters,... etc.) \\ \hline
\end{tabular}}
\end{table}

\subsection{\color{black}Data extraction}

\textcolor{black}{The lengthy data extraction process thoroughly examines selected research papers to identify and collect essential data. A quality assessment is conducted in the initial phase to ensure that the research papers align with our research questions, resulting in fewer papers for further analysis. The remaining papers undergo data extraction to obtain pertinent information, and Table \ref{tab:tabnew2} outlines the extracted metadata.}

\subsection{\color{black}Quality assessment}
\textcolor{black}{In our study, we conducted a quality assessment using a questionnaire that included ten questions related to the research questions and objectives. These questions covered a range of topics, including 1) the clarity and appropriateness of the research questions, 2) the comprehensiveness of the literature review, 3) the appropriateness and relevance of the datasets used, 4) the effectiveness of the classification methods used, 5) the depth and comprehensiveness of the discussions on proposed solutions, 6) the degree to which the proposed solutions are supported by experimental analysis, and 7) the clarity of the presentation of the results. By answering these questions, we \textcolor{black}{could} evaluate the quality and identify the advantages and disadvantages of each work. The formulation of the quality assessment criteria in Table \ref{tab:tabnew3} is based on eight questions that assess the quality of our research \textcolor{black}{concerning} the research questions.
}


\section{6G–IoT Intelligence: an overview}\label{sec:3}

For future 6G and beyond, there will be a heavy reliance on Network Intelligence (NI) with AI models representing its backbone \cite{letaief2021edge}, which ensures that network management will be \textcolor{black}{adequately} managed and fully automated. The AI models have shown considerable effectiveness in solving complex tasks, such as deriving complicated associations from massive, overlapping data, which will be even more required for future sophisticated networks and applications, such as IoE \cite{dang2020should}. In this section, we present a vision of the expected effectiveness of intelligence in 6G-IoT systems from three angles as illustrated in Figure \ref{fig:fig2}: 1) the architectural landscape, 2) technological prospects, and 3) security and privacy benefits.

\subsection{Architectural Landscape}
Using AI, specifically, DL provides an innovative path to engineer and improve 6G networks throughout the physical and core layers. Advances in 6G wireless theory and communication will also influence the advancement and expansion of AI in the form of new learning theories and architectures, creating a positive feedback loop, and transforming the wireless landscape, from connecting objects to connecting intelligence \cite{letaief2021edge}.

\subsubsection{Physical (PHY) Layer Enhancements}
During the last few years, investigations and efforts have been made to deploy AI in the physical layer of wireless communication networks, including User Equipment (UE) and at the cell edge \cite{ali20206g, chorti2022context, letaief2021edge}. Therefore, several problems with current communication systems remain unaddressed due to inaccurate models or non-linearity \cite{ali20206g}, such as reciprocity in frequency division duplexing (FDD), predicting channels, detecting and reducing interference. According to Ali \textit{et al.} \cite{ali20206g}, many of the physical layer's optimization problems, such as spectrum sensing, optimal beamforming formulation, and throughput maximization employing power control, are non-convex. Such issues may be addressed through dual decomposition techniques that require time-consuming iterative algorithms. On the other hand, DL techniques have significant capabilities in managing such challenges in real-time without sacrificing performance. In addition, AI offers a new way to design the 6G radio interface by further improving the radio environment and communication algorithms. Letaief \textit{et al.} \cite{letaief2021edge} proposed innovative solutions such as joint source-channel coding (JSCC), task-oriented communication, and semantic communication.

\textcolor{black}{A Reconfigurable Intelligent Surface (RIS) is a customizable surface design capable of manipulating electromagnetic (EM) wave reflections by adjusting electric and magnetic characteristics. The core idea behind RIS is to control the electromagnetic properties of the environment. This manipulation is inherently a physical phenomenon, which makes RIS a tool for the PHY layer.  Chen  \textit{et al.} \cite{chen2022reconfigurable}  delves into the emerging significance of reconfigurable intelligent surfaces in the context of 6G wireless communication systems, specifically their potential in data delivery and accurate positioning for IoT. RIS can adeptly control radio waves, so its role in 6G communication is growing. This article focuses on the burgeoning interest in RIS-assisted positioning, detailing its working principle, channel model, and characteristics.}

\subsubsection{Medium Access Control (MAC)}
ML supports an automated learning paradigm shift towards high-performance algorithms for addressing resource allocation challenges within wireless networks \cite{chorti2022context, dang2020should}. MAC layer-related tasks such as selecting and matching users for multiple-input multiple-output (MIMO), modulation, and handover control can also be optimized using AI \cite{dang2020should, nguyen2021security}. For instance, given that RL can handle combinatorial action spaces with multi-agent settings, RL is considered suitable for missions where the network can adapt to varying conditions while learning ideal strategies \cite{ali20206g}. Therefore, AI plays \textcolor{black}{a vital role} in 6G MAC layer optimization tasks for the following ML-based predictive tasks: 1) optimal resource allocation (e.g., in non-orthogonal multi-access (NOMA) and massive MIMO (mMIMO)), 2) ML-based predictive power management (e.g., traffic forecasting and prioritized packets separation), 3) FL-based mobility prediction (e.g., federated echo state network), 4) mobile data offloading, 5) link adaptation, and 6) caching \cite{dang2020should, ali20206g, letaief2019roadmap, yang2021federated}. 

\subsubsection{Core-Network and Services Intelligence}
One of the key benefits of using AI at the core and edge of the network, specifically by integrating training capabilities into the network nodes, is to enable intelligent data collection, processing, delivery, and utilization at the network edge. These benefits will improve the network service quality. Therefore, the inference method can produce reliable and cost-effective services \cite{letaief2019roadmap}. As one of several cases of this type, device-server co-inferencing can overcome existing traffic and processing constraints by spreading a sizeable DNN network across different edge node types (I.e., devices and servers) \cite{letaief2021edge}. In addition, future innovative self-organizing and self-repairing characteristics, autonomous connections between devices, and the human-centric network with enhanced intelligence within the 6G network translate directly into a wealth of other communication services \cite{dang2020should}.

\subsubsection{\textcolor{black}{6G architecture for energy-efficient communication}}
\textcolor{black}{
Efficient energy use in communication has garnered interest from various sectors, including industrial automation, healthcare, transportation, and more. Furthermore, the surge in AI-driven sixth-generation (6G) technologies aimed at advancing smart automation systems is now in the spotlight for scholars and businesses. The backbone of most advanced automation systems comprises IoT-based user terminal (UT) devices that deliver multimedia content — such as video, audio, image, and text — with precision and efficacy. Sodhro et al. \cite{sodhro2020toward} address the growing importance of energy-efficient communication in various sectors and the rise of AI-driven 6G technology in intelligent automation systems. These systems predominantly use IoT devices for high-quality multimedia content delivery. Recognizing that the traditional Quality of Service (QoS) doesn't adequately reflect user sentiment during multimedia transmission, the study focuses on Quality of Experience (QoE). The paper introduces a QoS-based joint energy and entropy optimization algorithm (QJEEO), formulates a 6G-influenced multimedia data structure for QoE evaluation, establishes a link between surveyed user scores and IoT device performance metrics, and devises a correlation model for QoS and estimated QoE perceptions. Experimental findings underscore that QoE can be assessed and aligned with QoS factors, specifically packet loss ratio and transfer delay, to enhance user satisfaction in energy-efficient, 6G-based multimedia transmissions.}

\subsubsection{\color{black}5G/6G IoT testbeds}
\textcolor{black}{The table \ref{tab:tabnew4} provides a list of 5G and 6G IoT testbeds that are crucial for testing and developing advanced 5G and 6G IoT technologies and applications. The testbeds are designed to test and develop 5G/6G-based IoT services and applications, such as smart transportation, energy management, healthcare, and industrial automation. The European Union has funded several testbeds listed, including 5G-TRANSFORMER, 5G-MoNArch, 5G-PICTURE, 5G-CORAL, and 5G-EmPOWER. Other funded testbeds include the 5G Open Innovation Lab in the USA, ENCQOR 5G in Canada, 5G City in Denmark, the 5G Alliance for Connected Industries and Automation in Germany, and the 6G platform in Germany. The testbeds feature network slicing, multi-access edge computing (MEC), cloud computing, virtualization, terahertz and sub-terahertz communications, and artificial intelligence (AI).}

\subsection{Supported Technological Prospects}
AI-backed context awareness within the network provides satisfying networking experiences that future users and applications require. \textcolor{black}{URLLC focuses on fulfilling the mission and safety-critical applications' demanding latency and dependability expectations}. However, the emergence of new applications, such as extended reality (XR) applications, requires reliability and latency requirements that are significantly more challenging than those previously defined in the 5G URLLC. Hence, an upgrade was envisioned called eXtreme URLLC (xURLLC). According to Park et al. \cite{park2020extreme}, xURLLC is based on three fundamental concepts, including 1) quicker and trustworthy data-driven ML-based predictability, 2) the utilization of both radio frequency and non-radio frequency modalities, and 3) collaborative communication and management co-design. The extensive KPIs expectations for future xURLLC and futuristic mission-critical applications, including IoE will require all aspects of innovative 6G-targeted AI-based methodologies and technologies, which can be roughly grouped into two classes: 1) In-Network Domains and 2) In-Application Domains.

\subsubsection{In-Network Domains}
In addition to introducing AI-based enhancements natively into existing network features, 5G/6G and beyond are also being introduced into \textcolor{black}{various} other intelligent technologies for efficient management and optimization. Network infrastructure virtualization is an emergent concept for present and future generations of networks. The network slicing \cite{abdel2022security, letaief2021edge} defines a networking architectural model permitting multiple isolated and virtualized logical networks on top of the pre-existing physical infrastructure. Each technology is designed to fit specific business conditions for varying applications. The concept is expected to be backed by other emerging technologies, including Software-Defined Networking (SDN) and Network Functions Virtualization (NFV) \cite{letaief2021edge, nguyen2021security}. The massive volumes of data processed worldwide by 2025 are expected to exceed 180 zettabytes \cite{ma2022trusted}. Hence, AI-powered predictive analytics utilize data to forecast future trends such as future traffic profiles, customer placement, behavior, and preferences. In addition, proactive caching is a recent solution to dramatically reduce peak traffic on the wireless core networks \cite{letaief2019roadmap}.

\subsubsection{In-Application Domains}
Using ML to enhance networking and wireless communications will fundamentally influence software development practices. \textcolor{black}{Considering the need for not only enormous speeds and extremely low latency for future mission-critical applications, including advanced healthcare systems (e.g., telemedicine), fully autonomous transportation systems, Industry 5.0, meta-universe, and augmented/extended/virtual reality applications, but also the network should be highly adaptive, dynamic, and context-aware which AI models enable}. \textcolor{black}{Therefore, opportunistic data transfer (ODT), a context-aware network optimization technique, can circumvent autonomous transport systems' data transfer challenge}. In addition, the ML-based network flow prediction module can pick network interfaces and program data transfers based on expected resource availability within a specified application-specific delay tolerance window \cite{ali20206g}.

\begin{table*}[h!]
\centering
\setlength{\tabcolsep}{2.5pt}
\renewcommand{\arraystretch}{1}
\caption{{\color{black}5G/6G Technologies for IoT applications.}}
\centering
\label{tab:tab5g6g}
\scriptsize
{\color{black}
\begin{tabular}{|p{1.2in}|p{1.2in}|p{4.2in}|}
\hline
\textbf{Technologies}                                                                                         & \textbf{Key Aspects}                                                 & \textbf{Description}                                                                                                                                                                                                                                                                                                                                              \\ \hline
\multirow{4}{*}{\begin{tabular}[c]{@{}l@{}}Enhanced Mobile \\ Broadband (eMBB) \cite{wang2023road}\end{tabular}}                  & High Data Rates                                                      & \begin{tabular}[c]{@{}l@{}}- Supports high data rates for bandwidth-demanding   applications such as 4K/8K video streaming,\\ virtual reality (VR), and   augmented reality (AR). Supports data rates of up to 20 Gbps for downloads\\    and 10 Gbps for uploads.\\ - Supports data rates of up to 20 Gbps for downloads   and 10 Gbps for uploads.\end{tabular} \\ \cline{2-3} 
                                                                                                              & Broad Coverage                                                       & \begin{tabular}[c]{@{}l@{}}- Ensures consistent user experience across different   environments — indoors, outdoors, in urban areas,\\  and in more remote   locations.\\ - The minimum guaranteed data rate should be 100 Mbps   for downloads and 50 Mbps for uploads.\end{tabular}                                                                             \\ \cline{2-3} 
                                                                                                              & Network Capacity                                                     & \begin{tabular}[c]{@{}l@{}}- Designed to handle high traffic density,   accommodating many simultaneous high-demand users \\ in a small area.\\ - Designed to handle traffic density of up to 10   Mbps/square meter.\end{tabular}                                                                                                                                \\ \cline{2-3} 
                                                                                                              & Mobility                                                             & \begin{tabular}[c]{@{}l@{}}- Offers high data rates and quality of service even   when the user is moving rapidly.\\ - Aims to provide latency of less than 4 ms for   time-critical communications, \\ and less than 20 ms for regular communications.\end{tabular}                                                                                              \\ \hline
\multirow{4}{*}{\begin{tabular}[c]{@{}l@{}}Ultra-Reliable Low-Latency \\ Communications (URLLC) \cite{siddiqui2023urllc}\end{tabular}} & Low Latency                                                          & \begin{tabular}[c]{@{}l@{}}- The delay between the sender initiating a data   transfer and the receiver getting it. \\ It's crucial for applications like remote surgery or autonomous driving where delays\\  can have severe   consequences.\\ - The URLLC Target is 1-millisecond round-trip   latency\end{tabular}                                            \\ \cline{2-3} 
                                                                                                              & High Reliability                                                     & - The ability of a system to function correctly,   without interruption, over a certain period or amount of data transfer.                                                                                                                                                                                                                                        \\ \cline{2-3} 
                                                                                                              & Availability                                                         & \begin{tabular}[c]{@{}l@{}}- The ability of a system to remain in a functional   state, even in the presence of faults or disruptions. \\ In URLLC, high   availability ensures that the system is functional and accessible when   needed.\end{tabular}                                                                                                          \\ \cline{2-3} 
                                                                                                              & Security                                                             & \begin{tabular}[c]{@{}l@{}}- Given the mission-critical nature of URLLC use   cases, robust data security measures are essential. \\ Security in URLLC   includes data integrity, and confidentiality.\end{tabular}                                                                                                                                               \\ \hline
\begin{tabular}[c]{@{}l@{}}Massive Machine Type   \\ Communications (mMTC) \cite{shen2023five}\end{tabular}                       & High-density network                                                 & \begin{tabular}[c]{@{}l@{}}- Supports the large number of devices (up to 1 million   per square kilometer), low data rates (tens of kbps),\\  and low-cost. \\ - Dealing with potential interference issues with   large-scale devices.\end{tabular}                                                                                                                  \\ \hline
Network Slicing   \cite{wang2023artificial}                                                                                            & eMBB Slice, URLLC Slice, and mMTC Slice                              & \begin{tabular}[c]{@{}l@{}}- Managing multiple slices over a shared physical   infrastructure. \\ - Providing Quality of Service (QoS) according to   slice requirements.\end{tabular}                                                                                                                                                                            \\ \hline
\multirow{2}{*}{Edge Computing \cite{zhou2023edge}}                                                                               & Data Offload                                                         & \begin{tabular}[c]{@{}l@{}}- Edge computing allows data to be processed closer   to the source, reducing\\  the amount of data that must be sent back to a   central server.\end{tabular}                                                                                                                                                                         \\ \cline{2-3} 
                                                                                                              & Bandwidth and Capacity                                               & \begin{tabular}[c]{@{}l@{}}- By processing data closer to the source, edge   computing can reduce the amount \\ of bandwidth required and increase network   capacity.\end{tabular}                                                                                                                                                                               \\ \hline
\multirow{3}{*}{\begin{tabular}[c]{@{}l@{}}Terahertz (THz) \\ Communications \cite{han2022terahertz}\end{tabular}}                    & High-speed wireless access and backhaul                              & \begin{tabular}[c]{@{}l@{}}- Using THz frequencies for ultra-high-speed wireless   communication.\\ - Beamforming, MIMO techniques, and short-range   communication can be used.\end{tabular}                                                                                                                                                                     \\ \cline{2-3} 
                                                                                                              & Ultra-high-resolution sensing                                        & \begin{tabular}[c]{@{}l@{}}- Exploiting THz frequencies for high-resolution   sensing applications.\\ - Frequency selection, short-range communication.\end{tabular}                                                                                                                                                                                              \\ \cline{2-3} 
                                                                                                              & Wireless chip-to-chip communication                                  & \begin{tabular}[c]{@{}l@{}}- Using THz communication for inter-chip data   exchange. \\ - Advanced semiconductor materials and nano-antenna   arrays.\end{tabular}                                                                                                                                                                                                \\ \hline
\multirow{2}{*}{\begin{tabular}[c]{@{}l@{}}Advanced AI and \\ Machine Learning \cite{liu2023machine}\end{tabular}}                  & Network Optimization                                                 & - Utilizing AI/ML for predicting and optimizing   network conditions.                                                                                                                                                                                                                                                                                             \\ \cline{2-3} 
                                                                                                              & Predictive Maintenance                                               & - Predicting failures and maintaining the network efficiently   using AI/ML.                                                                                                                                                                                                                                                                                      \\ \hline
\multirow{3}{*}{\begin{tabular}[c]{@{}l@{}}Integration of Satellite \\ and Terrestrial Networks \cite{zhu2022creating}\end{tabular}} & Satellite-backed eMBB (Enhanced Mobile Broadband)                    & \begin{tabular}[c]{@{}l@{}}- Satellite communications supporting high data-rate   services with \\ peak data rates up to 20 Gbps.\end{tabular}                                                                                                                                                                                                                    \\ \cline{2-3} 
                                                                                                              & Satellite-backed URLLC (Ultra-Reliable Low-Latency   Communications) & - Satellite networks providing ultra-fast and highly   reliable communication with a latency goal of around 1ms.                                                                                                                                                                                                                                                  \\ \cline{2-3} 
                                                                                                              & Satellite-backed mMTC (Massive Machine Type   Communication)         & \begin{tabular}[c]{@{}l@{}}- Satellite networks supporting high-density   communication between\\  machines that typically generate small amounts of data.\end{tabular}                                                                                                                                                                                           \\ \hline
\multirow{2}{*}{\begin{tabular}[c]{@{}l@{}}Advanced Positioning \\ and Sensing \cite{behravan2022positioning}\end{tabular}}                  & Advanced Positioning in 6G                                           & - Improved accuracy in device positioning, with expected   sub-meter or even centimeter-level precision.                                                                                                                                                                                                                                                          \\ \cline{2-3} 
                                                                                                              & Advanced Sensing in 6G                                               & \begin{tabular}[c]{@{}l@{}}- The ability for the network to perceive and   understand the surrounding environment \\ by using integrated sensors and AI   algorithms.\end{tabular}                                                                                                                                                                                \\ \hline
\end{tabular}}
\end{table*}

\subsection{\textcolor{black}{5G/6G peer-to-peer IoT communications}}
\textcolor{black}{With the advent of next-generation networking technologies such as 5G and the forthcoming 6G, we're about to witness an unparalleled transformation in IoT communications. These new-generation networks promise exponential speed, capacity, and latency improvements, which are vital for efficient and reliable IoT functionalities. This sub-section provides an exhaustive review of the emerging technologies in 5G and projected 6G networking that is poised to revolutionize peer-to-peer IoT communications. Table \ref{tab:tab5g6g} presents the new technologies in 5G/6G IoT communications.} 

\subsection{\textcolor{black}{5G IoT Technologies}}

\textcolor{black}{The following five technologies are integral to the functioning of the 5G IoT. We will delve into Enhanced Mobile Broadband (eMBB), which enhances data bandwidth and latency; Ultra-Reliable Low-Latency Communications (URLLC), crucial for critical applications; Massive Machine Type Communications (mMTC), designed to handle large-scale IoT deployments; Network Slicing, a feature that allows for effective resource allocation and quality management; and Edge Computing, a significant factor in reducing latency and network congestion. These technologies collectively revolutionize IoT, facilitating efficient, robust communication and creating various applications, from daily tasks to essential industrial processes.}

\begin{itemize}

\item \textcolor{black}{Enhanced Mobile Broadband (eMBB): This technology aims to provide significant improvements in data bandwidth and latency compared to 4G networks. It allows IoT devices to transfer data at much higher rates, making it ideal for high-definition video streaming, VR/AR applications, and real-time monitoring \cite{al2023resource}.}

\item \textcolor{black}{Ultra-Reliable Low-Latency Communications (URLLC): URLLC is a crucial part of 5G technology. It aims to provide ultra-reliable communication links with low latency, which is essential for critical applications such as autonomous vehicles, remote surgeries, and industrial automation \cite{she2023guest}.}

\item \textcolor{black}{Massive Machine Type Communications (mMTC): mMTC allows a large number of devices to be connected to the network simultaneously. This is crucial for large-scale IoT deployments, such as smart cities or large-scale agricultural monitoring, where thousands to millions of devices need to be interconnected \cite{hsu2023hyper}.}

\item \textcolor{black}{Network Slicing: Network slicing is a 5G feature that allows creating multiple virtual networks over a common physical infrastructure. This allows for better resource allocation and quality of service management, especially important in IoT applications where different devices may have vastly different requirements \cite{huang2023opportunistic}.}

\item \textcolor{black}{Edge Computing: While not a communication technology per se, edge computing is expected to play a significant role in 5G IoT. By moving computation and storage closer to the devices, edge computing reduces latency and network congestion \cite{xu2023edge}.}

\end{itemize}

\subsection{\textcolor{black}{6G IoT Technologies}}

\textcolor{black}{This subsection delves into the future of connectivity with 6G IoT technologies. As we prepare to step into the next generation of networks, we explore the transformative role of Terahertz (THz) Communications that promise unprecedented data rates; Advanced AI and Machine Learning set to manage the network infrastructure and process immense data; Integration of Satellite and Terrestrial Networks offering global coverage; Advanced Positioning and Sensing for precise location tracking and sensor data incorporation; and finally, Advanced Security and Privacy to address escalating concerns surrounding data protection. Together, these technologies encapsulate the advancements that 6G will bring, pushing the boundaries of IoT to new heights.}

\begin{itemize}
     
\item \textcolor{black}{Terahertz (THz) Communications: 6G is expected to utilize the terahertz frequency bands, enabling much higher data rates, potentially up to 100 Gbps or more \cite{song2021terahertz}.}

\item \textcolor{black}{Advanced AI and Machine Learning: AI and ML technologies are expected to play a major role in 6G networks, both in managing the network infrastructure and in processing the vast amounts of data generated by IoT devices \cite{ferrag2023generative}.}

\item \textcolor{black}{Integration of Satellite and Terrestrial Networks: 6G is expected to fully integrate satellite networks with terrestrial networks, providing truly global coverage and seamless handoff between different types of networks \cite{yu2023towards}.}

\item \textcolor{black}{Advanced Positioning and Sensing: 6G networks may include advanced positioning technologies, with accuracy down to the centimeter level, and may also incorporate data from various types of sensors into the network infrastructure \cite{alkhateeb2023deepsense}.}

\item \textcolor{black}{Advanced Security and Privacy: With the growing concerns about security and privacy, particularly in IoT applications, 6G is expected to incorporate advanced security and privacy features, both at the device level and at the network level \cite{mao2023security}.}

\end{itemize}
\subsection{Security and Privacy}
For 6G networks to successfully address the varying threats within the space-to-air-to-ground embedded network environment and novel technologies, AI is envisioned as the core driver for overcoming both security and privacy challenges \cite{nguyen2021security, abdel2022security, chorti2022context}. We classify intelligent security in 6G under two categories, namely, 1) Infrastructural Measures and 2) Additional Services. 

\subsubsection{Infrastructural Measures}
Integrating training and inference functionalities across multiple \textcolor{black}{network nodes helps} maintain privacy and confidentiality and ensures high security. Considering that the 6G infrastructure is designed for connected intelligence and will deploy AI at different network hierarchy levels, the AI-powered mini-cell can be used to block \textcolor{black}{specific} threats such as DoS, man-in-the-middle and information theft at the lowest levels before it reaches the targeted applications \cite{siriwardhana2021ai}. 

To preserve privacy, Over-the-Air FL is getting attention as a prominent approach for fast wireless data aggregation \textcolor{black}{by exploiting the overlay property} in multiple access channels, without the need for accessing the network nodes' private data \cite{letaief2021edge}. The context-aware security is a promising area of research where the goal is to obtain a network that can extract the threat level from the existing situation, use the context to determine the necessary security level, and match security levels to security parameters \cite{chorti2022context}. 

Another intelligent security concept predicted to play an \textcolor{black}{essential} role in future networks is the Moving Target Defense (MTD). MTD technologies can provide enhanced security flexibility by continuously and dynamically reshaping the underlying system on different layers during the execution cycle. This makes it difficult for adversaries to successfully exploit a continuously evolving, instantaneous, and unpredictable targeted system. For example, when an attack is noticed, the MTD module(s) can tell the SDN controller which way to re-program its data plane to reduce the extent of the attack \cite{nguyen2021security}. Furthermore, various security and privacy architectural mechanisms are proposed \cite{abdel2022security}, including \textcolor{black}{protecting sensitive customer information} such as location, for which privacy-aware ML-based offloading schemes can be used. In addition, biometrics-based authentication or a password-less utility can be adopted within the network's access control arrangements.

\subsubsection{Additional Services}
Apart from the native security intelligence expected for 6G, including the techniques mentioned above, Nguyen \textit{et al.} \cite{nguyen2021security} envisioned that the significant improvement in 6G security is the broad adoption of security-targeted AI-driven techniques and in-line deep packet inspection (DPI) capabilities within firewalls and network intrusion detection systems (NIDS). For example, the authors suggested that the 5G security gateway situated on the Access and Mobility Function (AMF) will require significant enhancements to its existing capabilities which they predict, among other things, in 6G will involve an integrated AI-based intrusion prevention engine. Therefore, other advanced technologies and services predicted to protect the security and privacy of future 6G mobile networks have virtually secured their positions. For example, Blockchain and Distributed Ledger Technologies (DLT) are expected to support next-generation networks, including 6G, \textcolor{black}{regarding} privacy, transparency, and immutability in various critical tasks. These tasks include access control, authentication, mutual trust, anonymization, and single point of failure perturbations avoidance at the infrastructure and application levels \cite{friha2021blockchain, dang2020should}. 

Current Blockchain designs and operational smartness security implementations are mainly realized in two forms, including 1) programmable software and 2) consensus mechanism. The programmable software is an intelligent contract \textcolor{black}{that automatically} performs operations according to predefined conditions. The consensus mechanism is a distributed algorithm that verifies the validity of submitted candidate blocks, with the verification process being different depending on the type of Blockchain (i.e., public, private, or consortium) and its implementation logic (e.g., puzzle-based, vote-based, etc.).

The 6G is expected to introduce ultra-high speeds (over 1Tbps \cite{DOCOMO}), large data volumes (5 zettabytes per month by 2030 \cite{ITU}), and a vast number of globally connected mobile devices (13.1 billion by 2023 \cite{cisco2020cisco}). AI-based security should be implemented in these operations to keep up with the 6G, and to identify anomalous patterns in blockchain transactions and potential vulnerabilities in smart contracts at higher speeds and low costs \cite{nguyen2021security}.

\subsection{\textcolor{black}{Highlights 5G/6G in Peer-to-Peer IoT Learning}}
\textcolor{black}{The integration of 5G/6G technology into IoT ecosystems will be transformative, providing a powerful impetus for peer-to-peer (P2P) learning among IoT devices \cite{ahammed2023vision}. We explore some of the unique aspects of 5G/6G that enhance P2P IoT learning.}

\subsubsection{\textcolor{black}{Network Slicing}} \textcolor{black}{5G/6G's network slicing capability allows for the creation of bespoke networks tailored to provide different levels of service for diverse types of devices. Such customization will permit high-priority IoT devices to learn from each other unimpeded by lower-priority devices, thereby ensuring the efficacy of the learning process. Wu et al. \cite{wu2022ai} discuss the use of AI-based solutions across the network slicing lifecycle, thereby allowing for intelligent management of network slices (termed as "AI for slicing"). Then, they explore network slicing solutions constructed to support emerging AI services, which involves creating AI instances and executing effective resource management (termed as "slicing for AI").}

\subsubsection{\textcolor{black}{Increased Speed and Capacity}} \textcolor{black}{At the forefront of 5G/6G's unique attributes is its significantly heightened bandwidth and data transfer rate, surpassing its predecessor (4G) by up to tenfold. 4G offers download speeds up to 1 Gbps, 5G offers download speeds up to 20 Gbps, and the still-conceptual 6G is expected to offer speeds up to 100 Gbps. This remarkable speed boost facilitates swift data exchange and processing, allowing IoT devices to communicate and learn from each other in real-time \cite{pei2023federated}.}

\subsubsection{\textcolor{black}{Ultra-Low Latency}} \textcolor{black}{Another distinct feature of 5G/6G is its ultra-low latency, which can be reduced to as low as one millisecond. This latency minimization is essential for IoT devices, permitting them to have rapid peer-to-peer learning \cite{liu2023machine}.}

\subsubsection{\textcolor{black}{Reliability and Availability}} \textcolor{black}{5G/6G will make strides in improving reliability and availability over its predecessors (i.e., 5G). IoT devices can thus rely on the network to maintain their connections and continue their learning processes without interruptions, assuring transparency in peer-to-peer learning \cite{ye2023spatial}.}

\subsubsection{\textcolor{black}{Edge Computing Support}} \textcolor{black}{5G has significantly enhanced edge computing, optimizing cloud computing systems by processing data close to the network's edge, where data generation occurs. This local data processing capability means IoT devices can learn from data on-site, reducing the need for long-distance data transmission and hastening peer-to-peer learning. Edge Intelligence uses AI methods, which are anticipated to be a crucial factor in \textcolor{black}{developing} future 6G networks, bolstering their performance, enabling new services, and introducing new functions \cite{peltonen20206g}.}

\subsubsection{\textcolor{black}{Energy minimization}} 
\textcolor{black}{Le \textit{et al.} \cite{le2022reconfigurable} apply reconfigurable intelligent surface (RIS)-aided wireless power transfer to improve battery life in federated learning (FL)-based wireless networks. Therefore, optimizing transmission time, power control, and the RIS's phase shifts show that the total transmit power can be minimized while meeting minimum harvested energy and transmission data rate requirements. Zarandi and Tabassum \cite{zarandi2021federated} present a federated deep reinforcement learning (DRL) framework using double deep Q-network (DDQN) to optimize multi-objective problems in IoT devices, minimizing task completion delay and energy consumption. The framework enhances scalability and privacy, and the results show faster learning speeds than federated DQN and non-federated DDQN. Zheng \textit{et al.} \cite{zheng2023drl} proposes an optimization strategy for wireless-powered multi-access edge computing networks to minimize total computation delay. Using a mixed integer programming model, the proposed strategy breaks down the problem into sub-parts for offloading decisions and power transfer duration optimization. In addition, the proposed strategy leverages deep reinforcement learning to handle time-varying channel conditions, which leads to near-minimal computation delays and low computational complexity.}

\begin{table*}[h!]
\centering
\setlength{\tabcolsep}{2.5pt}
\renewcommand{\arraystretch}{1}
\caption{{\color{black}Edge Learning-based Security and Privacy Solutions.}}
\centering
\label{tab:tabCyber}
\scriptsize
{\color{black}
\begin{tabular}{|p{0.5in}|p{0.2in}|p{0.6in}|p{0.66in}|p{0.66in}|p{0.4in}|p{0.6in}|p{1.4in}|p{1.4in}|}
\hline
\textbf{Framework} & \textbf{Year} & \textbf{Threat model} & \textbf{Mitigation solution} & \textbf{Learning mode} & \textbf{ML model} & \textbf{Datasets} & \textbf{Pros (+)} & \textbf{Open Issues (-)} \\ \hline
Truex \textit{et al.} \cite{truex2019hybrid} & 2019 & An honest-but-curious aggregator & Threshold Homomorphic Encryption & Federated Edge Learning & DT, CNN, SVM  &  Nursery Data Set  & + The proposed FL system remains resilient to an inference against potential adversaries  & - High computational and communication costs\\ \hline
Zhou \textit{et al.} \cite{zhou2020privacy} & 2020 & Collusion attacks & Paillier homomorphic encryption & Federated Edge Learning  & FNN  & Fashion-MNIST dataset  &  + The proposed scheme is formally verified to guarantee both data security and model security and resist collusion attacks & - The experiments are based only on a single dataset (Fashion-MNIST), which may not provide sufficient evidence of the scalability in other scenarios\\ \hline
Mothukuri \textit{et al.} \cite{mothukuri2021federated} & 2021 & IoT network attacks & Ensemble Learning & Decentralized Edge Learning & LSTM and GRU  & Modbus network data set  &  + The use of long short-term memory (LSTM) and gated recurrent units (GRUs) neural network models helps in achieving higher accuracy rates & - The experimental results presented in the article are based on a specific data set and may not necessarily generalize to other IoT networks or data sets\\ \hline
Cui \textit{et al.} \cite{cui2021security} & 2021 & DoS, U2R, R2L, and Prob attacks  &  Generative Adversarial Network & Decentralized Edge Learning &  CNN & KDD CUP 1999  & + Preserves the privacy of local model parameters while improving the utility of the anomaly detection model & - The proposed idea is evaluated with old dataset\\ \hline
Xu \textit{et al.} \cite{xu2022simple} & 2022 & N/A & Fixed-point quantization method & Federated Edge Learning & MLP  & MNIST dataset and CIFAR10 dataset  &  + Achieves higher accuracy and lower quantization error than other quantization methods & - The proposed idea is not evaluated with IoT dataset\\ \hline
Friha \textit{et al.} \cite{friha2022felids} & 2022 & IoT network attacks & Deep learning solution & Federated Edge Learning & CNN, RNN, DNN  &  CSE-CIC-IDS2018, MQTTset, and InSDN  &  + FELIDS can overcome the privacy issues associated with centralized machine learning models by using a federated learning approach & - Considers network traffic and does not take into account other factors such as device security or physical attacks\\ \hline
Gao \textit{et al.} \cite{gao2023sverifl} & 2023 & Malicious server may conduct dishonest data aggregation & Homomorphic encryption & Federated Edge Learning & CNN  & Fashion-MNIST dataset   & + It ensures the integrity of participant-uploaded parameters and the correctness of aggregated results from the server  & - The paper does not discuss the limitations or scalability of SVeriFL\\ \hline
Ouyang \textit{et al.} \cite{ouyang2023artificial} & 2023 & The privacy risks in the existing FL blockchain design patterns & Blockchain and Smart Contracts & N/A & N/A  &  N/A & + The framework is secure, private, and decentralized  & - Future research is needed to explore the scalability and performance of the framework in more complex scenarios\\ \hline
Friha \textit{et al.} \cite{friha20232df} & 2023 & Quantum-based Crypto-analysis Attacks  & A differentially private gradient exchange scheme & Decentralized Edge Learning &  DNN & Edge-IIoT dataset  & + The proposed system achieves a high level of performance, with comparable accuracy to the centralized learning approach  & The proposed system is vulnerable to adversarial attacks \\ \hline
Li \textit{et al.} \cite{li2023efficient} & 2023 &  Malicious local models & Dynamic Weighted Aggregation & Federated Edge Learning &   CNN &  CSE-CIC-IDS2018  &  + The client filtering and local model weighting strategy improves the global model's performance and reduces communication overhead & - Considers network traffic and does not take into account other factors such as device security or physical attacks\\ \hline
You \textit{et al.} \cite{you2023accuracy} & 2023 &  The vulnerability in the fairness of model reward & Differential privacy & Federated Edge Learning &  Pre-trained VGG &  CIFAR10, Rice, Fashion-MNIST, and CIFAR10 datasets & + The proposed FedAAC framework provides a scalable solution that can be extended to existing federated learning approaches  & The proposed system is vulnerable to adversarial attacks \\ \hline
Baucas \textit{et al.} \cite{baucas2023federated} & 2023 & Attacks against data privacy in wearable IoT devices & Blockchain Technology & Federated Edge Learning &  CNN & HAR dataset  &  +  The model accuracy shows the platform's ability to preserve the integrity of a predictive service & - The paper does not discuss the scalability of the proposed platform as well as the potential latency issues\\ \hline
Abou El Houda \textit{et al.} \cite{abou2023mitfed} & 2023 &  Network attacks  & SDN and Blockchain & Federated Edge Learning & CNN  & NSL-KDD dataset  & +  The paper shows that MiTFed achieves high accuracy and efficiency in detecting new and emerging security threats in both binary and multi-class classification  & - The dataset used in the performance evaluation is outdated\\ \hline
Chen \textit{et al.} \cite{chen2023ppt} & 2023 & Attacks in P2P networks & An enhanced Eschenauer-Gligor (E-G) scheme & Decentralized Edge Learning & CNN  & SMS Spam Collection  & + The proposed system can resist various security threats, preserve user privacy, and achieve better computation efficiency and prediction performance & - The proposed system is vulnerable to adversarial attacks\\ \hline
\end{tabular}\\
DT: Decision trees, CNN: Convolutional neural networks, SVM: Support Vector Machines, FCNN: Fully Connected Neural Network, LSTM: Long Short-Term Memory, GRUs: gated recurrent units, MLP: Multi-Layer Perceptron.}
\end{table*}

\begin{figure}
\centering
\includegraphics[width=0.5\textwidth]{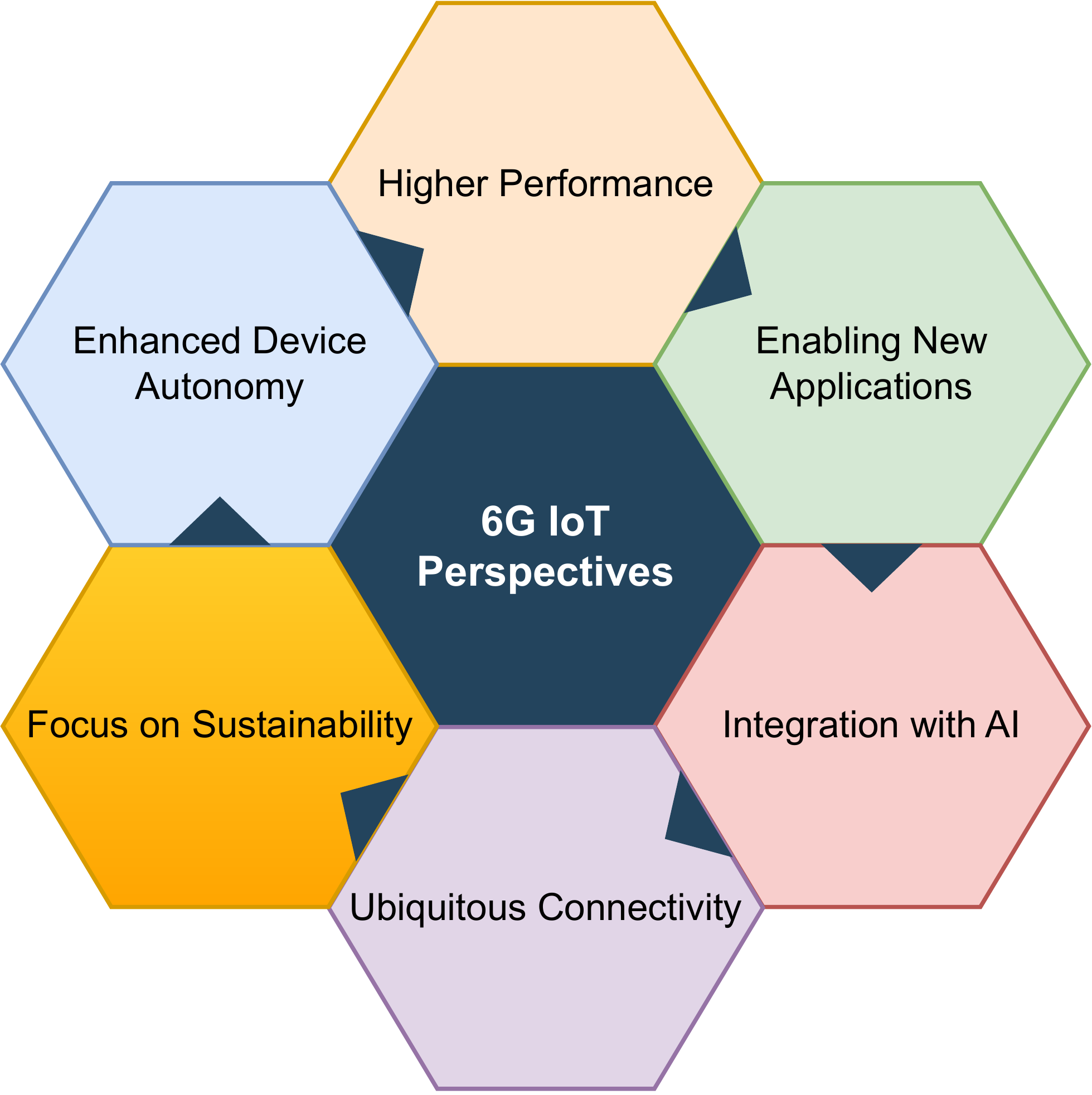}
\caption{\textcolor{black}{Perspectives on 6G IoT and Its Transformative Potential}}
\label{fig:Fignewr3}
\end{figure}

\subsection{{\color{black}Perspectives on 6G IoT}}
{\color{black}
The 6th generation (6G) of mobile networks is poised to bring transformative changes in IoT networks. As presented in Fig. \ref{fig:Fignewr3}, one of the significant perspectives is the possibility of achieving higher data rates, ultra-low latency, improved reliability, and massive connectivity, all of which will significantly enhance the performance of IoT applications \cite{wang2023road}. 6G IoT will enable new applications and services, from highly immersive augmented reality/virtual reality experiences to advanced Industry 4.0 applications, smart cities, and automated transportation systems \cite{chafii2023twelve}. The integration of ultra-dense networking and artificial intelligence will facilitate more intricate decision-making processes, which will set the stage for devices to function with an unprecedented level of autonomy \cite{zhu2023blockchain}.

Another critical perspective of 6G IoT is the realization of truly ubiquitous connectivity, reaching the most remote corners of the globe and even extending into space  \cite{issa2023blockchain}. Satellite networks, high-altitude platforms, and other novel communication paradigms will facilitate this ubiquitous coverage, ensuring no IoT device is left disconnected \cite{fang20215g}. Additionally, 6G is expected to champion the "green IoT" notion, focusing on energy efficiency and sustainability regarding device power consumption and network operations \cite{mao2021ai}. This will play a crucial role in the global efforts towards combating climate change and promoting sustainable growth. By leveraging the benefits of 6G, the IoT ecosystem could become more robust, intelligent, and comprehensive, leading to a more innovative and sustainable interconnected world \cite{qian2022distributed}. The possibility of edge computing combined with 6G's capabilities could mean that devices can process data and make decisions in real-time without necessarily needing to connect to a central hub or cloud \cite{yang2022federated}. 
}

\section{An Overview of ML-Associated and 6G-enabled IoT Security}\label{sec:2}

\textcolor{black}{In the context of ML-Associated and 6G-enabled IoT Security, a threat refers to potential danger or harm to the security of the system, while a vulnerability refers to a weakness or flaw in the system that can be exploited by a threat actor \cite{mitev2023physical}. Threats in ML-associated 6G-enabled IoT security inherit the threats associated with the previous generation communication networks including but not limited to data breaches, denial of service (DoS) attacks, unauthorized access to sensitive data, and tampering ML models \cite{fadlullah2022balancing}. On the other hand, vulnerabilities in ML-Associated and 6G-enabled IoT Security include outdated software, unsecured network connections, weak passwords, and insecure authentication mechanisms. These vulnerabilities can make it easier for a threat actor to attack the system \cite{sun2020machine}.}

\textcolor{black}{This section presents state-of-the-art reviews related to ML-related and 6G-enabled IoT security}. We classify them into two categories: 1) ML-based security for 6G-IoT systems, which incorporate studies on IoT security and the vision of future 6G networks, and 2) ML-associated threats, which provide an overview of works on threats facing ML-associated paradigms, namely centralized, FL and distributed learning. In Table \ref{tab:tabrelatedwork}, we provide a detailed comparison between our work and state-of-the-art studies.

\subsection{Machine Learning For 6G–IoT security}
ML is considered a key tool for robust security, especially for anomaly classification tasks. In this part, we focus on IoT and 6G from the perspective of ML-based security.

\subsubsection{\color{black}{Usage of ML Paradigms in 6G-enabled IoT Applications}}

\textcolor{black}{The table \ref{tab:tabpara} provides an overview of how the different machine learning paradigms can be used in 6G-enabled IoT applications. The different machine learning paradigms can be categorized into five paradigms, including, Supervised Learning, Unsupervised Learning, Reinforcement Learning, Semi-Supervised Learning, and Transfer Learning \cite{sun2020machine, kato2020ten, demirhan2023integrated}. Supervised learning is the most commonly used paradigm, where the model is trained on labeled data and learns to map inputs to outputs based on example pairs of input-output data. On the other hand, unsupervised learning involves training the model on unlabeled data and finding patterns and relationships in the data without explicit guidance. Semi-supervised learning combines both supervised and unsupervised learning, and the model is trained on labeled and unlabeled data to improve its performance. Reinforcement learning involves training the model to make decisions based on feedback from the environment and taking actions that maximize reward over time.}

\subsubsection{IoT-related security}
Hussain \textit{et al.} \cite{hussain2020machine} explored how ML and Deep Learning (DL) have impacted the loT ecosystem from a security and privacy perspective. The survey first presents the security and privacy concerns and challenges facing the IoT, including those obstacles related to the resource constraints associated with IoT devices and the attack vectors and security expectations. This is followed by highlighting various ML and DL mechanisms and their applicability to IoT security in various applications, such as forensic face recognition, cryptographic security character identification, and malicious code detection. In addition, the survey highlights the shortcomings of adopting ML techniques in IoT, including resource limitations, where IoT devices may not be suitable to support or execute sophisticated computational processing. Among other recommendations, the authors advocate that to overcome some of the limitations of ML approaches to IoT security, both DL and Deep Reinforcement Learning (DRL) theoretical frameworks should be further enhanced to allow adequate quantification of performances based on metrics such as computing complexity.   

In the same context, Zaman \textit{et al.} \cite{zaman2021security} provide a study regarding IoT security threats by layer and security schemes to address them. The layers involved are perception, network, transport, processing, and application, focusing on different IoT protocols related to each ecosystem layer. In response to the layer-based threats, rule-based (such as fuzzy logic) and AI-based layer response actions were presented, including ML (such as SVM) and DL (such as DNN and KNN) based systems, as well as performance evaluations of these AI-based layer-wise response actions.

In addition, Mohanta \textit{et al.} \cite{mohanta2020survey} suggest the use of emerging technologies such as AI, ML, and blockchain technologies to address existing security and privacy issues in IoT applications, such as jamming, DoS, and malicious nodes identification. The authors begin by highlighting layer-wise security issues in the IoT system and then answer the question of "How can these technologies be used to mitigate security threats in IoT?" in an overview. A study by Sarker \textit{et al.} \cite{sarker2022internet} offered a holistic view of IoT security intelligence, which is driven by ML and DL techniques that mine information out of raw data to smartly safeguard IoT systems against a range of sophisticated cyberattacks, including booting, sinkhole, cloud malware, access control attacks. Based on their study, the authors outline the corresponding future research directions and associated problems within the scope of their study. 
Xiao \textit{et al.} \cite{xiao2018iot} examined cyberattacks against IoT environments, such as eavesdropping, spoofing, and jamming. In addition, the authors identify several IoT-specific security techniques based on learning, such as malware identification, access control and secure offloading. The study discusses the challenges of state-of-the-art machine learning-based protection techniques, such as computing and communication overhead and partial state observation. Tahsien \textit{et al.} \cite{tahsien2020machine} discussed the layered IoT architecture. The importance of IoT security in terms of possible attacks under different types, such as physical and cyber attacks, attack surfaces including device perception and cloud applications, and the effects of such attacks including accessibility, integrity, and authorization are discussed in detail. In addition, prospective ML-based contributions to IoT security based on different ML and DL algorithms are presented.

\subsubsection{Vision For 6G networks}
In a prospective overview that outlines the principles of a 6G system, Saad \textit{et al.} \cite{saad2019vision} position 6G as a fundamental transformation of Self-Organizing Networks (SONs), whereby the network only adjusts its operations to particular environmental states, into a Self-Sustaining Network (SSN) capable of sustaining its Key Performance Indicators (KPIs) across the extraordinarily complex and highly dynamic operating environments arising from the rich application landscape of 6G. \textcolor{black}{The authors elaborate that AI}, specifically Reinforcement Learning (RL), addresses the goal of building SSNs that can independently sustain high KPIs and handle network resources, functionality, and oversight. Furthermore, in the same study, it is anticipated that the AI-based 6G features to be joined by a collaborative network intelligence located at the edge, resulting in a 6G system that can accommodate future services such as Massive URLLC, and may even be capable of replacing classical network frame structures. 

\textcolor{black}{Nguyen \textit{et al.} \cite{nguyen20216g} provides a review of the synergies between the 6G wireless communication networks and the IoT. It delves into the fundamental 6G technologies set to bolster IoT's future, such as edge intelligence, reconfigurable intelligent surfaces, multi-environmental communications, Terahertz communications, and blockchain. Distinctively, it offers an extensive exploration of 6G's potential impacts across five crucial IoT domains: healthcare, vehicular and autonomous driving, unmanned aerial vehicles, satellite-based IoT, and industrial IoT.}

Although the main constituents of the 6G architecture remain undefined and yet to be standardized, some aspects can be foreseen, and their associated threats are being discussed. For example, Siriwardhana \textit{et al.} \cite{siriwardhana2021ai} provided a future-oriented view of the immense role of AI in 6G network security, pinpointing upcoming research directions through its discussion of the AI-based security and privacy challenges along with some proposed possible solutions. The authors broadly categorize the 6G network threat landscape into two categories, namely, architectural and technological threats. The first entails attacks on the infrastructural level such as attacks on the User Plane Micro Services (UPMS), and the Control Plane Micro Services (CPMS). \textcolor{black}{In contrast,} the second entails attacks on embedded technologies such as attacks on ML, blockchain, cryptographic protocols, and Visible Light Communication (VLC). 

Additionally, the authors suggest several AI-based solutions that can address such threats, including edge-based FL for securing networks under massive data and device conditions. The limitations and threats of AI, including data injection/manipulation and logical corruption, were also discussed. In the same context, Nguyen \textit{et al.} \cite{nguyen2021security} discuss potential security and privacy issues around multiple levels of 6G, namely the physical, connection, and service levels. In addition to those vulnerabilities by inheriting from earlier communication technologies, 6G introduces additional threat engines from emerging radio technologies and attacks against pervasive intelligence. Predictions on protective measures against such threats are provided. These include AI-based security, Differential Privacy (DP), blockchain, real-time adaptive security, deep network slicing, and quantum-based cryptography.

\textcolor{black}{Vaezi \textit{et al.} \cite{vaezi2022cellular} presented the challenges and solutions associated with the proliferation of IoT networks, as devices ranging from sensors to robots increasingly interconnect. With the main objective of harnessing data insights for improved living standards and revenue generation, there is an emphasis on ensuring reliable, scalable, and timely connectivity across vast networks. The study offers an in-depth look at present and forthcoming communication methodologies tailored to IoT applications, notably within cellular and non-terrestrial realms. It sheds light on enhancements in wireless tech that bolster IoT connectivity in 5G and post-5G networks, especially in line with crucial 5G performance indicators such as energy efficiency, reliability, low latency, and connection density. \textcolor{black}{Azari \textit{et al.} \cite{azari2022evolution}  provided an in-depth survey of the evolution and increasing significance of Non-Terrestrial Networks (NTNs) in the context of 5G and the forthcoming 6G ecosystems. With technological advancements and cost-effective manufacturing, NTNs, when integrated with terrestrial networks (TNs), have broadened their scope of applications. The article delves into the features of this integration, exploring the new services, use cases, technological enablers, architectures, and higher-layer facets crucial for NTN assimilation.}}

\textcolor{black}{Guo \textit{et al.} \cite{guo2021enabling} addresses the increasing demands of contemporary IoT applications like augmented/virtual reality games, autonomous driving, and other intelligent technologies, which have outpaced the capabilities of the 5G networks. Recognizing the need to advance toward 6G networks to meet these demands, the article offers an extensive survey on 6G's potential to support a vast IoT ecosystem. The study outlines the drivers and limitations of 5G and the requirements of emerging IoT applications. It also presents visions for 6G, detailing its technical requirements, use cases, and trends. A novel 6G network architecture, termed space-air-ground-underwater/sea networks, enhanced by edge computing, is introduced. The paper additionally discusses innovative technologies like machine learning and blockchain in the context of 6G, elucidating their motivations, applications, and challenges.}

\subsection{Machine Learning Associated Threats}
Although ML-based solutions can provide reasonable defense capabilities, they are prone to various adversarial attacks. Motivated by the facts above, we present different studies on ML threats \textcolor{black}{concerning} three learning paradigms, as presented below:

\subsubsection{Centralized Edge Learning} Centralized learning is considered the first learning paradigm for ML, where data is gathered at one location for training. Despite its advantages, This approach also involves disadvantages and associated threats. For instance, Liu \textit{et al.} \cite{liu2018survey} review the security-related threats to ML and provide a systematic study on such threats in two dimensions: 1) the training stage and 2) the test/inference stage. Then, the authors classify existing protective techniques used for ML into four classes: assessment techniques, training phase precautions, inference-test stage defenses, and privacy and data protection precautions. In another study, Xue \textit{et al.} \cite{xue2020machine} present threat models where \textcolor{black}{adversaries target ML classifiers}. The analysis of the reasons behind the possibility of being attacked is presented. Therefore, security concerns are categorized under five categories: poisoning of the training set, backdooring through the training set, adversarial attacks, model theft, and inference attacks. In addition, \textcolor{black}{several} suggestions on ML-related security evaluations are also provided. 

Hu \textit{et al.} \cite{hu2021artificial} present the entire life-cycle of an AI-based operating environment as a roadmap to outline potential security-related threats occurring at each phase and subsequently elaborate on the corresponding countermeasures that can be taken. Oseni \textit{et al.} \cite{oseni2021security} and Liu \textit{et al.} \cite{Liu.comst.2021} focus on investigating adversarial attacks against AI-based systems, considering areas such as available methods for generating adversary samples. Oseni \textit{et al.} also expands on the mathematical engines of AI, particularly the emerging variants of reinforcement and federated learning \textcolor{black}{to} illustrate how vulnerabilities in AI models are exploited. In addition, the study considers several cyber defenses to help prevent AI systems against these types of threats, including data sanitization, robust statistics, defensive distillation, and gradient masking.

Kuzlu \textit{et al.} \cite{kuzlu2021role} provides a study in which concepts around IoT security are presented, \textcolor{black}{focusing} on attacks using and targeting AI. The authors \textcolor{black}{classify} AI attacks into three categories, namely 1) vulnerability scanning automation, which includes fuzzing and symbolic execution, 2) input attacks, and 3) data poisoning and fake data insertion, involving datasets and algorithm poisoning. Hao and Tao \cite{hao2022adversarial} review existing adversary evasion and poisoning attacks in smart grids. The types of adversarial examples are classified based on the threat aspect, including the attacker's influence, knowledge, specificity, computation and approach, and security breach. Multiple attack types per category are also provided, such as white box/blue box attacks, targeted/non-targeted attacks, and specific approaches such as gradient, decision, and transfer attacks. The authors propose six approaches for adequate checks to mitigate adversarial examples on image classification: gradient hiding, adjoint detection models, statistical methods, preprocessing methods, the ensemble of classifiers, and proximity metrics.

Chen \textit{et al.} \cite{chen2022machine} discussed the critical infrastructure frameworks monitored by the IoT and the associated security vulnerabilities with the main focus on Advanced persistent threat (APT) attack patterns. Specifically, the authors examine cutting-edge AI-based approaches to discover and successfully mitigate attacks on such networks. Among them, 14 AI-based approaches are selected according to their application frequency. Al-Rubaie \textit{et al.} \cite{al2019privacy} focused on different threats to ML privacy such as reconstruction, de-anonymization, and membership inference attacks. For privacy preservation, the authors consider multiple cryptographic techniques as a defense barrier to mitigate these attacks and preserve ML privacy in its different stages (i.e., data preparation, learning, and inference), including DP, homomorphic encryption (HE), garbled circuits, and secure processors.

\subsubsection{Federated Edge Learning} The introduction of FL provided a privacy-preserving paradigm for ML training. Although FL can ensure privacy preservation to some extent, there are FL-specific threats, and maintained privacy is only offered under certain circumstances. Mothukuri \textit{et al.} \cite{mothukuri2021survey} review specific privacy and security issues in federated learning that need to be addressed. \textcolor{black}{Their investigation suggests that}, overall, there are less significant specific privacy threats associated with federated learning than security threats. The authors state the highest priority security threats: communication blockages, poisoning, and backdoor attacks, whereas inference attacks pose one of the greatest threats to FL privacy. Liu \textit{et al.} \cite{liu2022threats} reviewed current threats and defense mechanisms in the FL domain across all stages of FL, namely data and behavioral auditing, training, and inference. \textcolor{black}{The authors discussed potential threats, related attacks, and available defenses for each stage}. Lyu \textit{et al.} \cite{lyu2020threats} review existing privacy threats and attacks in all categories of FL, namely Horizontal Federated Learning (HFL), Vertical Federated Learning (VFL), and Transfer Federated Learning (TFL). The authors also describe the privacy leakage issues in FL by pointing out that these issues can originate from the aggregator or individual participants. It is important to note that threats can be associated with insiders, outsiders, and malicious or semi-honest nodes. The attacks on FL are classified into two broad categories: 1) poisoning attacks and 2) inference attacks. \textcolor{black}{ Yang \textit{et al.} \cite{yang2019federated} present an overview of federated learning, a novel approach addressing challenges such as data privacy and security in AI development. The authors offer definitions, classifications, and potential applications of a secure federated learning framework, further discussing its successful application across different business domains. The authors advocate for a paradigm shift in AI, redirecting the focus from enhancing model performance (the predominant current focus) to exploring compliant data integration methods, thereby aligning with data privacy and security laws. Wahab \textit{et al.} \cite{wahab2021federated} provide a tutorial on Federated Learning and its related concepts, technologies, and learning approaches. The authors then categorize the literature into high-level challenges addressed by Federated Learning and further divide these into low-level challenges. This three-tier classification allows a deeper understanding of the topic and the methods to tackle particular problems. The paper also offers a set of desirable criteria and future research directions for each category of high-level challenges to assist the research community in designing innovative and efficient solutions.}

\textcolor{black}{Implementing blockchain as a ledger technology can help decentralize FL training without requiring a central server, improving security and scalability. This combination of FL and blockchain has led to the creation of a new paradigm called FLchain. FLchain potentially transforms Mobile-edge computing (MEC) networks into decentralized, secure, and privacy-enhancing systems. Zhu \textit{et al.} \cite{zhu2023blockchain} identify critical issues in FL that blockchain can address and categorizes existing system models into decoupled, coupled, and overlapped classes based on federated learning and blockchain integration. It compares the benefits and downsides of these models and investigates potential solutions to their limitations. Issa \textit{et al.} \cite{issa2023blockchain} proposes using blockchain technology and smart contracts to secure FL in IoT systems. It reviews blockchain-based FL methods and discusses current IoT security issues. The paper also covers IoT data analytics from a security perspective and the challenges and risks of integrating blockchain and FL in IoT. Open research questions are addressed, and a thorough literature review of blockchain-based FL approaches for IoT applications is provided. Nguyen \textit{et al.} \cite{nguyen2021federated} provides an overview of FLchain's fundamental concepts and explores its opportunities within MEC networks. Several challenges related to FLchain design are also identified, including communication cost, resource allocation, incentive mechanism, security, and privacy protection. Potential applications of FLchain in popular MEC domains such as edge data sharing, edge content caching, and edge crowdsensing are discussed. Therefore, Alazab \textit{et al.} \cite{alazab2021federated} present a review of various FL models developed to enhance authentication, privacy, trust management, and attack detection. The article also explores real-time use cases recently employing FL to preserve data privacy and enhance system performance. Ghimire and Rawat \cite{ghimire2022recent} survey the application of FL, a privacy-aware machine learning model, in enhancing the security of IoT systems. They compare centralized learning, on-site distributed learning, and FL, focusing primarily on security. The discussion also covers performance issues such as accuracy, latency, and resource constraints that might affect the overall functionality of IoT. In addition to evaluating current research efforts, challenges, and trends in the field, the authors consider future developments in this paradigm. The article provides readers with an in-depth understanding of FL's role in cybersecurity, outlining different security attacks and countermeasures.}

\textcolor{black}{Despite the potential of FL, non-IID (non-independent and identically distributed) data present on individual devices participating in the FL process pose significant issues. This statistical heterogeneity can hamper the model's performance and discourage user participation in FL. Ma \textit{et al.} \cite{ma2022state} discusses the challenges posed by non-IID data in the context of FL. The authors \textcolor{black}{review} the state-of-the-art solutions for non-IID problems, aiming to fill a gap in the literature and facilitate further implementation of FL. Boobalan \textit{et al.} \cite{boobalan2022fusion} propose an overview of the combination of FL and IIoT, addressing data privacy and on-device learning motivations, potential usage of machine learning, deep learning, and blockchain techniques for secure IIoT. It also explores the management of large and diverse data sets and discusses applications in industries like automotive, robotics, agriculture, energy, and healthcare. The upcoming deployment of billions of IoT devices, facilitated by faster Internet speeds from 5G/6G, will produce a massive amount of data, including potentially sensitive user information. This surge in data will escalate communication and storage costs and intensify privacy concerns within traditional, centralized cloud-learning systems for IoT platforms. \textcolor{black}{FL reduces costs and enhances user-level privacy by eliminating the need for data centralization. However, implementing FL in IoT networks is not without its challenges.} Zhang \textit{et al.} \cite{zhang2022federated} delves into the opportunities and hurdles of integrating FL in IoT platforms and its potential to enable a variety of IoT applications. It identifies and explores seven significant challenges of using FL in IoT platforms and underlines recent promising strategies for overcoming them.} In a study by Jere \textit{et al.} \cite{jere2020taxonomy}, the authors point out client-side attacks as the most significant threats to FL platforms since malicious FL clients are capable of tampering with and repositioning the model edges while developing it. In addition, the authors discuss different critical FL attacks, including Generative Adversarial Network (GAN) reconstruction, data poisoning, membership inference, and model inversion attacks. Some of the defense mechanisms reported in the paper include DP, robust aggregation, and outlier detection. The details of Federated Edge Learning are presented in Algorithm \ref{alg:alg1}.

\begin{figure*}
\centering
\includegraphics[width=0.8\textwidth]{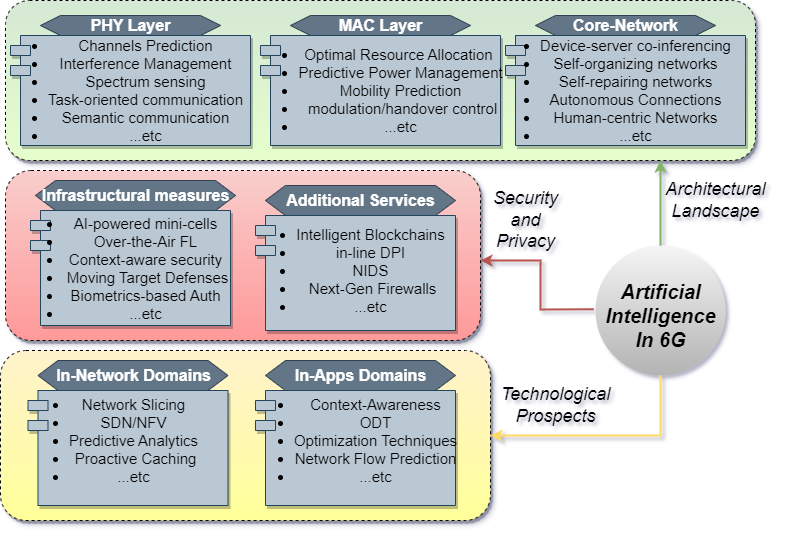}
\caption{\noteothmane{Predicted AI involvement in future 6G networks: 1) Architectural landscape: AI as an enabler of native intelligent functionality. 2) Security and privacy: AI as an embedded/additive defender; 3) Technology Prospects: AI as an enabling intelligence service for high-level layers.}}
\label{fig:fig2}
\end{figure*}

\subsubsection{\textcolor{black}{Distributed Learning}} This learning paradigm incorporates learning from multiple sources. Chen \textit{et al.} \cite{chen2021distributed} discussed a range of possible avenues for deploying a variety of distributed learning approaches on real-world wireless edge networks. The outlined approaches include FL, multi-agent RL, Federated Distillation (FD), and distributed inference. \textcolor{black}{The authors highlight the benefits and potential security and privacy issues these approaches may encounter, such as gradient leakage in FL}. In an attempt to answer the question of whether distributed learning is appropriate for wireless communications, Qian \textit{et al.} \cite{qian2022distributed} survey the state-of-the-art research on distributed learning for wireless communication, as well as the application cases, framework, algorithms, and other suitable alternatives for distributed learning. The research focuses on three layers: 1) physical, 2) medium access control (MAC), and 3) network layer. Furthermore, other emerging areas such as tensor and blockchain technologies are explored by reporting that these systems are prone to security attacks. 

Ma \textit{et al.} \cite{ma2022trusted} provide an information-exchange level security and privacy risk classification of distributed ML, which is organized according to the core phases of an ML workflow, namely preprocessing, learning, knowledge extraction, and result intermediation. The authors investigate and discuss possible risks involved in each level through an overview of current attack techniques, such as model poisoning/inversion attacks, inference attacks, label leakage, and data reconstruction attacks. The details of Distributed Edge Learning are presented in the Algorithm \ref{alg:alg2}. \textcolor{black}{When} a given client is involved in multiple FL sessions, it can communicate its real-time global status which is an array of every FL session status. The global status can be formalized as $GS_{C_{ID}} = [{S_{0}:Status}, ..,{S_{n}:Status}]$, where $C_{ID}$ presents the client ID, and $S_{ID}$ presents the FL session ID. The notations used in Algorithms \ref{alg:alg1} and \ref{alg:alg2} are presented in Table \ref{tab:notation}. 

\begin{table}
\centering
\caption{Notations used in Algorithms \ref{alg:alg1} and \ref{alg:alg2}}  
\label{tab:notation}
\setlength{\tabcolsep}{3pt}
\begin{tabular}{|p{1.2in}|p{1.8in}|} \hline
\textbf{Notation} & \textbf{Description}\\\hline
 $\eta$ &  Learning rate \\\hline
$Epo$ & Number of local epochs \\\hline
$Batch$ & Local minibatch size\\\hline
$K$ & Total clients number \\\hline
$State$ & State of the client \\\hline
$x$& Single sample\\\hline
$C$ & Fraction\\\hline
$R$ & Global rounds\\\hline
$S_{t}$ & A subset of selected clients\\\hline
$n$ & The number of local samples\\\hline
$f$& Model\\\hline
$\mathcal{P}$ & Pre-processed dataset\\\hline
$f_c(.)$& Loss function\\\hline
$InitializeModel(.)$& Random model initialization\\\hline
$ConnectedClients(.)$& Number of active clients required\\\hline
$RequestUpdate()$ & A function used to request the model updates \\\hline
$ReceiveUpdate()$& A function used to receive the model updates \\\hline 
$AggregateModels()$& A function used to aggregate  the model updates \\\hline
$UpdateModel()$& A function used to updates the model after the aggregation \\\hline
\end{tabular}
\end{table}

\begin{algorithm}[htp]
    \SetAlgoLined\DontPrintSemicolon
    \SetKwFunction{func}{}
    \SetKwFunction{proc}{}
    \textbf{Data}: $\eta$ , $Epo$, $Batch$, $K$, $k$. \\
    \SetKwProg{myfunc}{Edge Server}{}{}
    \myfunc{EdgeFedLearn \func{$K$, $C$, $R$}:}{
        $f_{1} \leftarrow InitializeModel()$ \;\\
        \For{$t = 1,..,R$}
        {
            $S_{t}$ $\leftarrow$ Subset(max($C\cdot K, 1$), $"random"$)\;\\
            \textbf{Parallel.}\For{$k$ $\in$ $S_{t}$ }
                {
                    $f_{t+1}^{k}$ $\leftarrow$ $ClientUpdate(f_{t}, k$)  
                }
            $f_{t+1}$ $\leftarrow$ $\sum_{k=1}^{K} \frac{n_{k}}{n} f_{t+1}^{k}$
        }
        Broadcast $f_{t+1}$ the updated model to clients
        }

    \setcounter{AlgoLine}{0}
    \SetKwProg{myproc}{IoT device}{}{}
    \myproc{ ClientUpdate \proc{$f$, $k$}:}{
    \nl $\mathcal{B}$ $\leftarrow$ Split($\mathcal{P}$, $Batch$) \;\\
    \nl \For{i = 1,..,$Epo$}
        {
            \For{$b$ $\in$ $\mathcal{B}$}
                {
                    $f$ $\leftarrow$ $f$ $-$ $ \eta	\nabla f_c(x, b)$ 
                }
        }
    Return $f$ to Edge Server
}
\caption{Federated Edge Learning}
\label{alg:alg1}
\end{algorithm} 

\begin{algorithm}[htp]
    \SetKwFunction{func}{}
    \SetKwFunction{proc}{}
    \textbf{Data}: $\eta$, $Epo$, $Batch$, $K$ , $k$, $State$\\
    $f_{1} \leftarrow InitializeModel()$ \;\\
    Broadcast $f_{1}$ the initial model to clients\\
    \While{$K$ $\neq$ length(ConnectedClients())}
            {
                continue
            }
    $RequestUpdate()$\\
    $ReceiveUpdate()$\\
    $AggregateModels()$ \Comment{State =1}\\
    $UpdateModel()$\Comment{State =2}\\
    \SetKwProg{myfunc}{$State = 1$}{}{}
    \myfunc{EdgeFedLearn \func{$K$, $C$, $R$}:}{
        \For{$t = 1,..,R$}
        {
            $S_{t}$ $\leftarrow$ Subset(max($C\cdot K, 1$), $"random"$)\;\\
            \textbf{Parallel.}\For{$k$ $\in$ $S_{t}$ }
                {
                    $f_{t+1}^{k}$ $\leftarrow$ $ClientUpdate(f_{t}, k$)  
                }
            $f_{t+1}$ $\leftarrow$ $\sum_{k=1}^{K} \frac{n_{k}}{n} f_{t+1}^{k}$
        }
        Broadcast $f_{t+1}$ the updated model to clients
        }

    \setcounter{AlgoLine}{0}
    \SetKwProg{myproc}{$State = 2$}{}{}
    \myproc{ClientUpdate \proc{$f$, $k$}:}{
    \nl $\mathcal{B}$ $\leftarrow$ Split($\mathcal{P}$, $Batch$) \;\\
    \nl \For{i = 1,..,$Epo$}
        {
            \For{$b$ $\in$ $\mathcal{B}$}
                {
                    $f$ $\leftarrow$ $f$ $-$ $ \eta	\nabla f_c(x, b)$ 
                }
        }
    Return $f$ to Client
}
\caption{Distributed Edge Learning}
\label{alg:alg2}
\end{algorithm} 

\begin{table*}[t!]
\centering
\setlength{\tabcolsep}{2.5pt}
\renewcommand{\arraystretch}{1}
\caption{{\color{black}Difference between Edge Learning in 4G IoT, Edge Learning in 5G IoT, and Edge Learning in 6G IoT.}}
\centering
\label{tab:cha6g}
{\color{black}
\begin{tabular}{|p{1in}|p{1.7in}|p{1.7in}|p{1.7in}|}
\hline
\textbf{Feature}              & \textbf{Edge Learning in 4G IoT} & \textbf{Edge Learning in 5G IoT} & \textbf{Edge Learning in 6G IoT} \\ \hline
Bandwidth                     & 10   Mbps                                    & 1   Gbps                                     & 10   Gbps                                    \\ \hline
Latency                       & 100   ms                                     & 1   ms                                       & 100   µs                                     \\ \hline
Data   Rate                   & 100   MBps                                   & 10   GBps                                    & 100   GBps                                   \\ \hline
Energy   Efficiency           & Medium                                       & High                                         & Very   High                                  \\ \hline
Network Density	& 1000-10,000 devices per square km	& 100,000-1,000,000 devices per square km	& 10 million+ devices per square km  \\ \hline
Network Connectivity	& Cellular network and Wi-Fi hotspots &	Cellular network and Wi-Fi hotspots, with an increased focus on dense urban areas	& Integrated satellite, airborne, and terrestrial networks \\ \hline
Spectrum Efficiency	& 2-3 bps/Hz &	10-20 bps/Hz	& 100 bps/Hz\\ \hline
Network Architecture	& Centralized	& Decentralized	& Autonomous\\ \hline
Applications &	Streaming, Video Conferencing, Gaming &	IoT, Autonomous Vehicles, Smart Cities	& IoT, Holographic Communications, Teleportation, Brain-Machine Interface \\ \hline
Interoperability              & Limited                                      & Improved                                     & Full   Interoperability                      \\ \hline
    Edge Computing Capability & Limited (Basic analytics, traditional ML) & Moderate (Advanced analytics, deep learning) & Advanced (Distributed computing, edge AI) \\ \hline
AI   Integration              & Limited                                      & Advanced                                     & Full   AI Integration                        \\ \hline
Virtual   Reality Integration & Limited                                      & Advanced                                     & Full   Virtual Reality Integration           \\ \hline
Autonomous   Devices          & Limited                                      & Advanced                                     & Full   Autonomous                            \\ \hline
Coverage & Limited (Urban areas) & Wide (Urban and rural areas) & Very wide \\ \hline
    Reliability & Medium (Packet loss, interference) & High (Low packet loss, low interference) & Ultra-high (No packet loss, no interference) \\ \hline
Semantic Communications &	Limited, with communication primarily based on IP addresses and port numbers	& Increased support for semantic communication through enhanced edge computing and network slicing	& Advanced support for semantic communication, including the integration of augmented reality and virtual reality \\ \hline
Holographic MIMO Surfaces  &	Not supported	  & Limited support for holographic MIMO  &	Full integration of holographic MIMO for enhanced communication and data transmission \\ \hline
Millimeter-Wave and Terahertz Bands	 & Not supported or limited support	 & Increased support for millimeter-wave and terahertz bands  &	Full integration of millimeter-wave and terahertz bands for enhanced communication and data transmission \\ \hline
Network Slicing	 & Not supported	 & Enhanced support for network slicing to provide customized services to different types of users	  & Full integration of network slicing for customized communication and data transmission services \\ \hline
Physical Layer	&  Orthogonal frequency division multiplexing (OFDM)	&  Orthogonal frequency division multiplexing (OFDM) with additional support for millimeter-wave and terahertz bands	&  Next-generation modulation and multiplexing techniques for enhanced data transmission and communication \\ \hline
Massive   MIMO                & Not Available                                    & Available                                    & Available                                    \\ \hline
uMUB                          & Not   Available                              & Available                                    & Available                                    \\ \hline
uHSLLC                        & Not   Available                              & Available                                    & Available                                    \\ \hline
Near Space Communications &	Not supported &	Not supported & Full integration for near space communication and data transmission \\ \hline
mMTC                          & Available                                    & Available                                    & Available                                    \\ \hline
uHDD                          & Not   Available                              & Not   Available                              & Available                                    \\ \hline
Key Technologies & LTE, WiMAX, VoLTE, MIMO & mmWave, Sub-6 GHz, Massive MIMO, Network Slicing, Edge Computing, IoT-optimized networks & Terahertz frequencies, quantum cryptography, AI-driven networks, holographic communication, 10x faster speeds
 \\     \hline
Use Cases & Smart homes, wearables, fleet management & Smart factories, autonomous vehicles, e-health & Smart cities, augmented reality, digital twins \\ \hline
\end{tabular}\\
Massive MIMO: Multiple Input Multiple Output, uMUB - Unlicensed Multiuser Beamforming, uHSLLC - Unlicensed High-Speed Low Latency Communication, mMTC - Massive Machine Type Communication, uHDD - Unlicensed High-Definition Driving Display.}
\end{table*}

\begin{table*}[]
\centering
\setlength{\tabcolsep}{2.5pt}
\renewcommand{\arraystretch}{1}
{\color{black}
\caption{\color{black}Comparison of Security Issues in 4G, 5G, and 6G IoT}
\centering
\label{tab:tabnewr31}
\begin{tabular}{|p{0.8in}|p{1.8in}|p{1.8in}|p{2in}|}
\hline
\textbf{Feature}   & \textbf{4G IoT}                                                                                                                        & \textbf{5G IoT}                                                                                                                                    & \textbf{6G IoT}                                                                                                                                            \\ \hline
Security Threats   & \begin{tabular}[c]{@{}l@{}}- Unauthorized access\\ - Device spoofing\\ - DoS attacks\\ - Privacy violations\\ - Man-in-the-middle attacks\end{tabular}               & \begin{tabular}[c]{@{}l@{}}- Network slicing vulnerabilities\\ - High-speed edge computing risks\\ - AI threats\\ - Privacy breaches\\ - IoT device heterogeneity\end{tabular}                                             & \begin{tabular}[c]{@{}l@{}}- Quantum threats\\ - Advanced AI threats\\ - Increased privacy threats due to more \\ data processing\\ - Distributed network attacks\\ - Autonomous system vulnerabilities\end{tabular}             \\ \hline
Security Solutions & \begin{tabular}[c]{@{}l@{}}- Authentication\\ - Encryption\\ - Firewalls\\ - Intrusion detection systems\\ - Network monitoring tools\end{tabular}                  & \begin{tabular}[c]{@{}l@{}}- Decentralized security\\ - On-device AI for real-time\\  threat detection\\ - Advanced encryption methods\\ - Security in network slicing\end{tabular}                                & \begin{tabular}[c]{@{}l@{}}- Quantum-resistant algorithms\\ - AI-based threat detection\\ - Robust privacy-preserving methods\\ - Decentralized and distributed security\\ - Threat intelligence systems\end{tabular}                 \\ \hline
Challenges         & \begin{tabular}[c]{@{}l@{}}- Bandwidth and latency limitations\\ - Less sophisticated encryption\\ - Centralized security\\ - Insufficient threat intelligence\end{tabular} & \begin{tabular}[c]{@{}l@{}}- Security in a distributed network\\ - Securing AI models\\ - Safeguarding high-speed data\\ - Complexity of network slices\\ - Interoperability issues\end{tabular}                & \begin{tabular}[c]{@{}l@{}}- Quantum resistance\\ - Privacy with high-volume data\\ - Low-latency and high-bandwidth \\ security requirements\\ - Ensuring security in autonomous systems\\ - Large-scale secure integration challenges\end{tabular} \\ \hline
Opportunities      & \begin{tabular}[c]{@{}l@{}}- Enhanced IoT connectivity \\ and data sharing \\ - Improved M2M communication\\ - Advanced cyber defense systems\end{tabular}                & \begin{tabular}[c]{@{}l@{}}- Data processing at the edge\\ - Reduced latency\\ - Improved efficiency, smart \\ and autonomous systems\\ - Network slicing\\ - Next-gen threat detection\end{tabular} & \begin{tabular}[c]{@{}l@{}}- Hyperconnectivity\\ - URLLC\\ - AI at the extreme edge\\ - Highly personalized services\\ - Quantum encryption\\ - Advanced AI-based security\end{tabular}                          \\ \hline
\end{tabular}
}
\end{table*}

\begin{figure}[t!]
\centering
\includegraphics[width=0.5\textwidth]{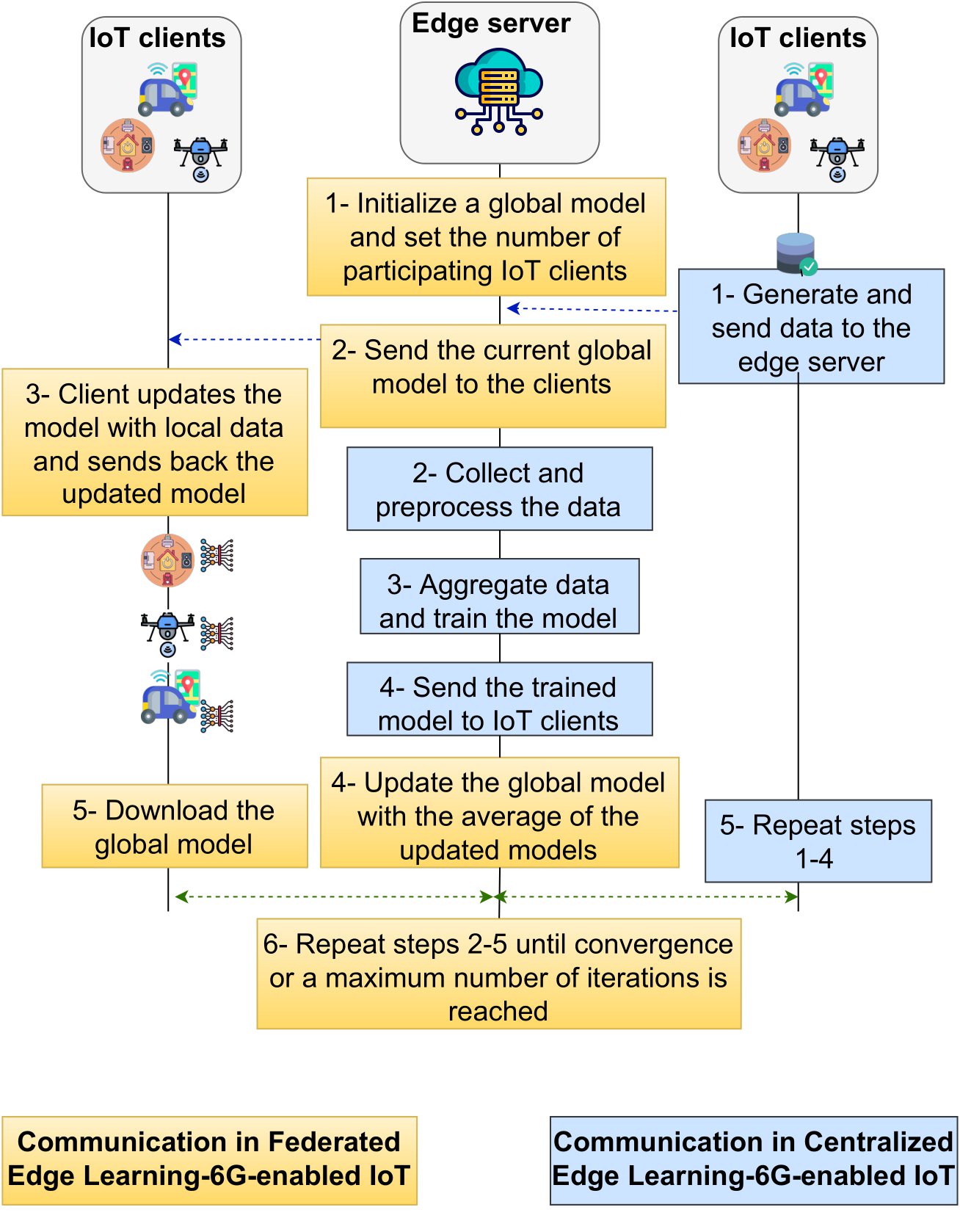}
\caption{\textcolor{black}{Network communication processes for Centralized Edge Learning-6G-enabled IoT and Federated Edge Learning-6G-enabled IoT.}}
\label{fig:figcomm}
\end{figure}

\subsection{\color{black}Comparative Analysis of Edge Learning and Security in 4G, 5G, and 6G-enabled IoT Environments}

\subsubsection{\textcolor{black}{Comparative Analysis of Edge Learning}}

Edge learning is a subset of edge computing, which aims to perform machine learning algorithms on data generated at the network's edge, i.e., closer to the source of data generation. \textcolor{black}{Figure \ref{fig:figcomm} illustrate the network architecture and communication processes for centralized edge learning-6G-enabled IoT and federated edge learning-6G-enabled IoT. Centralized edge learning and federated edge learning are two approaches to training machine learning models in 6G-enabled IoT, which differ in how they handle data. In centralized edge learning, all the data used to train the model is collected and stored in a central location, such as an edge server or cloud-based platform. The model is then trained based on this centralized dataset, which may be relatively large and diversified. On the other hand, federated edge learning refers to training machine learning models on distributed datasets across many devices. This approach can be beneficial when data privacy is a concern, as the data remains on the device and is not transmitted to an edge server. However, both approaches have advantages and disadvantages, and the suitable approach will depend on the specific needs and constraints of the current machine learning task for 6G-enabled IoTs~\cite{he2022acefl,ferrag2022edge}.}

\textcolor{black}{With the advent of IoT devices, there is an ever-increasing demand for real-time data processing and analysis, which is fueling the need for edge learning \cite{gong2022holographic}. Table \ref{tab:cha6g} compares the features of Edge Learning in 4G IoT, 5G IoT, and 6G IoT. We can observe a significant improvement in various aspects of each generation of technology. Moving from 4G to 6G, we observe significant improvements in latency, bandwidth, the number of devices supported, computational power, energy consumption, and cost per device. These improvements are expected to profoundly impact the applications of edge learning \cite{al2023edge,mao2023security,veith2023road}. For instance, the bandwidth, data rate, energy efficiency, and network density in 6G IoT are significantly higher compared to 4G IoT. Moreover, the latency in 6G IoT has improved to 100 microseconds compared to 100 milliseconds in 4G IoT. The level of AI and Virtual Reality integration in 6G IoT is also more advanced compared to 4G IoT. Furthermore, the level of autonomous devices has significantly improved in 6G IoT. Despite this, the availability of Massive MIMO remains constant throughout the three generations of the technology. However, other technologies like uMUB and uHDD are only available in 5G IoT and 6G IoT. The level of interoperability has also significantly improved in 5G IoT and 6G IoT.}

\textcolor{black}{In 4G-enabled IoT, the latency ranges from 50-1000 ms, which may not be suitable for applications that require real-time data processing. The bandwidth is limited to 1-100 Mbps, which may not be sufficient for handling large datasets. The number of devices supported is relatively small, which limits the scalability of the system \cite{alotaibi2023securing}. The computational power is also limited to less than 1 GFLOPS, which may not be enough for running complex machine learning algorithms.}

\textcolor{black}{In 5G-enabled IoT, the latency is significantly reduced to 1-10 ms, making it suitable for real-time applications. The bandwidth is also improved to 10-1000 Mbps, enabling the handling of larger datasets \cite{naser2022toward}.  The number of devices supported is increased to 10,000-1M, improving the system's scalability. The computational power is also increased to 1-10 GFLOPS, making it possible to run complex machine-learning algorithms~\cite{alexandropoulos2022time}.}

\textcolor{black}{In 6G-enabled IoT, the latency is reduced even further to less than 1 ms, enabling ultra-low latency applications. The bandwidth is also significantly increased to more than 1000 Mbps, enabling the handling of massive datasets \cite{wei2022multi}. The number of devices supported is increased to more than 1M, making it possible to handle large-scale IoT deployments. The computational power is also significantly increased to more than 10 GFLOPS, enabling the processing of complex machine learning algorithms. The energy consumption and cost per device also decrease significantly as we move from 4G to 6G, making it possible to deploy edge learning in resource-constrained environments \cite{chaccour2022seven}.}

\subsubsection{\textcolor{black}{Comparative Analysis of Security}}
\textcolor{black}{The progression of network generations from 4G to 5G, and now to 6G, has brought not only substantial improvements in terms of performance but also considerable changes in security requirements and mechanisms. While the core focus in 4G and 5G security has been \textcolor{black}{on protecting} user data and preventing attacks on the network infrastructure, 6G aims to take security to a new level of comprehensiveness and sophistication.}

\begin{figure}[t!]
\centering
\includegraphics[width=0.45\textwidth]{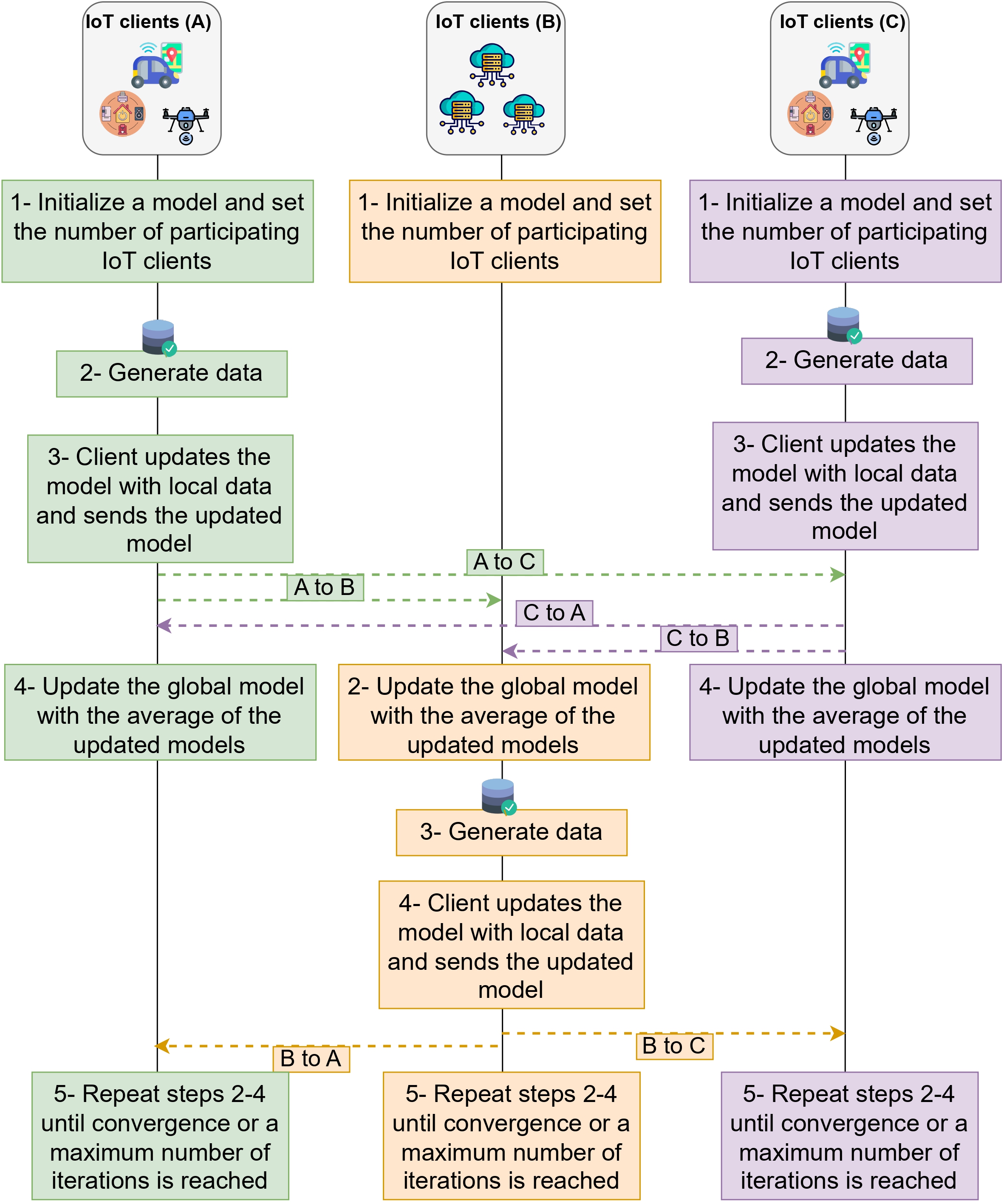}
\caption{\textcolor{black}{Network communication processes for Distributed Edge Learning-6G-enabled IoT.}}
\label{fig:figcommdist}
\end{figure}

\textcolor{black}{Table \ref{tab:tabnewr31} presents a comparative analysis of the security challenges and opportunities presented by 4G, 5G, and 6G in the context of IoT and edge learning. In the case of 4G IoT, the security threats are mainly centered around unauthorized access, device spoofing, DoS attacks, and privacy violations \cite{ferrag2018security}. Security solutions for these issues traditionally revolve around authentication, encryption, firewalls, and intrusion detection systems, with the main challenges being bandwidth and latency limitations, less sophisticated encryption methods, and a centralized security approach \cite{cao2013survey}.}

\textcolor{black}{As we transition into 5G IoT, edge learning becomes a pivotal part of the landscape. The complexity of network slicing, high-speed edge computing, and AI-powered threats contribute to the expanding threat horizon. Security solutions in 5G take on a more decentralized approach, utilizing on-device AI for real-time threat detection and mitigation \cite{boualouache2023survey}. However, managing security in a highly distributed network environment, safeguarding high-speed, high-volume data transmissions, and securing AI and machine learning models present significant challenges \cite{ramezanpour2023security,ghourab2023interplay}.}

\textcolor{black}{With the advent of 6G IoT, hyperconnectivity with ultra-reliable, low-latency communication and AI at the extreme edge offer highly personalized services based on user data. However, this level of connectivity also brings about advanced security threats, such as quantum computing threats and sophisticated AI-driven attacks \cite{mao2023security}. To counter these, the 6G IoT landscape will require quantum-resistant algorithms, advanced AI-based threat detection systems, and robust privacy-preserving methods \cite{ramezanpour2023security}. The primary challenges will revolve around maintaining privacy with high-volume data processing, ensuring quantum resistance, and addressing the extremely low-latency and high-bandwidth security requirements.
}

\subsection{\noteothmane{Practical Deployments of AI in Edge/Fog}}
\noteothmane{Although edge computing and fog computing are frequently used as synonyms, they are not identical concepts. Both emphasize the distribution of computing resources near the data's origin and usage; however, their methods differ. Edge computing involves processing data at or near the point of generation, like IoT devices, sensors, or other endpoints. In contrast, fog computing is a broader notion that expands the cloud computing paradigm to the network's periphery. This involves distributing computing, storage, and networking resources across multiple layers, from the edge devices to the cloud \cite{mukherjee2020security}. Effectively managing IT assets in cloud and fog/edge environments is a challenging task that necessitates a methodical decision-making process. The scarcity and heterogeneity of resources, in conjunction with the dynamic and diverse workloads and the unpredictable nature of advanced IT environments, have compounded the complexities of managing resources in such a landscape. Observing the infrastructure's behavior is crucial since comprehending the workload's behavior can mitigate the intricacy of the challenge and improve the outcome of a particular implementation \cite{iftikhar2022ai}. To overcome these challenges, adopting AI/ML-based solutions has gained traction, leveraging their ability to make sequential decisions and achieve optimal outcomes in this complex and ever-changing environment \cite{iftikhar2023hunterplus}. For example, in anticipating workload patterns or spatiotemporal impacts ahead of time for assisting in orchestrating resources \cite{liu2017adaptive}.}

AI-enabled fog and edge computing are being deployed in various industries, including healthcare, transportation, manufacturing, and retail. Edge AI technology has been developed and commercialized by various companies, and a wide range of industrial products are currently available or in development. Such implementations include edge AI chips, where companies such as NVIDIA \cite{NVIDIAEDGE}, Intel, and Qualcomm have developed specialized chips designed to run AI algorithms on edge devices. These chips are used in various applications, including autonomous vehicles, drones, and smart cameras. Also, Edge AI software platforms, where several companies, such as AWS IoT Greengrass, Google Cloud IoT Edge \cite{googleEdge}, and Microsoft Azure IoT Edge, have developed software platforms that enable developers to deploy AI models on edge devices. These platforms provide tools for building, training, and deploying machine learning models. In addition, edge AI cameras with built-in AI capabilities are becoming increasingly popular in industrial and commercial settings \cite{morishita2021cmos}. These cameras are used for facial recognition, object detection, and anomaly detection. Also, edge AI sensors, where IoT sensors with built-in AI capabilities are also becoming more common. These sensors are used in predictive maintenance, asset tracking, and environmental monitoring applications. Furthermore, Edge AI robots, where robots with onboard AI capabilities are being used in various industrial applications, including warehouse automation, manufacturing, and agriculture. In terms of business models, AI-enabled fog and edge computing are being deployed through various models, including software as a service (SaaS), platform as a service (PaaS), and infrastructure as a service (IaaS). For instance, in the healthcare industry, companies like Philips offer edge computing for remote patient monitoring \cite{Philips}. Similarly, in the manufacturing industry, companies such as GE offer predictive maintenance through their Predix platform \cite{GE}.

\noteothmane{Current research has shown that AI-enabled fog/edge computing can significantly reduce latency and energy consumption in various applications. In a recent paper by Hua \textit{et al.} \cite{hua2023edge}, the authors examined the mutually supportive feedback loop between AI and edge computing. On the one hand, the distributed nature of the edge/fog paradigm triggers fluctuating workloads for different edge devices depending on location and temporal conditions, making the implementation of edge computing very challenging due to unexpectedness and lack of certainty. In this case, AI optimization challenges are of significant value.}

\noteothmane{On the other hand, AI needs a fair amount of computing power and appropriate energy backing for learning. However, devices often fail to meet both of these requirements. One solution is to do all the heavy lifting in the cloud, which poses new challenges, including a shortage of bandwidth and elevated latency when dealing with a wide range of different AI models running on a wide range of endpoint devices. The emergence of edge/fog computing enables AI to be distributed close to end devices and users at the edge and endpoint with some processing and storage capabilities, addressing the demands for high network steadiness and minimal latency. In another study conducted by Joshi \textit{et al.} \cite{joshi2023enabling}, an emerging concept of edge intelligence, highlighted by the authors, was the so-called "all-in-edge" tier, in which the formation and inference of AI models are carried out only by the edge servers. Research on AI-driven optimization of edge/fog systems can be roughly classified into two major categories: optimization-related challenges (e.g., offloading, resource allocation \cite{cui2023deep}, energy consumption \cite{pan2023joint}) and privacy/security-related challenges \cite{hua2023edge}. For instance, DRL has been used \textcolor{black}{to learn the network dynamics in the work of effectively}. Cheng \textit{et al.} \cite{cheng2019space}. The authors proposed a DRL-based offloading method based on an embedded space-air-ground network to minimize energy and latency overhead.}


\begin{table*}[htbp]
  \centering
\setlength{\tabcolsep}{2.5pt}
\renewcommand{\arraystretch}{1}
\caption{{\color{black}A list of 5G/6G IoT testbeds.}}
\scriptsize
\centering
 \label{tab:tabnew4}
    {\color{black}
\begin{tabular}{|p{1in}|p{1.6in}|p{0.4in}|p{1.5in}|p{0.3in}|p{0.8in}|p{0.2in}|}
\hline
\textbf{Testbed}                                              & \textbf{Description}                                                                                                            & \textbf{Location} & \textbf{Features}                                                                             & \textbf{Focus} & \textbf{Funding}               & \textbf{Ref.} \\ \hline
5G-TRANSFORMER                                                & A European testbed for 5G-based IoT services and applications.                                                                  & Europe            & Network slicing, cloud computing, multi-access edge computing (MEC), and virtualization.      & 5G IoT         & EU-funded                      &    \cite{5gppp}      \\ \hline
5G-MoNArch                                                    & A mobile network architecture testbed for 5G IoT applications.                                                                  & Europe            & Millimeter Wave (mmWave) Technology, MEC, cloud computing, and virtualization.                                    & 5G IoT         & EU-funded                      &  \cite{l2}   \\ \hline
5G-PICTURE                                                    & A testbed for 5G-based IoT services and applications, including remote surgery and augmented reality.                           & Europe            & Network slicing, MEC, cloud computing, and virtualization.                                    & 5G IoT         & EU-funded                      &       \cite{l3}    \\ \hline
5G-CORAL                                                      & A testbed for 5G-based IoT services and applications, including smart transportation and energy management.                     & Europe            & Virtualization, cloud computing, MEC (\textcolor{black}{Multi-Access Edge Computing}), and network slicing.                                    & 5G IoT         & EU-funded                      &   \cite{l4}     \\ \hline
5G-EmPOWER                                                    & A testbed for 5G-based IoT services and applications, including smart cities and healthcare.                                    & Europe            & Network slicing, MEC, edge computing, and virtualization.                                    & 5G IoT         & EU-funded                      &    \cite{l5}       \\ \hline
6G Flagship                                                   & A Finnish research program focused on developing the technology and applications for 6G-based IoT.                              & Finland           & Terahertz and sub-terahertz communications, AI, and edge computing.                           & 6G IoT         & Industry and government-funded &    \cite{l6}      \\ \hline
5G Open Innovation Lab                                        & An open innovation platform for developing 5G-based IoT solutions.                                                              & USA               & Network slicing, MEC, cloud computing, and virtualization.                                    & 5G IoT         & Industry-funded                &     \cite{l7}    \\ \hline
ENCQOR 5G                                                     & A Canadian testbed for 5G-based IoT services and applications, including smart cities and healthcare.                           & Canada            & Massive MIMO, MEC, cloud computing, and virtualization.                                    & 5G IoT         & Government and industry-funded &     \cite{l8}     \\ \hline
5G City                                                       & A Danish testbed for 5G-based IoT services and applications, including smart transportation and energy management.              & Denmark           & Millimeter Wave (mmWave) Technology, MEC, cloud computing, and virtualization.                                    & 5G IoT         & Industry-funded                &    \cite{l9}    \\ \hline
5G Alliance for Connected Industries and Automation (5G-ACIA) & A German-led initiative focused on developing 5G-based IoT solutions for industrial automation and control.                     & Germany           & Network slicing, MEC, cloud computing, and virtualization.                                    & 5G IoT         & Industry-funded                &   \cite{l11}       \\ \hline
6g-platform                                      & The German platform for future communication technologies and 6G.              & Germany      & Terahertz and sub-terahertz communications, AI, and edge computing.                           & 6G IoT         & Government-funded              &     \cite{l12}      \\ \hline
\end{tabular}}
\end{table*}

\section{{\color{black}Edge Learning for Cyber Security} \label{sec:edge}}
{\color{black}

Edge learning, a variant of machine learning, leverages the power of edge computing—processing data close to its source, thereby reducing latency and bandwidth usage. When applied to cybersecurity, this technology provides swift, real-time detection and mitigation of cyber threats, thereby boosting the overall resilience of the systems. This section aims to discuss edge learning and its application in cybersecurity comprehensively.

\subsection{Differential privacy-based solutions}

Differential privacy-based solutions \cite{hassan2019differential} can protect sensitive data in the context of IoT security by adding random noise to the data, making it difficult for attackers to identify individual users or extract sensitive information. This technique can effectively prevent privacy breaches and unauthorized access to data, which are major concerns in IoT security. Moreover, differential privacy can help build trust between users and IoT systems by providing transparency and assurance that their data is safeguarded. 

Truex \textit{et al.} \cite{truex2019hybrid} presents a new approach to address the privacy and accuracy trade-offs associated with existing federated learning systems. The authors highlight the vulnerabilities of existing federated learning systems and propose a solution that combines differential privacy and secure multiparty computation. The authors' approach combines these two techniques that enable the reduction of noise injection as the number of parties increases without sacrificing privacy while maintaining a pre-defined rate of trust. The system is designed to withstand three potential adversaries: an honest but curious aggregator, colluding parties, and outsiders. The system assumes secure communication channels between each party and the aggregator and uses the threshold variant of the Paillier encryption scheme. This scheme ensures the privacy of individual messages sent to the aggregator, preventing any set of parties from decrypting ciphertexts. The proposed FL system remains resilient to an inference against potential adversaries who may be service users. The experimental results validate the system's effectiveness and superiority over state-of-the-art solutions, making it a scalable and reliable approach to federated learning. Zhou \textit{et al.} \cite{zhou2020privacy} proposes a privacy-preserving federated learning scheme in fog computing that enables fog nodes to collect Internet-of-Things (IoT) device data and complete the learning task. The proposed scheme addresses the issues of the uneven distribution of data and the large gap of computing power, which affects the efficiency of training and model accuracy in federated learning. The scheme leverages differential privacy to resist data attacks and uses a combination of blinding and Paillier homomorphic encryption to secure model parameters. The proposed scheme is formally verified to guarantee data security and model security, and resist collusion attacks. 

Friha \textit{et al.} \cite{friha20232df} proposed a decentralized, secure, and Differentially Private (DP) Federated Learning (FL)-based IDS (2DF-IDS) system for securing smart industrial facilities. The proposed system offers high performance in identifying different types of cyber attacks in an Industrial IoT system while mitigating the risks associated with conventional FL approaches. Additionally, the system offers improved overall performance compared to other competing FL-based IDS solutions, particularly under strict privacy settings. You \textit{et al.} \cite{you2023accuracy} proposed a Federated Adaptive Accuracy Controlling (FedAAC) framework, which dynamically controls the model accuracy to match the contribution of existing participating clients. To achieve this, an Accuracy Degrading algorithm is designed to obtain a decaying version of the accuracy of the global model by executing the accuracy degrading task specified by the server on the client side. To address the unbalance between the client and the central server for the model reward, assurance is introduced to ensure clients' contributions always match the model reward they receive. The differential privacy mechanism is also introduced into the FedAAC implementation, and the client-level differential privacy approach is improved for the scenarios targeted by FedAAC.

Chen \textit{et al.} \cite{chen2023ppt} proposes a decentralized global model training protocol, named PPT, that addresses security, privacy preservation, and robustness requirements in the context of Federated Learning in P2P networks. The paper proposes solutions to security threats and privacy preservation using the symmetric cryptosystem, secure key distribution, and random noise generation. PPT also adopts game theory to resist collusion attacks and has elaborate communication efficiency and dropout-robustness designs. Extensive experiments show that PPT outperforms Google's Secure Aggregation and LDP-based FL methods in computation efficiency and prediction performance, but the proposed system is vulnerable to adversarial attacks.

\subsection{Ensemble Learning-based solution}

Ensemble Learning \cite{mohammed2023comprehensive} is a machine learning technique that combines multiple individual models to create a more robust and accurate prediction model. When it comes to IoT security, Ensemble Learning can be a valuable tool to enhance the security of IoT devices and networks. Mothukuri \textit{et al.} \cite{mothukuri2021federated} proposed a decentralized, federated learning approach to enable anomaly detection on Internet of Things (IoT) networks. This approach enables on-device training, eliminating the need to transfer network data to a centralized server and ensuring optimal results in predicting intrusion in IoT networks. The proposed approach utilizes long short-term memory (LSTM) and gated recurrent units (GRUs) neural network models to train the machine learning model on the Modbus network data set. The article also provides an overview of LSTM and GRU networks, architecture, and real-time implementations. The proposed approach is beneficial in securing data privacy at end devices, achieving optimal results, and being computationally inexpensive.
 
\subsection{GAN-based solution}

Generative Adversarial Networks (GANs) \cite{dunmore2023generative} have been applied successfully in various fields, including computer vision and natural language processing. In recent years, researchers have started exploring their use in IoT security. One potential application of GAN-based solutions is to generate realistic attack scenarios and create diverse datasets for training security models. GANs can also generate synthetic data that mimics real-world sensor data. This can be used to detect anomalies and prevent security breaches. Furthermore, GANs can generate adversarial examples to test the robustness of IoT security systems. Cui \textit{et al.} \cite{cui2021security} presented an approach for decentralized and asynchronous Federated Learning (FL) for Internet of Things (IoT) anomaly detection. The proposed approach uses a modified Generative Adversarial Network (GAN) model called DP-GAN, which has an additional perceptron, the DP identifier (DPI), to generate differential noise, while approximating the raw data to a better degree. The proposed method also integrates blockchain to achieve global aggregation and improve system reliability. The paper demonstrates the effectiveness of the proposed approach in terms of accuracy, privacy protection, and efficient convergence, as shown by the experiments on benchmark datasets.

\subsection{Fixed-point quantization method-based solution}
\textcolor{black}{Fixed-point quantization is a method used in digital signal processing to reduce numerical data's storage requirements and computational complexity}. In IoT security, this technique can reduce the amount of data transmitted over the network, minimizing the risk of data leakage and unauthorized access. Xu \textit{et al.} \cite{xu2022simple} proposes Federated Learning with Minimum Square Quantization Error (FedMSQE), an approach to address the challenges of deploying popular neural network models on Internet of Medical Things (IoMT) devices. The paper focuses on the security of neural network model transmission and reducing the network scale. The proposed method uses fixed-point quantization with stochastic rounding to achieve a minor quantization error for each client in the FL setting. The authors demonstrate through numerical experiments that their proposed algorithm achieves higher accuracy and lower quantization error than other quantization methods and can be applied to any deep neural network on any single device. FedMSQE improves the security of FL and reduces the transmission cost while maintaining high accuracy.

\subsection{Deep Learning-based solution}

By analyzing large datasets of IoT device behavior and network traffic, deep learning algorithms can detect patterns and anomalies that may indicate malicious activity \cite{ferrag2020deep}. Friha \textit{et al.} \cite{friha2022felids} proposed FELIDS, a federated IDS based on deep learning for mitigating cyberattacks on agricultural IoT infrastructures. The authors identify gaps in the literature, including outdated or contextually inappropriate datasets, privacy issues for centralized models, and limited threat models that only address a subset of the attack vectors. FELIDS uses three deep learning classifiers - DNNs, CNNs, and RNNs - and evaluates their performance using three recent real-world traffic datasets. The paper provides a detailed performance evaluation and comparative analysis between FELIDS, the centralized machine learning model, and state-of-the-art works. The paper uses three recent real-world traffic datasets, namely: CSE-CIC-IDS2018, MQTTset, and InSDN, to evaluate the performance of each classifier.

\subsection{Homomorphic encryption-based solution}
Homomorphic encryption \cite{marcolla2022survey} can protect the algorithms and models used in IoT applications, preventing attackers from reverse engineering or tampering with these crucial components. Gao \textit{et al.} \cite{gao2023sverifl} proposed a Homomorphic encryption-based solution, named SVeriFL, that provides a solution to the privacy and security challenges in federated learning, ensuring data integrity and correctness of uploaded parameters and aggregated results. It also allows for consistency verification among multiple participants. The scheme employs a dynamic successive verification mechanism, running throughout the FL training process without impacting the original FL performance. The security analysis and experiments demonstrate the effectiveness of the proposed system with moderate computational cost.

\subsection{Blockchain and smart contracts-based solution}

\textcolor{black}{Integrating} blockchain and smart contracts can offer a robust and decentralized solution for IoT security  \cite{ferrag2021performance}. By leveraging blockchain technology, data transactions can be recorded and verified by a distributed network of nodes, ensuring that data integrity is maintained and tampering is prevented. Ouyang \textit{et al.} \cite{ouyang2023artificial}  introduced a privacy framework for FL based on blockchain and smart contracts. The framework offers complete privacy services for off-chain federations that want to use on-chain FL. The proposed framework is implemented and analyzed based on two prevalent blockchain projects, Ethereum and inter-plenary file systems (IPFS). The experiments show that the framework has acceptable collaboration costs and has advantages in privacy, security, and decentralization. Additionally, the framework can enable automatic on-chain identification and autonomous FL of machine clusters of IoT devices or distributed participants. Baucas \textit{et al.} \cite{baucas2023federated} proposes a fog-based IoT platform that uses federated learning and blockchain technology to address the challenges of preserving patient data privacy and improving the security of wearable IoT devices. The platform enforces decentralized servers and resource reallocation to enhance adaptability and sustainability. Federated learning \textcolor{black}{preserves} patient data privacy, while blockchain technology provides access control and a cryptographic structure to improve security. The platform's testbed effectively simulates and evaluates the proposed implementation, and the model accuracy shows the platform's ability to preserve the integrity of a predictive service.

\subsection{Dynamic Weighted Aggregation-based solution}

The dynamic Weighted Aggregation-based solution is a promising approach for addressing the security challenges associated with IoT. \textcolor{black}{This solution integrates multiple security measures and assigns weights to each mechanism based on their importance and effectiveness in a given context}. The weights are then dynamically adjusted based on the changing security environment and the behavior of IoT devices. Li \textit{et al.} \cite{li2023efficient} introduces DAFL, a dynamic weighted aggregation federated learning system for intrusion detection. The proposed approach reduces the effect of poorly performing local models on the global model during training, which improves the global model's performance in intrusion detection and reduces communication overhead. The paper presents the system design and key implementation details of DAFL and evaluates its performance against existing approaches. The experimental results demonstrate that DAFL achieves better detection performance with lower communication overhead. Still, the proposed system does not consider other factors such as device security or physical attacks.

\subsection{Software-defined networking}
Software-defined networking (SDN) can offer a robust solution for addressing the security risks associated with IoT \cite{rafique2020complementing}. SDN-based security solutions allow for fine-grained access control and enforcement policies, \textcolor{black}{protecting} IoT devices and networks from unauthorized access. Abou El Houda \textit{et al.} \cite{abou2023mitfed} proposes a framework called MiTFed that enables multiple SDN domains to collaboratively build a global intrusion detection model using FL, blockchain, and SDN technologies. The framework achieves high accuracy and efficiency in detecting new and emerging security threats while preserving the privacy of collaborators, making it a promising solution to address security threats in large-scale distributed networks. Further research is needed to assess the scalability of the proposed framework and address potential performance issues due to the use of blockchain technology. The dataset used in the performance evaluation is outdated.}


\begin{figure*}
\centering
\includegraphics[width=0.9\textwidth]{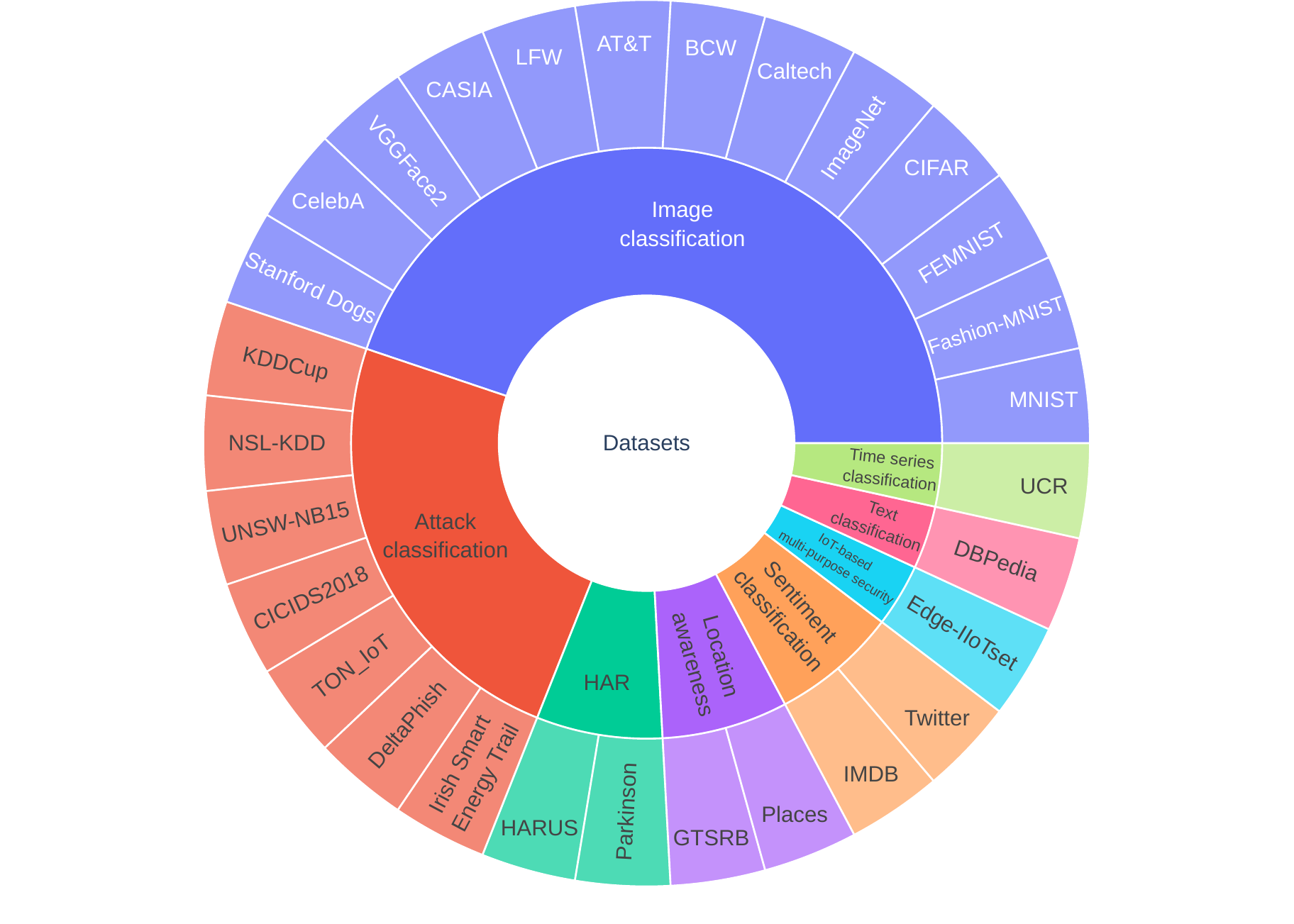}
\caption{Datasets used for machine learning security implementations and evaluations}
\label{fig:datasets}
\end{figure*}

\begin{table*}[h!]
\centering
\setlength{\tabcolsep}{2.5pt}
\renewcommand{\arraystretch}{1}
\caption{Datasets used as a benchmark experiment in the evaluation of Machine learning vulnerabilities}
\centering
\label{tab:tab2}
\begin{tabular}{|l|l|l|l|l|l|l|}
\hline
\textbf{Datasets}     & \textbf{Attack category} & \textbf{Attack type}           & \textbf{Machine learning} & \textbf{Learning mode} & \textbf{Targeted ML}       & \textbf{Systems}                           \\ \hline
\multirow{2}{*}{Edge-IIoTset dataset}	& Sybil attacks 	& Miscellaneous	& LPA & Centralized & Attack classification&	\cite{sadineni2022provnet}\\ \cline{2-7}
                                & Sybil attacks	& Miscellaneous & LightGBM, BiLSTM & Centralized & Attack classification&	\cite{ghourabi2022security}\\ \hline

Reuters-21578 dataset & Backdoor attack          & RNN backdoor attacks           & RNN  & Centralized                     & Text classification        & \cite{chen2021mitigating} \\ \hline
\multirow{2}{*}{KDDCup dataset}	& Sybil attacks	& Sybil-based poisoning attack	& SofT &	Distributed	& Attack classification&	\cite{fung2018mitigating}\\ \cline{2-7}
& Poisoning attacks	& Federated poisoning attack &	SqueezeNet&	Federated&	Attack classification&	\cite{li2021lomar}\\ \hline
\multirow{5}{*}{CIFAR-10 dataset}      & Backdoor attack          & Imperceptible backdoor pattern & DNN & Centralized                      & Image classification       & \cite{xiang2021reverse}   \\ \cline{2-7}
	& Poisoning attacks	&Clean-label poisoning attack&	ResNet and VGG-16&	Centralized	&Image classification&	\cite{zheng2021first}\\ \cline{2-7}
&	Poisoning attacks&	Clean-label poisoning attack&	Deep k-NN&	Transfer&	Image classification&	\cite{aghakhani2020bullseye} \\\cline{2-7}
& Poisoning attacks &	Federated poisoning attack &	ResNet &	Federated &	Image classification &	\cite{zhou2021deep} \\ \cline{2-7}
&Poisoning attacks&	Federated poisoning attack&	ResNet-18&	Federated&	Image classification&	\cite{zhang2020poisongan}\\ \hline
\multirow{6}{*}{MNIST dataset}         & Backdoor attack          & Distributed backdoor attack    & LeNet5 & Centralized                  & Image classification       & \cite{guo2021resisting}   \\ \cline{2-7}
&Poisoning attacks&	Poisonous label attack&	CNN&	Centralized&	Image classification&	\cite{liu2021poisonous}\\ \cline{2-7}
         & Poisoning attacks &	Federated poisoning attack   & MLP & Federated                  & Image classification       & \cite{uprety2021mitigating}   \\ \cline{2-7}
& Poisoning attacks	& Federated poisoning attack &	CNN	& Distributed	& Image classification&	\cite{cao2019understanding}\\ \cline{2-7}
& Sybil attacks	& Sybil-based poisoning attack	& SofT &	Distributed&	Image classification&	\cite{fung2018mitigating} \\ \cline{2-7}
&	Poisoning attacks &	Federated poisoning attack &	LetNet &	Federated &	Image classification &	\cite{zhou2021deep} \\ \hline
Caltech-101 dataset   & Synergetic attack        & Synergetic attack              & ResNet & Centralized                   & Image classification       & \cite{liu2021synergetic}  \\ \hline
UCR archive           & Adversarial attack       & blue-box adversarial attack   & FCN and ResNet  & Centralized&  Time series classification & \cite{yang2022tsadv}      \\ \hline
HARUS Dataset&	Poisoning attacks&	Federated poisoning attack&	FML	&Federated	&HAR	&\cite{sun2021data}\\ \hline
CASIA dataset	& Poisoning attacks	& Generative poisoning attack &	FaceNet	& Centralized	& Face recognition &	\cite{chen2021deeppoison}\\ \hline
TON\_IoT dataset & Poisoning attacks	& Federated poisoning attack	& KNN, LR, RF & Federated & Attack classification &	\cite{khan2022federated}\\ \hline
VGGFace2 dataset	&Sybil attacks	&Sybil-based poisoning attack	& SqueezeNet &	Distributed	& Face recognition&	\cite{fung2018mitigating}\\ \hline
LibriSpeech   dataset & Adversarial attack       & blue-box adversarial attack   & DNN                       & Centralized& Speech Recognition         & \cite{biolkova2022neural} \\ \hline
\multirow{4}{*}{Fashion-MNIST}
& Byzantine attacks&	Local model poisoning attack&	ResNet20&	Federated&	Image classification&	\cite{fang2020local}\\ \cline{2-7}
&Poisoning attacks&	Federated poisoning attack&	CNN&	Federated&	Image classification&	\cite{tolpegin2020data}\\ \cline{2-7}
& Sybil attacks	& Sybil-based collusion attack &	CNN &	Federated &	Image classification &	\cite{xiao2022sca} \\ \cline{2-7}
& Dropping Attack & Targeted dropping attack & CNN & Federated &  Image classification & \cite{severi2022network}  \\ \hline
 AG’s news dataset &	Adversarial attack &	White-box adversarial attack &	CharCNN-LSTM & Centralized&	Text classification	& \cite{ebrahimi2017hotflip} \\ \hline
NSL-KDD dataset	& Adversarial attack & White-box adversarial attack & DNN	& Centralized& Attack classification &	\cite{wang2022iwa} \\ \hline
Parkinson Dataset&	Poisoning attacks&	Federated poisoning attack&	FML	&Federated&	HAR&	\cite{sun2021data}\\ \hline
\multirow{2}{*}{FEMNIST dataset}	& Poisoning attacks &	Federated poisoning attack &	DNN	& Federated & Image classification &	\cite{tabatabai2022exploration} \\ \cline{2-7} 
& Poisoning attacks	& Federated poisoning attack &	N/A &	Federated &	Image classification &	\cite{shejwalkar2022back}\\ \hline
CICIDS2018 dataset &	Poisoning attacks	& Clean label attack &	 CNN-FL &	Federated &	Attack classification &	\cite{zhang2022secfednids} \\ \hline
\multirow{2}{*}{IMDB dataset} &	Poisoning attacks	& Federated poisoning attack	& SVM	& Federated & Sentiment classification &	\cite{doku2021mitigating}\\ \cline{2-7}
&	Poisoning attacks&	Label flipping attack&	BiLSTM &	Federated&	Sentiment classification&	\cite{jebreel2022defending}\\ \hline
AT\&T dataset	& Poisoning attacks &	Federated poisoning attack & CNN &	Federated &	Face recognition &	\cite{zhang2019poisoning}\\ \hline
GPS trajectory dataset&	Poisoning attacks&	Data aggregation attack&	N/A&	Federated&	Travel mode detection &\cite{zhao2021garbage}\\ \hline
Osteoporotic Fracture &	Poisoning attacks	&Data poisoning attack	&DNN&	Centralized&	Multivariate numerical&	\cite{wu2021indirect} \\ \hline
LFW dataset	& Poisoning attacks	& Generative poisoning attack &	FaceNet	& Centralized	& Face recognition &	\cite{chen2021deeppoison}\\ \hline
ImageNet10 dataset &	Backdoor attacks &	Clean label backdoor attack	& DNN &	Centralized &	Image classification &	\cite{ning2021invisible}\\ \hline
UNSW-NB15 dataset &	Poisoning attacks &	Label flipping attack &	 CNN-FL	& Federated	& Attack classification &	\cite{zhang2022secfednids} \\ \hline
Stanford Dogs dataset &	Poisoning attacks&	Data poisoning attack&	ResNet-13 &	Centralized	& Image classification &	\cite{wu2021indirect}\\ \hline
Irish Smart Energy Trail &	Poisoning attacks	& Data poisoning attack &	FFNN, GRUs, AEA	& Centralized &	Electricity theft detection &	\cite{takiddin2020robust}\\ \hline
Breast Cancer Wisconsin &	Byzantine attacks &	Local model poisoning attack &	ResNet20 &	Federated &	Image classification &	\cite{fang2020local} \\ \hline
GTSRB dataset & Combined attacks  & Trojaning attack & DNN & Centralized & Sign recognition &  \cite{qiu2021mt} \\ \hline
DBPedia-14 dataset & Dropping attacks & Targeted dropping attack & CNN & Federated &  Text classification & \cite{severi2022network}  \\ \hline
DeltaPhish dataset & Adversarial attacks & Grey-Box adversarial attack & 13 classifiers & Centralized & Phishing website detection & \cite{apruzzese2022mitigating} \\ \hline
\end{tabular}\\
LPA: Label Propagation, LightGBM: light gradient-boosting machine, biLSTM: Bidirectional LSTM, RNN: Recurrent Neural Network, DNN: Deep Neural Network, LeNet5: DNN architecture for recognizing the handwritten and machine-printed characters, ResNet: Residual neural network, FCN: Fully Convolutional Network, MLP: Multilayer Perceptron, CNN: Convolutional neural network, LSTM: Long short-term memory, I3D: inflated 3D convolutional network, CharCNN-LSTM:  Consists of a 2-layer stacked LSTM width 6 for temporal convolutions, SVM: Support Vector Machine, KNN:  k-nearest neighbor, LR: linear regression, RF: random forest, FML: Federated multitask learning, SofT: A single layer fully-connected softmax for classification, BiLSTM: Bidirectional Long/Short-Term Memory, VGG-16: Very deep convolutional networks for large-scale image recognition, FFNN: feed-forward neural networks, GRUs: gated recurrent units, AEA: deep auto-encoder with attention.
\end{table*}

\section{Datasets for \textcolor{black}{Edge Learning}}\label{sec:44}
Following upon a thorough analysis of recent contributions in the literature, we provide a detailed report on the datasets used by the scientific community for experimenting and evaluating ML techniques on cyber attacks, which is provided in Table \ref{tab:tab2}. Based on the content of each dataset, we classify them into the following nine main categories: \textcolor{black}{1) IoT-based multi-purpose security, 2) Attack classification, 3) Electricity theft detection, 4) Image classification, 5) Time series classification, 6) Human Activity Recognition, 7) Sentiment classification, 8) Location awareness, and 9) Text classification}, as illustrated in Figure \ref{fig:datasets}.

\subsection{IoT-based multi-purpose security}
In recent years, addressing IoT security has received much focus from academia, industry, and governments, mainly motivated by the expanding attack surface from existing threats across the IoT networks. The availability of sophisticated, easy-to-run, and easy-to-find exploit scripts and specialized utility \cite{ferrag2020deep}. IoT-based security defenses using AI (specifically anomaly-based approaches) have provided a good line of defense against these threats, and the efforts provided by the security community deserve mention \cite{ferrag2021federated}. However, given that one of the most \textcolor{black}{critical} factors in providing an effective final model is the data. This means that it is safe to say that the final model is well-founded if the data is well-grounded. Although different datasets have been proposed to design, train, and evaluate AI-based security mechanisms for IoT-related environments, the majority of them are either context-specific (e.g., a specific protocol), constrained threat model (limited types and quantities of attacks), or collected within a shallow generation framework (either real or virtual) \cite{ferrag2022edge}. Others, however, have been designed as IoT-based multi-purpose security-related datasets such as the Edge-IIoTset dataset. 
\begin{itemize}
    \item \textit{Edge-IIoTset dataset \cite{ferrag2022edge}:} is a comprehensive and realistic dataset collected within multiple IoT/IIoT environments, intended to be used by ML-based cybersecurity models for training and assessment under both training approaches (centralized and FL). Data are collected from a variety of sources, including notifications, logs, and network traffic. A total of 61 features were selected from 1176 encountered characteristics. In particular, the dataset is produced with a designed IoT/IIoT test bed featuring a broad spectrum of interconnected devices, sensors/actuators, protocols, and multi-layer implementations. The generation framework is designed in a multi-layer approach comprising seven layers, including 1) the IoT/IIoT perception layer where there exist more than 10 types of devices including flame sensors, ultrasonic sensors, and water level detection sensors, 2) the edge layer where various brokers and controllers are implemented 3) the fog layer where various platforms are installed, such as IoT hubs and digital twins 4) the SDN layer 5) the NFV layer 6) the blockchain layer, and 7) the application layer. In addition to general-purpose protocols such as HTTP, the dataset includes a range of IoT and IIoT-specific protocols such as MQTT, CoAP, and Modbus. The dataset includes 14 attack classes: DoS/DDoS, man-in-the-middle, malware, and injection attacks. ProvNet-IoT proposed in \cite{sadineni2022provnet}, is a forensic provenance-based scheme for examining IoT-targeted network-level assaults. The system applies progressive provenance within the network to illustrate various events within different attack strategies and generate forensic proofs. The authors employed various community detection algorithms including Label Propagation (LPA). Experimental results on the Edge-IIoTset dataset assessed and demonstrated the potential of ProvNet-IoT to recognize specific selective artifacts in order to generate credible proof in forensic examinations. In the same context (i.e., IoT security), but in a different space, namely, smart health system security, Ghourabi \textit{et al.} \cite{ghourabi2022security} proposed a two-tiered security system. The first is an IDS for smart medical devices, including the Internet of Medical Things (IoMT), and the second is a malware detection system for general-purpose devices. The proposed system is built \textcolor{black}{using} an optimized LightGBM and a transformer-based model. Evaluations under the Edge-IIoTset reached accuracies as high as 99\%.
\end{itemize}

\subsection{Attack classification}
ML techniques provide valuable tools for cybersecurity solutions to detect intrusions, identify malware and implement mitigation schemes. Various datasets have been proposed to assist researchers in designing, developing, and evaluating their cybersecurity defense methods, including:
\begin{itemize}
    \item \textit{KDDCup dataset \cite{KDDCup}:} since 1999, it has been the most widely used dataset for assessing anomaly detection systems during that decade. It was built on the foundation of roughly $4$ GB of tcpdump data from seven weeks of actual network traffic (collected for the DARPA'98 IDS evaluation program). Within the training dataset, there are about 4.9 million individual connection vectors, with each holding 41 features with labels of either normal or attack. The attack types include DoS, User to Root (U2R), Remote to Local (R2L), and probing attack. Although this dataset is considered outdated, various modern studies continue using it. For example, the work of Fung \textit{et al.} \cite{fung2018mitigating} proposed a defense mechanism against Sybil-based poisoning attacks called FoolsGold, which is used for securing FL-based learning. The proposed system places no restriction on the number of expected attackers, nor does it necessitate additional information beyond the learning process while making minimal assumptions regarding the clients and their data. For the KDDCup dataset, the authors trained a single layer fully-connected softmax for the classification task, using 5 clients performing 5 attacks on the same target class. In the same context, Li \textit{et al.} \cite{li2021lomar} proposed LoMar, a Local Malicious Factor defense algorithm for addressing poisoning attacks on FL. The proposed algorithm is based two main phases. The first phase rates each remote client's model updates by scaling relative distribution across their neighbors using a kernel density estimation technique. In the second phase, a statistically optimal trigger threshold is derived to differentiate between malicious and safe updates. For the KDD Cup dataset, the authors use 2 classes with over two hundred thousand data samples as their source and target label, while setting up three malicious clients with flipped samples. The target label accuracy reported is 99.1\%. 
    \item \textit{NSL-KDD dataset \cite{tavallaee2009detailed}}: is a proposed dataset to overcome specific issues associated with the KDDCup dataset, including failure to define the attacks clearly. It consists of two main parts: 1) KDDTrain+ with 125,973 records and 2) KDDTest+ with 22,544 records, which are created from the KDDCup dataset, with four main categories of attacks, namely R2L, Prob, DoS, and U2R. A recent work by Wang \textit{et al.} \cite{wang2022iwa} proposed IFPA and IUA, two integration-based adversarial generation techniques. These techniques have the potential to conduct DNN-targeted white-box attacks. The IFPA is designed for cases where a specific number of points are to be disturbed, while the IUA is designed for cases where no perturbation point number is requested. The performance of the NSL-KDD dataset achieves an overall acceptable accuracy of 78.96\% on the test set.  
    \item \textit{UNSW-NB15 dataset \cite{moustafa2015unsw}:} developed in the Cyber Range Lab at the Australian Cyber Security Centre (ACCS) as a hybrid of real-world and synthetic attack models. It holds over 2.5 million data instances with 49 features mined from 100 GB of raw traffic. \textcolor{black}{It includes nine attack types}: DoS, shellcode, reconnaissance, generic, exploits, and worms. The work by Zhang \textit{et al.} \cite{zhang2022secfednids} proposed \textit{SecFedNIDS}, a poisoning attacks-robust FL-based NIDS. To effectively shadow poisonous traffic data and avoid its participation in further local training, SecFedNIDS employs a novel detection method for poisonous data based on the similarity of class paths. The authors followed a layer-wise suitability propagation technique to retrieve the classpath from the clean traffic data and transmit it to poisoned clients to assist in discriminating the poisoned data. The results showed that SecFedNIDS increases accuracy in case of poisoning attacks on the UNSW-NB15 dataset by up to 48\%.
    
   \item \textit{CICIDS2018 dataset \cite{sharafaldin2018toward}:} generated by the Canadian Institute for cyber security \cite{sharafaldin2018toward} using the profiles notion containing comprehensive descriptions of attacks and detailed abstract distribution models for applications, protocols, or lower-level network entities. In addition to the Benin profile, the dataset includes different attack types such as DoS/DDoS, botnet, infiltration, web, and brute-force attacks. This dataset has been widely used for evaluating Host and network IDSs in recent years.
  
    \item \textit{TON\_IoT dataset \cite{toniot}:} generated by collecting and analyzing heterogeneous data collections from IoT and IIoT domains. The experiment setup to generate this dataset uses several VMs with different operating systems to support interconnectivity between IIoT, cloud, and edge/fog layers. Different attack types can be found in this dataset including DoS/DDoS, ransomware, and web apps attacks, targeting IoT gateways and systems across the IoT/IIoT domain. Khan \textit{et al.} \cite{khan2022federated} proposed \textit{DepoisoningFSL}, a data poisoning defense federated split learning for edge computing. When the proposed system is compared to the KNN-based semi-supervised defense (KSSD) mechanism with different data poisoning rates ( up to 25\%), it presents an overall increased accuracy.
    
    \item \textit{DeltaPhish dataset \cite{corona2017deltaphish}:} was gathered from active phishing URLs retrieved online from the PhishTank feed from Oct 2015 to Jan 2016. The authors gathered and verified more than 1000 phishing pages manually. Within each phishing page, they subsequently harvested the related homepage of the hosting domain. Three to five legitimate pages are collected and manually approved using hyperlink analysis in the HTML code. 5,511 distinct web pages are included, out of which 1,012 are phishing pages. The authors in \cite{apruzzese2022mitigating} proposed a collection of Grey-box attacks on phishing detectors that a phishing adversary can use. The attacks vary according to the adversary who knows the specific phishing detector. In addition, the authors introduced an associated defense algorithm named Protective Operation Chain (POC), which is based on a mix of random feature picking. The feature linkages are exploited to minimize the attacker's guess of the target's phishing detector. Experiment evaluations on different public datasets including the DeltaPhish dataset showed that the proposed algorithm is robust to attacks on 13 different classifiers.
\end{itemize}

\subsection{Electricity theft detection}
Electricity theft is a significant concern for utility systems as it causes high economic losses. Furthermore, shifting from control to smart devices may introduce highly sophisticated attacks that are harder to defend against. This requires AI-based defense mechanisms to mitigate the risks, which in turn requires Benin profile data to differentiate the pattern of attacks. Examples of such datasets include:
\begin{itemize}
    \item \textit{Irish Smart Energy Trail:} collected by the Sustainable Energy Authority of Ireland, where the electricity records originate from 3,000 residential units' smart meters that performed 30-minute readings for 18 months. As a result, there are 25,000 reports per customer. This dataset is used by Takiddin \textit{et al.} \cite{takiddin2020robust} as a benign profile dataset to quantify the effect of data poisoning attacks on smart grids. The authors evaluated the capabilities of customer-specific and generalized baseline detectors for detecting data poisoning and power theft attacks.
    \end{itemize}

\subsection{Image classification}
Image classification represents a mighty task for assessing the architectures and modern methodologies within computer vision. Image classification is a core discipline concerned with understanding what an image looks like in its entirety, and the objective is to classify the given image with a specific category. Various datasets have been proposed to evaluate such security architectures and systems, including:

\begin{itemize}
    \item \textit{MNIST dataset \cite{MNIST}:} built from the NIST Special Databases (1 and 3), and containing binary images of numbers and handwritten characters. This dataset is widely used for training various AI-based image processing systems and can be considered the default dataset for image classification tasks around the globe. A vast amount of hand-written data is embedded in the dataset with over 70,000 images of 0 to 9 hand-written digits, which have been standardized in size and focused in a pixel-square grid. Within each image is an array (28 × 28 × 1) of varying floating-point numbers that depict grayscale shades of intensity. This dataset has been used widely for evaluating different edge-located security schemes in recent works \cite{guo2021resisting, liu2021poisonous,uprety2021mitigating,cao2019understanding,fung2018mitigating,zhou2021deep}.
    \item \textit{Fashion-MNIST	dataset \cite{xiao2017fashion}:} Based on the broad success that the MNIST dataset has achieved,  various works have attempted to take a similar approach. For example, Fashion-MNIST is designed to become a full-scale substitute for the original MNIST dataset when evaluating ML-based algorithms. The dataset shares the identical image size (70,000), the same data format (28x28 Grey-scale imagery), and the identical training (60,000 images) and testing (10,000 images) setup. However, instead of handwritten digits, Fashion-MNIST comprises fashion items that fall under ten categories, each with 7,000 images per category. This dataset is also used to evaluate edge-located security schemes such as in \cite{fang2020local,tolpegin2020data,xiao2022sca,severi2022network}.
    \item \textit{FEMNIST dataset \cite{caldas2018leaf}:} is a federated version of the MNIST dataset containing 341873 samples from over 3300 writers as FL clients. With distinguishable writing styles, individual authors' data is Non-Independent and Identically Distributed (Non-IID). Each client in the FEMNIST dataset has approximately 100 samples with 10 classes. This dataset has been used to evaluate edge-enabled FL-based security systems recently \cite{tabatabai2022exploration,shejwalkar2022back}.
    \item\textit{CIFAR-10 and CIFAR-100 datasets \cite{krizhevsky2009learning}:} represent labeled subsets from the 80 million tiny images dataset. The CIFAR-10 dataset comprises 60,000 color images of 32x32 split into 10 classes, each with 6,000 images per class. It includes 50,000 training and 10,000 test images. The dataset is partitioned into 5 training batches and 1 test batch, each containing 10000 images. The testing batch contains 1000 images randomly picked from every class. Each training batch contains the remaining images randomly, although some may have more images within 1 class than others. In addition, the training batches hold 5000 images from every class. The work by Xiang \textit{et al.} \cite{xiang2021reverse} used this dataset to benchmark their proposed backdoor attacks defense mechanism.
    \item \textit{ImageNet10 dataset \cite{russakovsky2015imagenet}:} is a hand-picked 10-class dataset derived from the ImageNet dataset, which provides high-resolution images divided into 1000 classes and formed only from the 320x320 core regions of the original images. The ImageNet10 classes were chosen to be optically distinct and to reflect natural objects. This dataset is used in different works for benchmarking and other datasets such as MNIST and CIFAR. \cite{ning2021invisible}
    \item \textit{Caltech-101 dataset:} is regarded to be a challenging dataset for different works in the experimental parts \cite{liu2021synergetic}. The dataset includes 101 classes and 1 background scene class. Every class contains between 40 and 800 images and there are a total of 9146 images in this dataset.
    \item \textit{Stanford Dogs dataset \cite{khosla2011novel}:} It was constructed from imagery and annotations from ImageNet for categorizing fine-grained images. The dataset features a collection of images representing 120 breeds of dogs worldwide. This dataset includes 20,580 images with 120 categories. This dataset can be used with other datasets such as MNIST to benchmark edge learning systems \cite{wu2021indirect}.
    \item \textit{Breast Cancer Wisconsin dataset \cite{wolberg2011uci}:} was developed to identify whether a person has breast cancer. It contains 569 examples (Benign: 357 and Malignant: 212), each having 30 features for the person's cell nuclei characteristics. This dataset was used by \cite{fang2020local} to evaluate ML-targeted data poisoning attacks, which can pose a great danger to human lives in smart healthcare applications. 
    \end{itemize}

\begin{figure*} 
\centering
\includegraphics[width=0.8\textwidth]{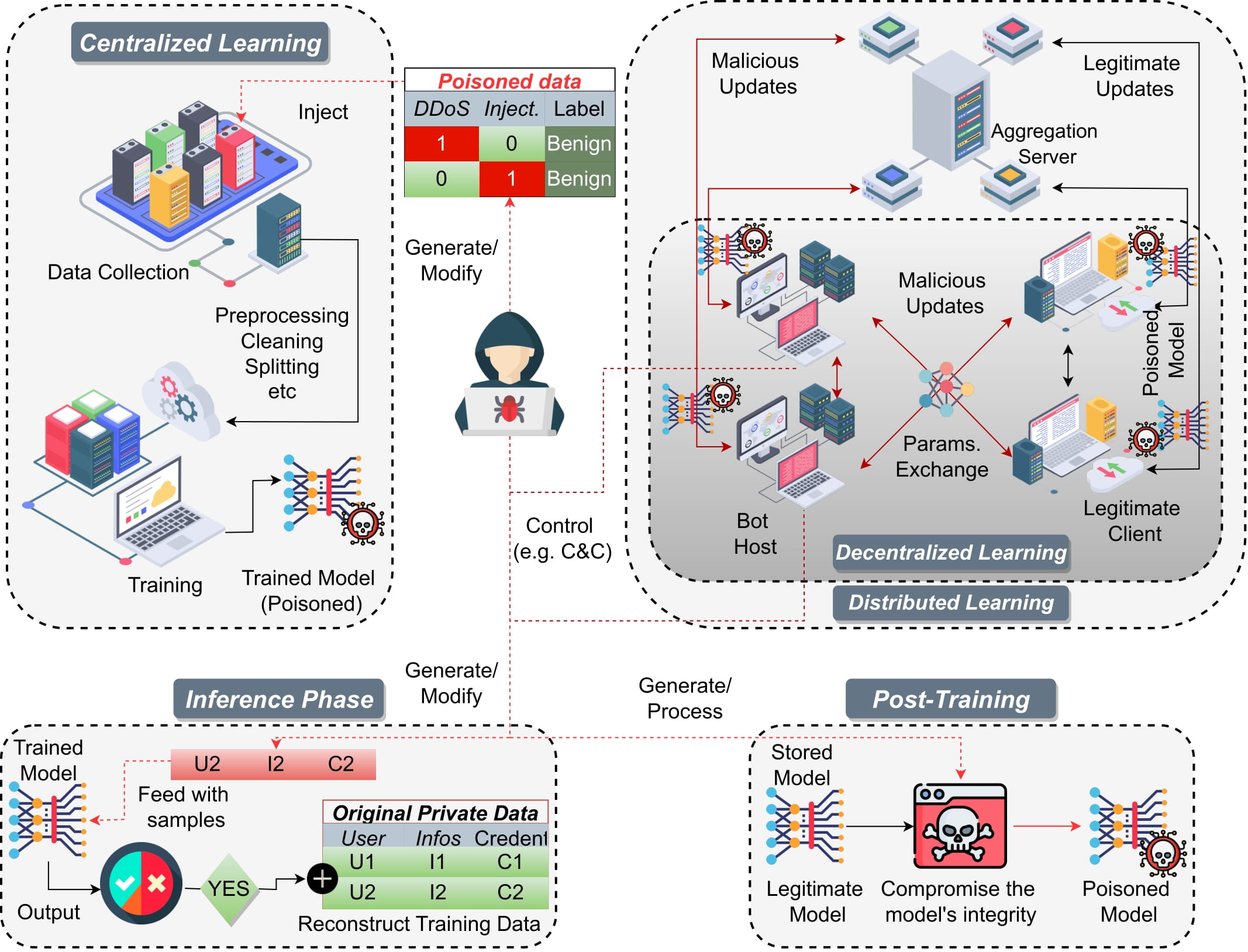}
\caption{Various attacks on ML classifiers under different contexts}
\label{fig:fig-attacks}
\end{figure*}

\subsubsection{Facial recognition}
Facial recognition refers to  \textcolor{black}{identifying or verifying} a person's identity based on their face. It catches, processes, and matches different features based on the details of the person's face. Different datasets are used for such tasks, including: 
\begin{itemize}
 \item \textit{AT\&T dataset \cite{cambridge2002database}:}	also known as the Database of Faces. The dataset consists of 400 grayscale (92×112) images of human faces taken from 40 individuals, with 10 images for each person. For certain individuals, the images were shot at varying times, differing in lighting and facial expressions. The work of Zhang \textit{et al.} \cite{zhang2019poisoning} evaluated a GAN-based poisoning attack on this dataset.
 \item \textit{LFW dataset \cite{huang2008labeled}:}	The Labeled Faces in the Wild is a collection of face photographs conceived to investigate the unconstrained face recognition task. It has over 13000 face images gathered from the web. Individual faces are labeled with their respective names. 1680 of the pictured persons possess two or more separate photos. This dataset was also used for generating poisoned training samples in \cite{chen2021deeppoison}.
 \item \textit{CASIA dataset \cite{li2013casia}:} collected in the period 2007-2010 by the Institute of Automation, Chinese Academy of Sciences (CASIA), including 725 face images. The dataset was designed for Near-infrared vs. Visible light (NIR-VIS) face recognition, with 640×480 resolutions and with 1 to 22 VIS and 5 to 50 NIR face images per person. This dataset was also used in \cite{chen2021deeppoison}.
\item \textit{VGGFace2 dataset \cite{cao2018vggface2}:} comprises more than 3M images featuring 9131 famous people across a wide range of ethnic groups. All images are downloaded from Google and exhibit considerable background, age, and lighting variation. Overall, the entire dataset is relatively balanced (40.7\% for females and 59.3\% for males). It includes 80-843 images for each individual. Fung \textit{et al.} \cite{fung2018mitigating} used the VGGFace2 dataset for evaluating a poisoning Sybil's defense mechanism.
 \item \textit{CelebA dataset \cite{liu2015deep}:} constructed by labeling selected images from the CelebFaces dataset, including 10000 celebrities, with 20 images for each one. And all photos are annotated with 40 face attributes and five key points. EdgeConnect is a two-stage adversarial scheme composed of an edge generator tracked by an image completion system, which can be used for scene generation. The scheme proposed in \cite{nazeri2019edgeconnect} is evaluated using the CelebA dataset.
\end{itemize}

\subsection{Time series classification}
Time series classification involves ML techniques designed to analyze several classes of labeled time series data to forecast or categorize which class a new dataset falls into.
\begin{itemize}
    \item \textit{UCR time series classification archive \cite{UCRArchive}:} was launched in 2002 and has since become an extensive resource for the time series data mining field. While the original version of the archive contained 16 data sets, the archive has seen periodic updates over the years. A major expansion  \textcolor{black}{occurred} in 2015 when the archive expanded to 85 datasets.  \textcolor{black}{Recent} works on adversarial attacks and defenses against time series classification (TSC) problems have been proposed as discussed in \cite{yang2022tsadv}. 
    \end{itemize}

\subsection{Human Activity Recognition}
Human Activity Recognition (HAR) has broad applicability in light of the worldwide use of interconnected sensing devices such as IoT devices, smartphones, and cameras, as well as its capability to collect data on human behavior. In addition, progress in AI has dramatically improved the extraction of profoundly hidden intelligence for precise identification and interpretation. Different datasets were proposed for HAR tasks including:
\begin{itemize}
    \item \textit{HARUS dataset	\cite{anguita2013public}:} gathered from a group of 30 volunteers with ages ranging between 19-48 years wearing a smartphone. The data about their activities are split into two classes (dynamic and static) with each having 3 activities. It includes a total of 10299 samples with 561 attributes. The work in \cite{sun2021data} evaluated their proposed bilevel optimization framework, which is intended to compute optimal poisoning attacks on FL.
    \item \textit{Parkinson dataset	\cite{little2007exploiting}:} created to identify people with Parkinson's disease. It consists of 5,875 observations from a series of biomedical readings of 31 people's voices, 23 of whom have Parkinson's disease. This dataset  \textcolor{black}{evaluates} the framework in \cite{sun2021data}.
\end{itemize}

\begin{table}[h!]
\centering
\setlength{\tabcolsep}{2.5pt}
\renewcommand{\arraystretch}{1}
\caption{Attacks on federated ML algorithms.}
\centering
\scriptsize
\label{tab:tab7}
\begin{tabular}{|p{0.6in}|p{0.6in}|p{0.6in}|p{1.2in}|}
\hline
\textbf{Attack}            & \textbf{ML phases} & \textbf{Attack Goal} & \textbf{Attack Description}                                                                                                                      \\ \hline
FL model extraction attack & Post-Deployment     & \textcolor{black}{Obtaining the model weights or AI architecture}                        & Malicious clients creating local models that   are functionally equivalent to target models of honest clients                             \\ \hline
FL evasion attack          & Post-Deployment     & Bypassing classifiers                       & Malicious clients generally will not modify   the target model of honest clients but will cheat the model to generate false   predictions \\ \hline
FL inversion attack        & Post-Deployment     & Private data search                         & Malicious clients require knowledge of   labels to retrieve training data from honest clients                                             \\ \hline
FL inference attack        & Post-Deployment     & Private data search                         & Identify whether a data point was used in a   training set of honest clients                                                              \\ \hline
FL poisoning attack        & Pre-Deployment      & Disrupt the functioning the aggregate model & Malicious clients inject malicious data in a   training set of honest clients                                                            \\ \hline
FL drooping attack         & Pre-Deployment      & Disrupt the functioning of the aggregate model & Malicious clients drop the network traffic   from selected honest clients                                                                 \\ \hline
\end{tabular}

\end{table}

\subsection{Sentiment classification}
Sentiment analytics is a self-learning approach that analyzes the sense of meaning in text, ranging from positive to negative. Using ML tools that are trained with instances of sentiment in text, they automatically learn to recognize sentiment with no human intervention.
\begin{itemize}
    \item \textit{IMDB dataset \cite{maas2011learning}:} consists of a large dataset of different reviews of movies for sentiment binary classification. Specifically, the dataset is a collection of 50000 movie reviews and their associated binary labels (positive or negative). This dataset is also used to benchmark edge learning targeted attacks such as poisoning and backdoor, and their associated defenses \cite{doku2021mitigating,jebreel2022defending}.
    \item \textit{Twitter dataset \cite{dong2014adaptive}:} consists of a dataset for target-dependent Twitter sentiment analysis which is manually annotated. A total of 6940 tweets is collected, with 6,248 tweets for the training set and 692 tweets for the test set. The sentiment labels include negative, neutral, and positive, with class distribution being 25\%, 50\%, and 25\%, respectively.
\end{itemize}

\subsection{Location awareness}
Location awareness designates the responsiveness of a particular system in ascertaining its position. This research field is significant for autonomous mobile operations. Available datasets for such tasks include: 
\begin{itemize}
    \item \textit{GTSRB dataset \cite{stallkamp2011german}:} for (German Traffic Sign Recognition Benchmark), which is a dataset that contains more than 50000 traffic sign images with 43 classes. It was constructed in 2010 from 10 hours of footage recorded using a Prosilica GC 1380CH camera (1360×1024 pixels), during daytime drives on various types of roads in Germany. The dataset is populated with over 1700 instances of road signs, with available image resolutions varying between 15 × 15 and 222 × 193 pixels. A framework for defending against Trojans in DNNs is proposed in \cite{qiu2021mt}. The proposed framework uses a moving target defense strategy given multi-dimensional training with random dimensions selection and was evaluated on this dataset.
    \item \textit{Places database \cite{zhou2017places}:} is designed to provide AI systems with high-level visual perception training, including background, recognition, and prediction tasks. It includes over 10M images covering over 400 specific scenes,  \textcolor{black}{ranging} from 5000 to 30000 training images per class. In addition, this dataset is used  \textcolor{black}{to evaluate} the EdgeConnect system discussed above \cite{nazeri2019edgeconnect}.
\end{itemize}

\subsection{Text classification}
Manual text classification involves analyzing the text material and making classifications accordingly, generally conducted by an annotator. Because text is among the widest kinds of unstructured data, it is difficult and costly to parse, comprehend, arrange, and classify textual data.  \textcolor{black}{Automated text classification categorizes} documents into pre-specified classes, usually using Natural Language Processing (NLP) and ML algorithms. Diverse parties have provided datasets for facilitating such tasks, like:
\begin{itemize}
    \item \textit{DBPedia datasets \cite{lehmann2015dbpedia}:} was conceived to mine organized content from Wikipedia's entries. It consists of an archive of information that provides hierarchical and classification data for more than 4.23M instances covering 685 classes. Variants of this data collection are a widely used reference for evaluating NLP/text classification projects, giving a valuable baseline for multi-class/multi-label hierarchical text classification. For instance, the DBPedia-14 version which consists of 560000 instances from 14 ontological categories (including film and animal classes), was employed in \cite{severi2022network} where the authors studied the impact of network-level adversaries on FL training and dropping attacks amplification.
\end{itemize}
\begin{figure*}
\centering
\includegraphics[width=\textwidth]{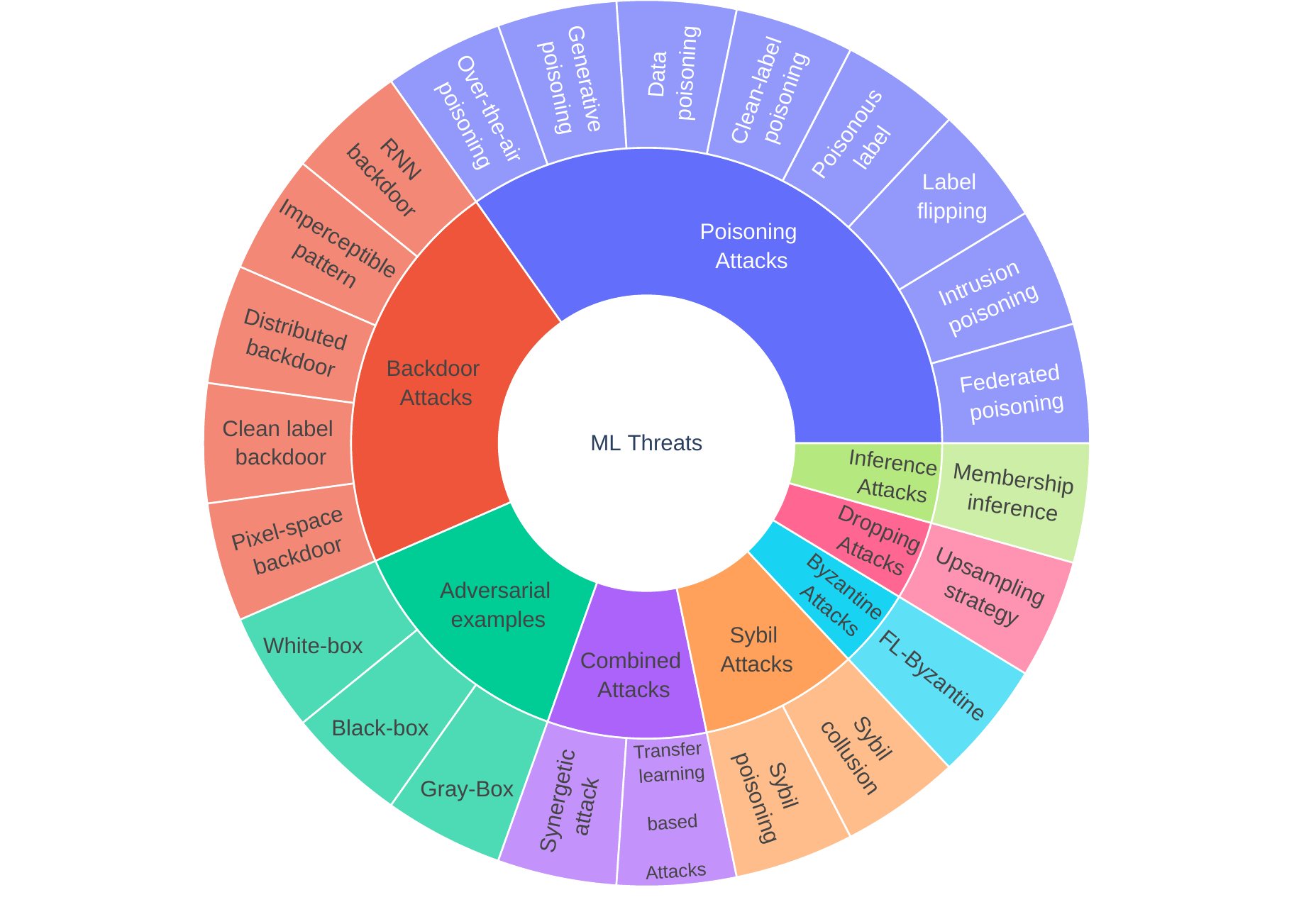}
\caption{Classification of machine learning vulnerabilities.}
\label{fig:fig10}
\end{figure*}

\begin{table*}[h!]
\centering
\setlength{\tabcolsep}{2.5pt}
\renewcommand{\arraystretch}{1}
\caption{Machine learning vulnerabilities.}
\centering
\label{tab:tab3}
\scriptsize
\begin{tabular}{|p{0.8in}|p{1.7in}|p{0.8in}|p{0.3in}|p{1.7in}|p{0.8in}|p{0.6in}|} \hline 
\textbf{Attack category} & \textbf{The attacker's knowledge} & \textbf{Attack type} & \textbf{Attack mode} & \textbf{The attacker's goal - Vulnerabilities} & \textbf{Targeted ML} & \textbf{Mitigation solution} \\ \hline 
\multirow{5}{*}{Backdoor attacks} & - Knowledge of the trigger (a small patch)
 & RNN backdoor attacks regarding text classification & CL, FL, DL & The adversary takes control of the global model training operation  & Text classification &  \cite{chen2021mitigating} \\ \cline{3-7} 
 & - Knowledge of some poisoned data based on the trigger & Imperceptible backdoor pattern & CL, FL, DL & The adversary poisons the training set with a small ensemble of pictures sourced from a source class (or classes) & Image classification & \cite{xiang2021reverse} \\ \cline{3-7}
 & -Inject maliciously the poisoned data to the victim to train a deep model with & Distributed backdoor attack & CL, FL, DL & Backdoor attack performed by several different collaborative attackers & Image classification & \cite{guo2021resisting} \\ \cline{3-7} 
 & - The trigger is stored as a secret by the attacker and only then exposed at the time of the test & Pixel-space backdoor attack & CL, FL, DL & Manipulate the pixel data values of the image to poison the images in the training dataset & Image classification & \cite{arshad2021pixdoor}\\ \cline{3-7}
 & - The attacker poison training examples and change their labels & Clean label backdoor attack & CL, FL, DL & Requiring zero-knowledge of the target model & Image classification & \cite{ning2021invisible} \\ \hline 
Combined attacks & Attack through jointly training neural network classifiers & Synergetic attack & CL, FL, DL & Combines adversarial examples and Trojan backdoor & Image classification &  \cite{wang2019neural, gao2019strip} \\ \hline 
\multirow{3}{*}{Adversarial examples}
 & blue-box attack: the configuration of the target ML models is unavailable to adversaries & blue-box adversarial attack  & CL, FL, DL & The adversary has restricted information about the model but certainly does not have knowledge of the model's parameters & - Time series classification\newline - Speech Recognition\newline - Video Recognition & -  \cite{yang2022tsadv}\newline -  \cite{biolkova2022neural}\\ \cline{3-7}
 & - White-box attack: the adversary has access to all the information of the target ML model. & White-box adversarial attack & CL, FL, DL & White-box attacks require  \textcolor{black}{complete} knowledge of the model being targeted, including its training method, architecture, and parameter values & - Text classification\newline - Attack classification &  \cite{ebrahimi2017hotflip}\newline -  \cite{wang2022iwa} \\ \cline{3-7}
 & - Grey-box attack: the adversary has some knowledge about the target ML model & Grey-box adversarial attack & CL, FL, DL & Requires partial knowledge instead of full knowledge &  Face Recognition &  \cite{wang2021similarity}\\ \hline 
\multirow{9}{*}{Poisoning attacks}
 & - There are two distinct attack scenarios blue-box and white-box attacks & Federated poisoning attack & FL, DL & Falsifying machine learning training data to generate unwanted results in a decentralized mode. The adversary's goal is to modify the parameters learned and manipulate the training data on their device & - Image classification\newline - Sentiment classification\newline - Attack classification\newline - Human Activity Recognition & - \cite{tabatabai2022exploration, cao2019understanding, tolpegin2020data }\newline - \cite{doku2021mitigating}\newline - \cite{khan2022federated}\newline - \cite{cao2019understanding}\newline - \cite{sun2021data} \\ \cline{3-7} 
 &  & Intrusion poisoning attack & CL, FL, DL & The application of poisoning attacks in the intrusion detection datasets & Attack classification & \cite{venkatesan2021poisoning} \\ \cline{3-7}
 & - White-box attacks: the attacker is assumed to know the trained parameters, the learning algorithm, the feature values, and the training set
 & Label flipping attack & CL, FL, DL & Attackers can flip the labels of some samples from one class to another, for example from attack to benign & - Sentiment classification \newline - Attack classification& -\cite{jebreel2022defending} \newline - \cite{zhang2022secfednids}\\ \cline{3-7} 
 &  & Poisonous label attack & CL, FL, DL & Injects fake images with the poisonous label in the training dataset than modifying directly the label of the images & - Image classification\newline  & \cite{liu2021poisonous} \\ \cline{3-7}
 & - blue-box attacks,  the attacker is assumed to know the learning algorithm and feature set, and collect a substitute data set but does not know the training set as well as the trained parameters
 & Clean-label poisoning attack & CL, FL, DL & The poison patterns are generated by inserting undetectable changes that will cause the model to misbehave in reaction to particular target inputs & - Transfer learning \newline - Attack classification & - \cite{aghakhani2020bullseye} \newline - \cite{zhang2022secfednids}\\ \cline{3-7} 
 &  & Data poisoning attack & CL, FL, DL & The attacker attempts to infect the training data by inserting well-formed samples to impose a harmful model on the learner & - Image classification\newline - Electricity theft detection & - \cite{wu2021indirect}\newline - \cite{takiddin2020robust} \\ \cline{3-7}
 & - In FL settings, the attacker's ultimate goal is to modify the source distribution of the raw data based on the corruption of the aggregate model by uploading the parameters of the poisoned local model & Generative poisoning attack & CL, FL, DL & Deploying a GAN model from the attacker's side to impersonate other participants' training dataset patterns & - Image classification \newline - Face recognition &  - \cite{zhang2020poisongan} \newline - \cite{chen2021deeppoison}\\ \cline{3-7}
 &  & Over-the-air spectrum poisoning attack & CL, FL, DL & Confuse the sender into making poor transmission decisions (i.e., an evasion attack) or to tamper with the sender's recycling process (i.e., a causal attack) & - Communication systems  &  \cite{sagduyu2019adversarial}
\\ \hline 
\multirow{2}{*}{Sybil attacks} & - It assumed that the attackers exploit the Sybils to initiate poisoning attacks against federated learning & Sybil-based poisoning attack & FL, DL & Sybils conduct federated learning poisoning attacks by delivering status updates that lead the distributed model to a poisoned common target. & - Face recognition\newline - Image classification  &  - \cite{fung2018mitigating}\newline - \cite{shejwalkar2022back} \\ \cline{3-7}
 &  & Sybil-based collusion attack & FL, DL & To complete local poisoning training, the attacker employs the label flipping scheme & - Image classification & \cite{xiao2022sca} \\ \hline 
Byzantine attacks & \textcolor{black}{The attacker's knowledge can be classified into three levels: (i) no knowledge, (ii) partial knowledge, and (iii) complete knowledge}& Local model poisoning attack & FL, DL & Affect the integrity of the learning phase in the training process & - Image classification & \cite{fang2020local}  \\ \hline
Dropping Attack & The adversary drops contributions from the clients in every round & Targeted dropping attack & FL, DL & By monitoring more rounds of the FL system; the adversary can drop the network traffic from selected clients & - Text classification & \cite{severi2022network}  \\ \hline
\end{tabular}\\
CL: Centralized Learning, FL: Federated Learning, DL: Distributed Learning
\end{table*}

\subsection{\textcolor{black}{Application of Dataset Types in 6G IoT Security}}

\textcolor{black}{The categories of datasets, when employed in machine learning models, substantially bolster security in edge learning environments. IoT-based multi-purpose security datasets enable the detection of unusual cyber activities in real-time, safeguarding IoT networks \cite{ahmetoglu2022comprehensive}. Attack classification data can inform the creation of sophisticated intrusion detection systems to protect against known attack vectors \cite{chakraborty2023intelligent}. Electricity theft detection through anomalous usage patterns ensures the integrity of the power grid, which indirectly benefits the security of all connected devices. Image classification allows for local object detection and facial recognition, which is crucial for many security applications in edge devices \cite{badr2023novel}. Time series classification can pinpoint sudden changes in network traffic indicative of a cyber attack \cite{shabani2023augmented}. Human Activity Recognition can identify abnormal or unauthorized activities in secure areas \cite{priyadarshini2023human}. Sentiment classification aids in spotting harmful or threatening digital communications \cite{chaparro2020sentiment}. Location awareness is crucial in verifying user authenticity based on their usual geographic patterns, while text classification can help identify spam or phishing attempts \cite{wang2022survey}.}

\subsection{\textcolor{black}{Highlights of Peer-to-Peer Features in Edge Learning across Various Dataset Types}}

\textcolor{black}{In all various dataset types, the peer-to-peer feature of edge learning is captured by ensuring that each device learns from its unique local data and shares the model's insights with the rest of the network. This maintains data privacy and security, utilizes the computational resources of edge devices, and enables learning from diverse and heterogeneous datasets.}

\begin{itemize}

\item \textcolor{black}{Multi-purpose security: IoT devices gather data on potential security threats, learning from it and sharing insights or models with other devices while maintaining the privacy of raw data.}

\item \textcolor{black}{Attack classification: In network infrastructures, devices collect and process data on different network attacks, updating learning models based on individual experiences and sharing these models with other devices.}

\item \textcolor{black}{Image classification: Devices like cameras or smartphones classify images locally, share the learned models, and keep the original images private to maintain privacy.}

\item \textcolor{black}{Time series classification: Sensors that generate time-series data classify it locally and share their models, adhering to data privacy and decentralized learning principles.}

\item \textcolor{black}{Human Activity Recognition (HAR): Wearable devices like smartwatches learn to recognize human activity from user data and share their learning models, maintaining the privacy of the raw data.}

\item \textcolor{black}{Sentiment classification: Devices (e.g., smartphones) classify sentiment on local data (such as social media posts and messages) and share the learned models while keeping the actual data private.}

\item \textcolor{black}{Location awareness: GPS-enabled devices learn from location data and share their models with other devices in the network, improving location awareness without revealing individual location data.}

\item \textcolor{black}{Text classification: Each device performs text classification tasks on its unique text data, sharing the learned models to collectively improve text classification tasks, keeping the original text data private.}

\end{itemize}

\section{Machine learning vulnerabilities}\label{sec:4}
Machine learning systems are vulnerable to attack, which limits the application of machine learning, especially in 6G–IoT Networks \cite{rosenberg2021adversarial}. There are six attacks on federated ML algorithms, including FL model extraction attack, FL evasion attack, FL inversion attack, FL inference attack, FL poisoning attack, and FL drooping attack, as presented in Table \ref{tab:tab7}. To visualize some of these threats, Figure \ref{fig:fig-attacks} illustrates various attacks against different ML paradigms including centralized, distributed, and decentralized learning during different model phases, including training, post-training, and inference phases. And while there are different strategies for targeting AI models, for instance in centralized learning by poisoning the data on which the model trains directly, in decentralized learning with poisoned updates, the result remains the same: a compromised model.  \textcolor{black}{This section provides a comprehensive classification} of threats against ML, as presented in Figure \ref{fig:fig10}. In addition, a thoughtful and comprehensive review of ML vulnerabilities based on the provided classification and related to three main aspects, namely, the knowledge already acquired by the attacker, the type of attack employed, and the final objective, which is provided in Table \ref{tab:tab3}.

\subsection{{\color{black}Pre/Post-deployment attacks}}
{\color{black}Table \ref{tab:tab7} lists the attacks that can be launched on federated machine learning (FL) algorithms along with their goals and descriptions. These attacks can be categorized based on the phase they occur - pre-deployment or post-deployment. Pre-deployment attacks target the aggregate model \textcolor{black}{before being} deployed to honest clients. Post-deployment attacks occur after the aggregate model has been deployed to honest clients. The following six attacks compromise the privacy of the private data of honest clients \cite{sun2021data, bagdasaryan2020backdoor,jere2020taxonomy,ferrag2023poisoning}:}

\begin{itemize}
    \item  {\color{black}FL model extraction attack: This attack occurs in the post-deployment phase, where malicious clients aim to extract private data from honest clients. The attackers create local models that are functionally equivalent to the target models of honest clients to extract data.}

    \item  {\color{black}FL evasion attack: This attack occurs in the post-deployment phase, where malicious clients aim to bypass the classifiers of honest clients. The attackers do not modify the target model of honest clients but instead cheat the model to generate false predictions.}

    \item  {\color{black}FL inversion attack: This attack occurs in the post-deployment phase, where malicious clients aim to search for the private data of honest clients. The attackers require knowledge of labels to retrieve training data from honest clients.}

    \item  {\color{black}FL inference attack: This attack occurs in the post-deployment phase, where malicious clients aim to identify whether a data point was used in the training set of honest clients. The attackers try to extract private data from honest clients.}

    \item  {\color{black}FL poisoning attack: This attack occurs in the pre-deployment phase, where malicious clients aim to disrupt the functioning of the aggregate model. The attackers inject malicious data into the training set of honest clients to manipulate the model.}

    \item  {\color{black}FL drooping attack: This attack occurs in the pre-deployment phase, where malicious clients aim to disrupt the functioning of the aggregate model. The attackers drop the network traffic from selected honest clients to manipulate the model.}
\end{itemize}

\subsection{Backdoor attacks}
{\color{black}Backdoor attacks involve manipulating training data or models to trigger specific behaviors in the model during the testing phase. \textcolor{black}{Different backdoor attacks} are presented, including RNN backdoor attacks, imperceptible backdoor pattern attacks, distributed backdoor attacks, pixel-space backdoor attacks, and clean-label backdoor attacks.}

\subsubsection{RNN backdoor attacks}
Backdoor attacks in deep neural networks are based on two assumptions, namely, the adversary takes control of the global model training operation and the second assumption is that the adversary only has control over particular training data. \textcolor{black}{RNN backdoor attacks on 6G could secretly embed malicious behaviors in network models, potentially compromising data integrity and user privacy. As 6G relies heavily on AI-driven solutions for efficient network management and data processing, such attacks can degrade network performance and security. Consequently, they pose risks to the expected ultra-reliable, low-latency communication features of 6G.} Chen \textit{et al.} \cite{chen2021mitigating} proposed a mitigating system against backdoor attacks in the text classification based on Backdoor Keyword Identification. The proposed system can identify the poisoning patterns in the training dataset with no trust data and no knowledge of the backdoor release. The following text classification datasets are used in the performance evaluation with the LSTM models, Reuters-21578 dataset, 20 newsgroups, DBpedia ontology, and IMDB. The results show good results \textcolor{black}{regarding} identification precision and recall of poisoning samples. To remove poisoning data, the authors proposed the following formula to filter the keywords in a dictionary and detect the suspicious keyword, which is most susceptible to be a backdoor keyword:

\begin{equation}
{Ident}_{backdoor}=\overline{f\left(k\right)}\cdot {log}_{10}gen\_sam\cdot {log}_{10}\frac{x}{gen\_sam}
\end{equation}

\textcolor{black}{ $\overline{f\left(k\right)}$ is the average score for the keyword $k$ in the dictionary. The score could be based on various metrics such as the frequency of the keyword, the similarity between the keyword and other words in the dictionary, or other measures. ${log}_{10}gen\_sam$ uses a logarithmic function of the number of samples that produce the keyword $gen\_sam$. The purpose is to adjust the weight of the keyword based on the number of samples in which it appears. ${log}_{10}\frac{x}{gen\_sam}$ represents the logarithm of the reciprocal of frequencies, which is used to normalize the weight of the keyword. The $x$ value is a fixed parameter that is higher than 0. By combining these components, the formula calculates an identifier $Ident_{backdoor}$ for each keyword, which can be used to detect suspicious keywords that are more likely to be backdoor keywords, i.e., the higher the value of $Ident_{backdoor}$, the more suspicious the keyword. }

\subsubsection{Imperceptible backdoor pattern}
In image classification, the attacker poisons the training set with a small ensemble of pictures sourced from a source class (or classes), incorporated with a backdoor feature, and then tagged to a target class. \textcolor{black}{The imperceptible backdoor pattern in 6G can subtly introduce vulnerabilities, enabling unauthorized access or control. This compromises the integrity and security of ultra-reliable, high-speed 6G networks. As a result, it jeopardizes user privacy, data confidentiality, and the dependability of critical applications.} Xiang \textit{et al.} \cite{xiang2021reverse} proposed a mitigation technique based on imperceptible backdoor patterns. The proposed technique can identify if the training set is poisoned and precisely recognizes the target class and training images in which the backdoor pattern is integrated. At the end of the process, the proposed technique performs reverse engineering to provide an estimate of this backdoor pattern used by the attacker. The CIFAR-10 dataset is used as a benchmark experiment, and the results show a reduced rate compared to state-of-the-art to no more than 4.9\%.

\textcolor{black}{The poisoned dataset used for training the victim classifier consists of a combination of the ${Dataset}_{backdoor}$ and the benign ${Dataset}_{train}$. The ${Dataset}_{backdoor}$ is a set of examples intentionally modified with a backdoor pattern to trick the classifier into predicting a specific target class when presented with an image from a particular source class. The ${Dataset}_{backdoor}$ is defined as follows:}

\begin{multline}
{Dataset}_{backdoor}=\{(f\left(x;{pattern}^*\right),y)|x\sim P_{{Source}^*},\\ \ y={Class}^*\}
\end{multline}

\textcolor{black}{$f\left(x;;{pattern}^*\right)$ is the embedding function that maps an image $x$ to a feature vector, and ${pattern}^*$ is the backdoor pattern that is added to the image. The resulting image $f\left(x;;{pattern}^*\right)$ is then labeled as ${Class}^*$, which is the target class that the backdoor is designed to trigger. The source class(es) $P_{{Source}^*}$ determines which images the backdoor is applied to. If $P_{{Source}^*}$ contains multiple classes, the backdoor can be triggered by images from any of those classes. The poisoned dataset is used to train the victim classifier, which learns to classify images from both \textcolor{black}{benign and poisoned datasets}. However, since the poisoned dataset contains examples with backdoors, the victim classifier can be manipulated to produce incorrect predictions on images that contain the backdoor pattern.}

\begin{figure*}
\centering
\includegraphics[width=0.9\textwidth]{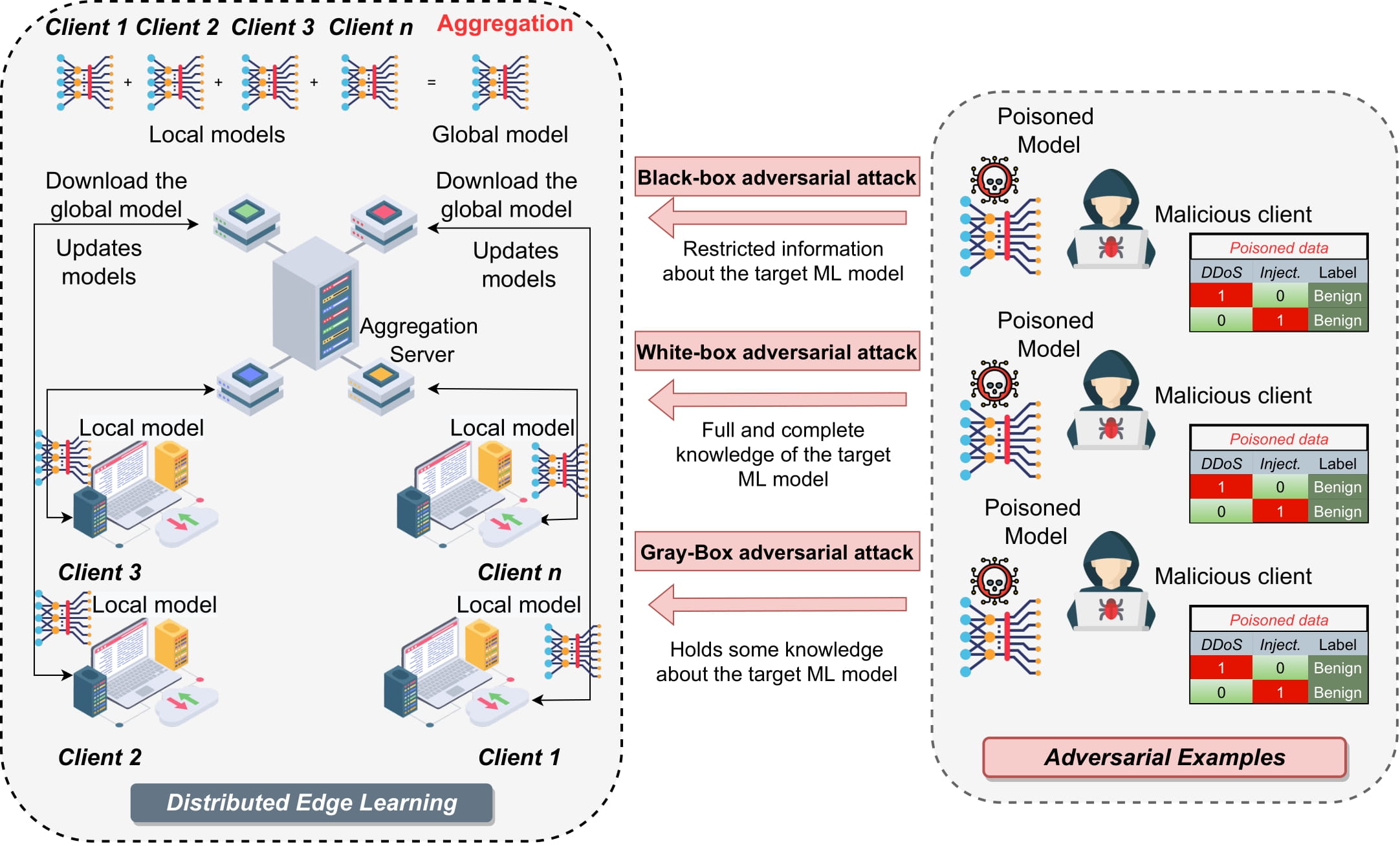}
\caption{Adversarial examples in \textcolor{black}{Edge Learning}.}
\label{fig:adv}
\end{figure*}

\subsubsection{Distributed backdoor attack}

Backdoor attacks performed by several collaborative attackers, i.e., distributed backdoor attacks, can reach very successful rates and are very difficult to detect. \textcolor{black}{A distributed backdoor attack in 6G networks could exploit the advanced and decentralized nature of the system, subtly introducing malicious functionalities without detection. By compromising multiple nodes or devices, the attack can gain control over data flow, causing disruptions or unauthorized data access.} Based on a dynamic norm clipping approach, Guo \textit{et al.} \cite{guo2021resisting} proposed a security system to defend against distributed backdoor attacks in federated learning. The proposed system is evaluated with the following four public datasets: MNIST, FMNIST, CIFAR-10, and Tiny Imagenet. The results show that the proposed system can reduce the attack access rate by 84.23\% compared to the state of the art. With the distributed backdoor attack, \textcolor{black}{some attackers constitute} a group with identical malicious purposes. More precisely, all attackers conduct the backdoor attack on their local models independently using their specific local backdoor trigger ${Trig}_i$ and the same target label ${Label}_{target}$. The attacker's corresponding goal $i$ is defined as follows :

\begin{multline}
G_i\left(w\right)=\sum_{j\in {Data}^i_{infected}}{\left[f(w;;x^i_j+{Trig}_i,{Label}_{target})\right]}+ \\ \sum_{j\in {Data}^i_{uninfected}}{\left[\left[f(w;;x^i_j+y^i_j)\right]\right]}
\end{multline}

Where $w$ represents the global model parameters,  ${Data}^i_{infected}$ is the infected image dataset of attacker $i$, ${Data}^i_{uninfected}$ is the uninfected image dataset of attacker $i$, $x^i_j$ represents input from the local dataset and $y^i_j$ represent the corresponding label sampled from the local dataset. \textcolor{black}{The objective of each attacker is to maximize the misclassification rate of the global model on a specific target label ${Label}_{target}$ when presented with inputs that contain a specific backdoor trigger ${Trig}_i$. The goal $G_i\left(w\right)$ of attacker $i$ is defined as the sum of two terms: the first term represents the misclassification rate of the infected images from attacker $i$ on the target label ${Label}_{target}$, while the second term represents the misclassification rate of the uninfected images from attacker $i$ on their true labels. The function $f(w;x,y)$ represents the output of the global model with parameters $w$ on input $x$ and label $y$. The double semicolon notation $w;;x^i_j+{Trig}_i$ indicates that the backdoor trigger ${Trig}_i$ is added to the input $x^i_j$ before it is fed into the global model.}

\subsubsection{Clean label backdoor attack}
\textcolor{black}{A clean-label backdoor attack on 6G could subtly embed malicious triggers in legitimate data during the system's training phase. Once deployed, adversaries can exploit these triggers to manipulate the 6G network's behavior, undermining its security and reliability. Such attacks could severely compromise the advanced functionalities and services anticipated in 6G ecosystems.} Ning \textit{et al.} \cite{ning2021invisible} proposed a blue box clean label backdoor attack, which requires zero knowledge of the target model.\textcolor{black}{ To} detect and eliminate, the authors propose the following two methods: 1) Supervised poison sample detection and 2) Unsupervised poison sample detection. For Backdoor injection, the adversary summarized another dataset by combining the clean database with a small part of the poisoned dataset as follows:

\begin{multline}
{Dataset}_{new}={Dataset\ }_{clean}\cup {Dataset}_{poisoned} 
\end{multline}

The adversary uses this dataset during training and verifies that \textcolor{black}{a combined dataset does not influence other properties}. \textcolor{black}{The backdoor injection attack involves adding malicious data to the training dataset, which can be used to exploit the model during inference. In this case, the adversary has combined a clean dataset with a small part of the poisoned dataset to create a new dataset for training. The poisoned dataset contains inputs and labels specifically chosen to trigger a backdoor in the model. To train the model using this dataset, the adversary minimizes a loss function that includes clean and poisoned data. The goal is to find the parameters of the model that minimize the total loss on both datasets. The loss function is typically based on cross-entropy, which measures the difference between the predicted probabilities and the true labels.}

The availability of a poisoned dataset gives the following loss function:

\begin{multline}
{F_l=min}_{model}\sum^n_{i=0}{Loss(model,\left(x^i,y^j\right))}+\\\sum^m_{j=0}{Loss(model,\left(x^{i'},y^{j'}\right))}
\end{multline}


 \textcolor{black}{$model$ represents the parameters of the model. $Loss(model,\left(x^i,y^j\right))$ represents the loss incurred by the model when \textcolor{black}{predicting} input $x^i$ with label $y^j$ from the poisoned dataset. $Loss(model,\left(x^{i'},y^{j'}\right))$ represents the loss incurred by the model when \textcolor{black}{predicting} input $x^{i'}$ with label $y^{j'}$ from the clean dataset. $n$ is the size of the poisoned dataset. $m$ is the size of the clean dataset.}

\subsubsection{Pixel-space backdoor attack} 
\textcolor{black}{Pixel-space backdoor attacks are a type of adversarial attack where an attacker modifies the pixel values of an image in a way that is imperceptible to the human eye but causes misclassification by the target model. \textcolor{black}{A Pixel-space backdoor attack subtly modifies images in a way undetectable to the human eye but significantly alters the outcome of machine learning models. In the context of 6G, which heavily relies on AI-driven applications and ultra-reliable communications, such attacks could mislead autonomous systems or degrade the quality of service. The integrity and security of 6G-enabled applications, like autonomous vehicles or telemedicine, could be compromised, risking user safety and trust.} Arshad et al.'s pixel-space backdoor attack \cite{arshad2021pixdoor} is a type of data poisoning attack where the attacker injects malicious samples into the training dataset to cause the model to learn a backdoor trigger. In this attack, the attacker flips certain bits of the pixel values of the training images, which can be done without significantly changing the \textcolor{black}{image's appearance}. These modified images are then inserted into the training dataset, along with their original labels. During training, the model learns to associate the backdoor trigger with the target label, allowing the attacker to trigger the attack by adding the trigger to any input image. Arshad \textit{et al.} evaluated their backdoor attack strategy on the MNIST digit dataset, which consists of 28x28 Grey-scale images of handwritten digits. They found that their attack was successful in 90\% of the cases, meaning that the target model misclassified the poisoned images with the intended target label. Therefore, it's important to note that the success of a backdoor attack depends on several factors, including the size and quality of the training dataset, the choice of the backdoor trigger, and the strength of the attacker's evasion techniques. Therefore, while Arshad et al.'s attack achieved a high success rate on the MNIST dataset, it may not be as effective on other datasets or against more robust models with stronger defenses against adversarial attacks.}

\begin{table*}[h!]
\centering
\setlength{\tabcolsep}{2.5pt}
\renewcommand{\arraystretch}{1}
\caption{Threat models of adversarial examples generation.}
\label{tab:av}
\begin{tabular}{|p{1.2in}|p{1.2in}|p{1in}|p{1.7in}|p{1.7in}|} 
\hline
\textbf{Attacks examples}                           & \textbf{Attacks for the specific tasks}   & \textbf{White/blue/Grey box}                     & \textbf{Adversarial Example Generation}                               & \textbf{Methods}                                                                                            \\ 
\hline
Miscellaneous
  attacks                              & Classification             & Feature-space
  white-box attacks            & The XGBoost can
  be used as the adversarial feature extraction model & Miscellaneous
  malware detectors based on OpCode n-gram features \cite{li2020adversarial}  \\ 
\hline
Adversarial
  transformation networks                & Classification             & Trained in a white-box
  or blue-box manner & Adversarial
  Autoencoding (AAE) and Perturbation ATN (P-ATN)         & Deterministic/stochastic
  defense models \cite{lee2022graddiv}                             \\ 
\hline
Fooling
  gradient-based attacks                     & Classification             & blue/Grey-box attack                             & Generating adversarial audio files                                    & Fooling Deep structured prediction models \cite{cisse2017houdini}                                                                     \\ 
\hline
Universal perturbations and antagonistic network                                                & Classification             & blue/Grey-box attack                             & Generating rogue images                                                                    & Breaking high-performance image classifiers \cite{sarkar2017upset}                                                                                                             \\ 
\hline
Carlini and
  Wagner Attacks (CW)                    & Classification             & blue/Grey-box attack                          & Generating adversarial examples based on distance metrics                                                                     & Defensive distillation  \cite{carlini2017towards}                                                                                                     \\ 
\hline
One Pixel
  Attack                                   & Classification             & Semiblue-Box Attack                                           & Generating adversarial examples using differential evolution                                                                & Covariance matrix adaptation evolution strategy       \cite{su2019one}                                                                                                      \\ 
\hline
Jacobian-Based
  Saliency Map Attack (JSMA)          & Classification             & White-box attack                                          & Generating adversarial samples using the mapping between inputs and outputs of deep learning                                                                    & Predictive measure of distance     \cite{papernot2016limitations}                                                                                                       \\ 
\hline
Basic 
  Least-Likely-Class Iterative Methods        & Classification             & White-Box Attack                                           & Generating adversarial images based on the linearization of the cost function                                                                     & Iterative least likely method \cite{kurakin2018adversarial}                                                                                                           \\ 
\hline
Fast Gradient
  Sign Method (FGSM)                   & Classification             & White-Box Attack                                           & Generating adversarial examples using a family of fast methods                                                                      & ~  Explaining and harnessing adversarial examples \cite{goodfellow2014explaining}                                                                                                          \\ 
\hline
L-BFGS  attack                           & Classification             &  White-Box Attack                                            & Generating adversarial examples using a box-constrained L-BFGS algorithm                                                                   & Intriguing properties of neural networks \cite{szegedy2013intriguing}                                                                                                         \\ 
\hline
Adversarial vulnerability of facial attributes                                    & Classification/Recognition & White-Box Attack                                           & Generating natural adversarial images                                                                   &   Fast flipping attribute technique                  \cite{rozsa2019facial}                                                                                       \\ 
\hline
Attacks on
  Semantic Segmentation object Detection & Image Segmentation & blue/Grey/White-box attack                                         & Adversarial target generation                                                                    & Universal adversarial perturbations \cite{hendrik2017universal}                                                                                                     \\ 
\hline
Strategically-timed attack             & Classification/Recognition & blue/Grey/White-box attack                                            & Combining a planning algorithm and a generative model                                                                     &  Adversarial attack on deep reinforcement learning systems  \cite{lin2017tactics}                                                                                                     \\ 
\hline
Attack on
  Recurrent Neural Networks                & Classification/Recognition & blue/Grey/White-box attack                                            & Creating adversarial sequences                                                                &  Fast gradient sign method and Forward derivative method \cite{papernot2016crafting}                                                                                                           \\ 
\hline
\end{tabular}
\end{table*}

\begin{figure*}
\centering
\includegraphics[width=0.9\textwidth]{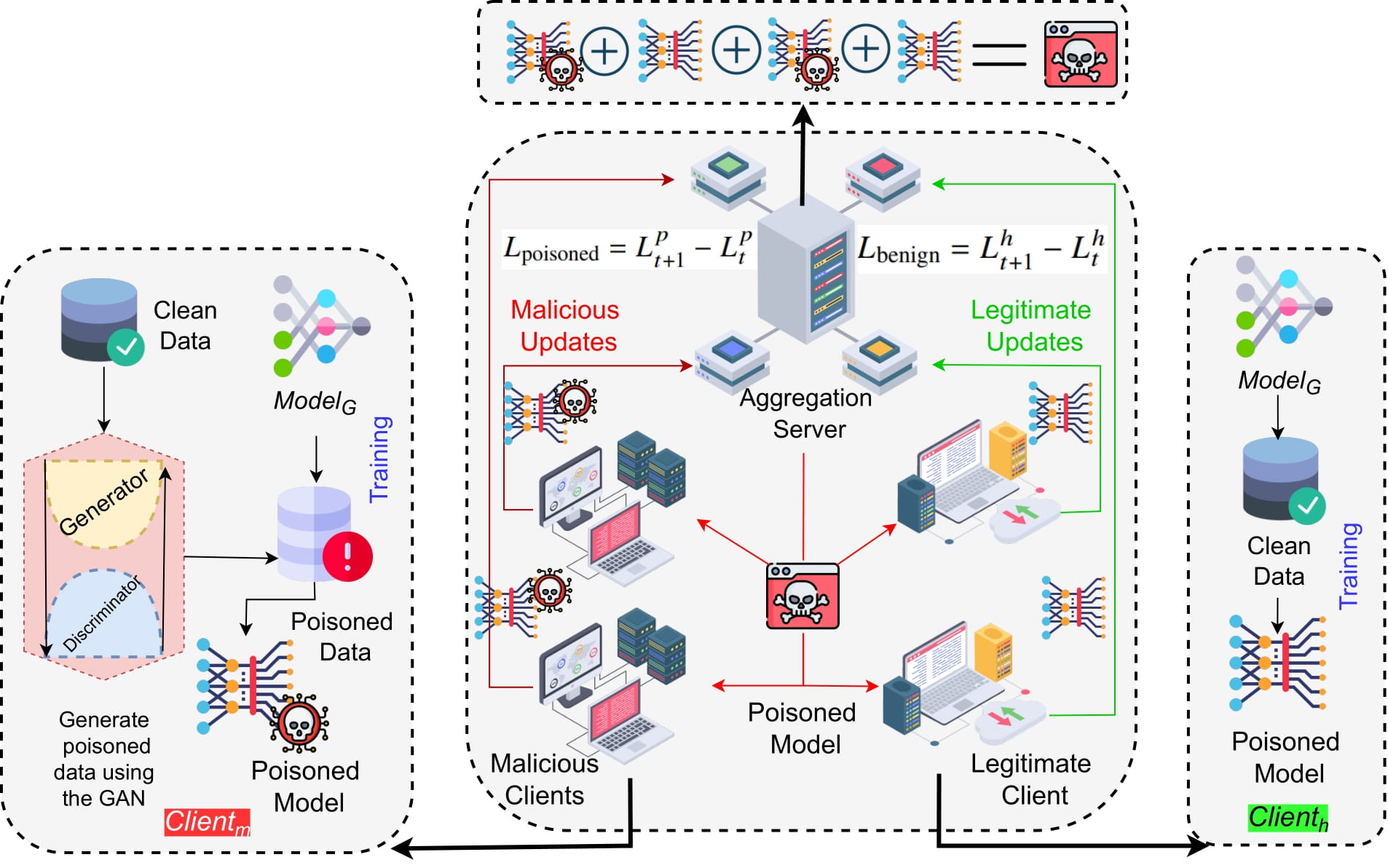}
\caption{FL model poisoning attack using generated poisoned samples}
\label{fig:gan}
\end{figure*}

\subsection{Adversarial examples}

\textcolor{black}{Adversarial examples refer to inputs specifically crafted to deceive machine learning models. Different adversarial attacks are presented, including blue-box attacks, white-box attacks, and Grey-box attacks.} Akhtar and Mian \cite{akhtar2018threat} proposed a classification of adversarial examples on deep learning into two categories: 1) Attacks for Classification and 2) Attacks Beyond Classification/Recognition. The Attacks for Classification include miscellaneous attacks, adversarial transformation networks (ATNs), Houdini, Upset and ANGRI, Universal Adversarial Perturbations, Deepfool, Carlini and Wagner Attacks (C\&W), One Pixel Attack, Jacobian-Based Saliency Map Attack (JSMA), Basic \& Least-Likely-Class Iterative Methods, Fast Gradient Sign Method (FGSM), and Box-Constrained L-BFGS. Attacks Beyond Classification/Recognition include Attacks on Face Attributes, Semantic Segmentation \& Object Detection, Deep Reinforcement Learning, Recurrent Neural Networks, and Attacks on Autoencoders and Generative Models. Table \ref{tab:av} presents the threat models of adversarial examples generation.  Figure \ref{fig:adv} presents the adversarial examples in \textcolor{black}{Edge Learning}. Therefore, in our work, we propose a classification of adversarial examples into three categories: 1) White-box adversarial attacks, 2) blue-box adversarial attacks, and 3) Grey-Box adversarial attacks.

\subsubsection{White-box adversarial attacks}
White-box attacks require \textcolor{black}{complete} knowledge of \textcolor{black}{the targeted model}, including its training method, architecture, parameter values, and, in most cases, its data training. \textcolor{black}{A white-box adversarial attack on 6G networks exploits knowledge of the system's architecture and algorithms to introduce crafted perturbations, undermining the accuracy and efficiency of data-driven applications. Such attacks can disrupt the ultra-reliable, low-latency communication fundamental to 6G, potentially compromising critical services and applications.} To manipulate discrete text structure at its one-hot representation, Ebrahimi \textit{et al.} \cite{ebrahimi2017hotflip} propose white-box adversarial examples, named HotFlip, which is applied for text classification. The HotFlip system uses the gradient \textcolor{black}{concerning} a one-hot input representation and can produce conflicting patterns with character substitutions ("flips"). The HotFlip also provides access to add and remove transactions by displaying them as character substitution sequences. The performance evaluation with AG’s news dataset and CharCNN-LSTM architecture shows that the adversary selects the flip transaction about 80\% of the time and prioritizes suppression over insertion by two-to-one. Wang \textit{et al.} \cite{wang2022iwa} propose white-box attacks based on the integrated gradient for fooling deep neural networks. The proposed attacks are evaluated with the following three datasets: CIFAR-10, MNIST, and NSL-KDD. \textcolor{black}{One of the most popular algorithms for generating white-box adversarial examples is the Fast Gradient Sign Method (FGSM) introduced by Goodfellow \textit{et al.} in 2014 \cite{goodfellow2014explaining}. The FGSM generates adversarial examples by taking the gradient of the loss function \textcolor{black}{concerning} the input and adding a perturbation to the input in \textcolor{black}{the gradient direction}. The formula for generating adversarial examples using FGSM is as follows:}

\begin{equation}
\textcolor{black}{x_{adv} = x + \epsilon\cdot sign(\nabla_{x} J(\theta, x, y))}
\end{equation}

\textcolor{black}{Where $x$ is the input, $y$ is the true label, $J$ is the loss function, $\theta$ is the model parameters, $\epsilon$ is a small constant that controls the magnitude of the perturbation, and $sign$ is the sign function.}

\subsubsection{blue-box adversarial attacks}
\textcolor{black}{
A blue-box adversarial attack targets 6G's advanced machine learning algorithms, exploiting vulnerabilities to mislead or manipulate network functions. By subtly modifying input data, attackers can cause 6G systems to misinterpret or misclassify information, potentially compromising network efficiency, security, and reliability.} Blue-box attacks provide a trained model with conflicting examples (during testing) \textcolor{black}{produced without} knowledge of the trained model. In particular cases, it is supposed that the adversary has restricted information about the model but does not know its parameters. Yang \textit{et al.} \cite{yang2022tsadv} propose a blue-box method called TSadv, which attacks deep neural networks on time series. For solving the constrained optimization problem, the TSadv method uses a gradient-free attack strategy. The performance evaluation with the University of California Riverside (UCR) time series dataset shows good success rate, the average number of iterations, and mean squared error. Therefore, Biolkov{\'a} and Nguyen \cite{biolkova2022neural} propose a blue-box adversarial attack in speech recognition. The proposed attack uses a neural predictor that approximates \textcolor{black}{the decision boundary's length, resulting }in a wrong audio signal transcription. The experiments on the LibriSpeech dataset show that the proposed attack can achieve a better success rate \textcolor{black}{than} BayesOpt \cite{ru2019bayesopt} and SignOpt \cite{cheng2019sign}.\textcolor{black}{ One of the most popular algorithms for generating blue-box adversarial examples is the Boundary Attack introduced by Brendel and Bethge in 2019 \cite{brendel2019approximating}. The Boundary Attack generates adversarial examples by iteratively exploring the decision boundary of the model. The formula for generating adversarial examples using Boundary Attack is as follows:}

\begin{equation}
\textcolor{black}{x_{adv} = x + \alpha\cdot\frac{g}{\left | g \right |_{p}}}
\end{equation}

\textcolor{black}{Where $x$ is the input, $\alpha$ is a small constant that controls the step size, $g$ is the gradient of the loss function \textcolor{black}{concerning} the input, and $\left | g \right |_{p}$ is the $l_{p}$ norm of the gradient.}

\subsubsection{Grey-box adversarial attacks}

\textcolor{black}{A grey-box adversarial attack targets 6G networks by exploiting partial knowledge of the system's architecture and parameters. Leveraging advanced machine learning techniques and the inherent complexities of 6G, attackers can introduce carefully crafted perturbations to disrupt network functionality.} Grey-box adversarial attacks involve generating adversarial examples with limited knowledge of the targeted model. The attacker has access to some, but not all, of the model's internal information. For example, the attacker may know the model's architecture but not the weights or biases. \textcolor{black}{The goal of a grey-box attack is to find an adversarial example that can deceive the targeted model into producing incorrect output, even though the attacker does not fully know the model's internal workings. This is often done by leveraging the gradients of the model concerning the input, which can be computed even if the attacker does not know the exact values of the model's parameters \cite{chen2023ai}. Grey-box attacks are particularly relevant in real-world scenarios, where attackers may not have access to the full details of a deployed model. However, they may still be able to mount effective attacks. Defenses against grey-box attacks typically involve techniques such as input sanitization \cite{zhao2020bridging} or model hardening, which aim to make it harder for attackers to find effective adversarial examples. One of the most popular algorithms for generating Grey-box adversarial examples is the Transferability Attack introduced by Papernot \textit{et al.} in 2016 \cite{papernot2016transferability}. The Transferability Attack generates adversarial examples by exploiting the transferability property of adversarial examples, meaning that adversarial examples generated for one model can also fool another. The formula for generating adversarial examples using Transferability Attack is as follows:}

\begin{equation}
\textcolor{black}{x_{adv} = x + \epsilon\cdot sign(\nabla_{x} J(\theta_{target}, x, y))}
\end{equation}

\textcolor{black}{This equation represents the process of generating an adversarial example for a given input image $x$, concerning a target class $\theta_{target}$ and a true label $y$, by adding a small perturbation $\epsilon$ to the original image $x$ in the direction of the sign of the gradient of the target class probability $J(\theta_{target}, x, y)$ concerning the input image $x$.}

In most cases, blue-box attacks are viewed as more difficult since the attacker has less information to operate with. Nevertheless, Grey box attacks can still be successful if the offensive party has sufficient knowledge of the pattern to produce successful adversarial examples.

\subsection{Combined attacks}
\textcolor{black}{Combined attacks are a combination of adversarial examples and Trojan backdoors, which can be used to bypass defenses in the model.}

\subsubsection{Synergetic attack}
A synergetic attack consists of a combination of backdoor and adversarial examples against neural network classifiers. \textcolor{black}{
Synergetic attacks in 6G exploit the interdependence of heterogeneous technologies, combining vulnerabilities across multiple layers of the network. By targeting the intricate integration points in 6G's collaborative systems, these attacks can disrupt seamless connectivity, compromising both service quality and security. Consequently, the complexity of 6G necessitates advanced security measures to counter such multifaceted threats.} Liu \textit{et al.} \cite{liu2021synergetic} proposed a synergetic attack, named AdvTrojan. The AdvTrojan is enabled based only on the model being infected through a backdoor during training, and the entries are thoughtfully disrupted. The AdvTrojan uses two stages that operate in parallel to progressively relocate the targeted input through the decision-making boundary to the adversary's objective class. The model is injected with a Trojan backdoor \textcolor{black}{during} training in the initial stage. The Trojan backdoor is enabled \textcolor{black}{during inference} by incrementing the relevant target inputs using the pre-specified Trojan trigger. In the second stage, the targeted entry is completed by a prudent mixture of some adversarial disruption and the Trojan horse trigger. In the second stage, the targeted entry is completed by a prudent mixture of the Trojan horse trigger and some adversarial disruption. The adversary disruption enhances the Trojan horse trigger to exchange the entry label in the class \textcolor{black}{the adversary targets}. In practical terms, the Trojan horse trigger forwards the entry to an artificial position in the space of the entry closest to the boundary of the model's decision. Afterward, the adversarial disruption makes the terminal push by shifting the forwarded example through the boundary of the decision, initiating the pre-established backdoor. The combined attacks can be defined as follows:

\begin{equation}
{Classifier}_{\theta \uparrow }\left(x\right)=\left\{ \begin{array}{c}
y_{target},\ \ {\rm if}\ x\ {\rm contains\ Trojan\ trigger}\ t \\ 
{Classifier}_{\theta \downarrow }\left(x\right),\ \ {\rm otherwise} \end{array}
\right.
\end{equation}

Where $y_{target}$ is the attacker's target, ${Classifier}_{\theta \uparrow }\left(x\right)$ is the classifier with normal behavior,  ${Classifier}_{\theta \downarrow }\left(x\right)$ is the Trojan-infected classifier, and $x\ $refers to the general input, that can be benign or malicious.

\subsection{Poisoning attacks}
\textcolor{black}{Poisoning attacks involve manipulating the training data to produce a biased or inaccurate model. Different types of poisoning attacks are presented, including federated poisoning attacks, intrusion poisoning attacks, label flipping attacks, poisonous label attacks, clean-label poisoning attacks, data poisoning attacks, and generative poisoning attacks \cite{ferrag2023poisoning,sun2021data}.} An attacker will penetrate a machine learning system and inject false or incorrect information into the database. Over time, as the algorithm learns from this falsified data, it will produce unwanted and potentially damaging outcomes \cite{fang2021data,kravchik2021poisoning}. However, data poisoning attacks can be divided into two major classes: those attacking availability and others attacking integrity. Availability attacks are frequently not particularly advanced but extensive, inserting at least as many bad data items as possible into a database. Once an attack is successful, the machine learning algorithm will be incorrect. Attacks against the integrity of machine learning are more complicated and could be more dangerous.

\subsubsection{Federated poisoning attack}

Federated learning uses machine learning in a decentralized mode that does not directly access \textcolor{black}{clients' private data}. However, Federated learning suffers from many issues, such as adversarial machine learning-related security attacks, high communications costs between clients and the server, and high accuracy \cite{cao2019understanding,tolpegin2020data,sun2021data,fung2018mitigating,shejwalkar2022back,zhou2021deep,chen2021deeppoison,li2021lomar}. \textcolor{black}{
Federated poisoning attacks can manipulate the aggregated model in 6G by introducing malicious updates from compromised devices. This could degrade the performance and reliability of 6G network services that rely on machine learning. As 6G networks are anticipated to be more AI-driven, such attacks may pose significant threats to network security and functionality.}To address these issues, Tabatabai \textit{et al.} \cite{tabatabai2022exploration} present a federated learning algorithm based on evolutionary methods. The proposed clustering algorithm collects clients in many clusters, each with a selected model at random to exploit the capabilities of individual models. Then, the clusters are built in a repeated procedure with the worst cluster being dropped at every repetition process until \textcolor{black}{only one cluster is} available. Specific clients are kicked out of the clusters for corrupted data usage or poor overall performance during all iterations. The winning clients are operated in the upcoming iteration. The cluster that remains with the surviving clients is then utilized for the best FL model training.  The Fast Gradient Sign Method (FGSM) is adopted for producing adversarial examples to create a federated poisoning attack. 

To mitigate data poisoning attacks, Doku and  Rawat \cite{doku2021mitigating} proposed a security system in a federated learning mode. \textcolor{black}{The proposed system uses a Support Vector Machine }to generate the total error a training dataset produces. The performance evaluation with the IMDB review sentiment dataset shows that the proposed system achieved an accuracy score of $0.72$. Uprety and Rawat \cite{uprety2021mitigating} proposed a technique for mitigating poisoning attacks that focuses on the nodes' reputation in the training process. The average reputation score of every client is computed with the method of probability distribution beta. The benchmark MNIST dataset is used in the performance evaluation, and the security analysis demonstrates that the classification model accuracy improved from $88$\% to $93$\% over $100$ rounds of communication after eliminating the attacker nodes by the aggregation servers. The generation of poison attacks in FL, as illustrated in Figure \ref{fig:gan}, can be defined by the following steps \cite{zhang2019poisoning}:

\begin{itemize}
\item Step 1: Send the global model ${Model}_G$ to honest clients ${Client}_h$ and malicious clients ${Client}_m$.

\item Step 2: The malicious clients ${Client}_m$ generate samples of targeted class.

\item Step 3: The malicious clients ${Client}_m$ assign wrong label to generated samples.

\item Step 4: The malicious clients ${Client}_m$insert poison data to the local dataset ${Dataset}_{Local}$.

\item Step 5: The malicious clients ${Client}_m\ $calculate the poisoned update $L_{{\rm poisoned}}=L^p_{t+1}-L^p_t$.

\item Step 6: The honest clients ${Client}_h$ calculate the benign update $L_{{\rm benign}}=L^h_{t+1}-L^h_t$.

\item Step 7: Update the local update $L_{{\rm poisoned}}$ and $L_{{\rm benign}}$ to the server.
\end{itemize}

\begin{figure*} 
\centering
\includegraphics[width=0.9\textwidth]{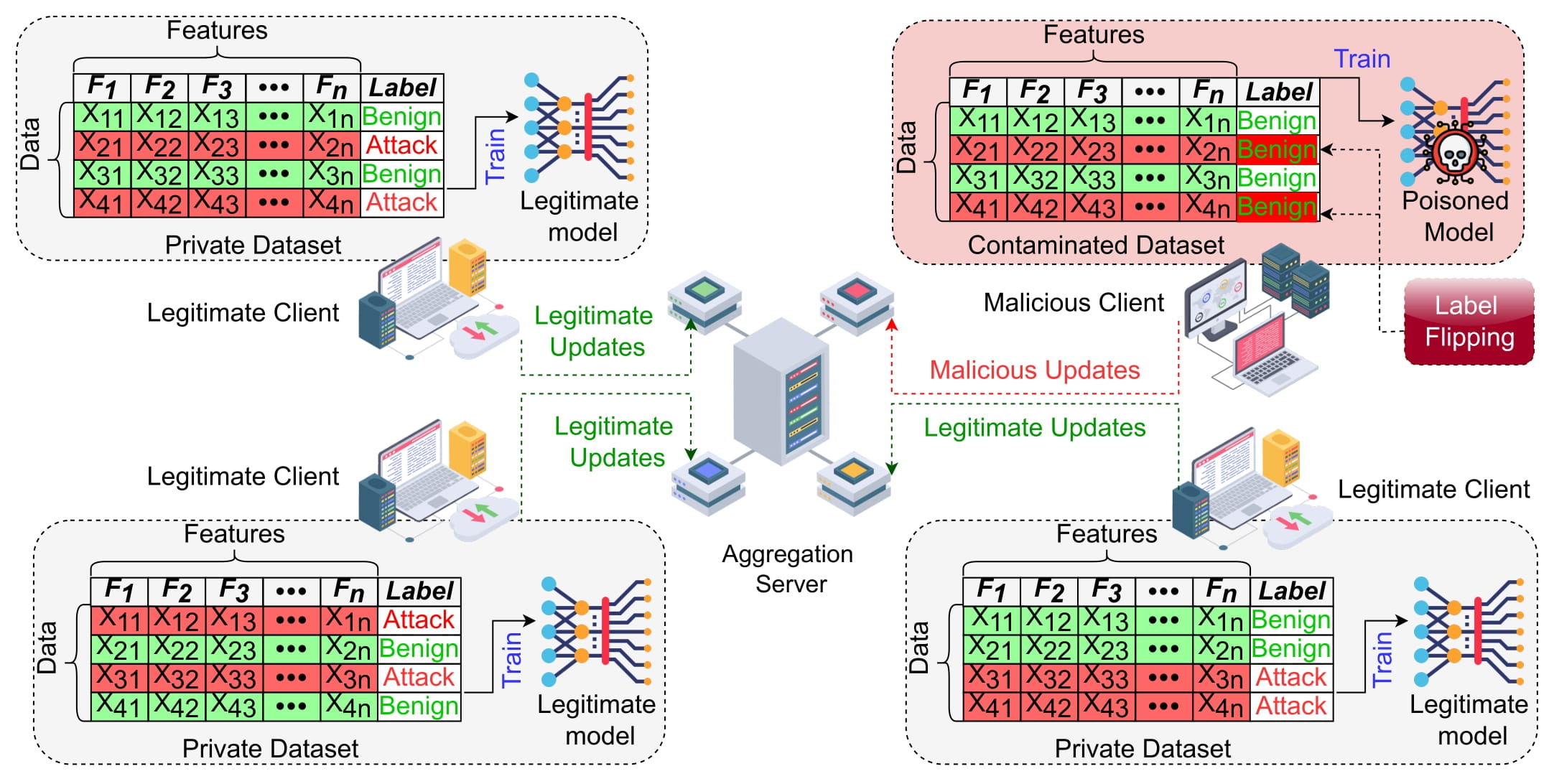}
\caption{FL-targeted label flipping attack}
\label{fig:fig-flip}
\end{figure*}

\subsubsection{Intrusion poisoning attack}

\textcolor{black}{Intrusion poisoning attacks can compromise 6G networks by feeding misleading data into the network's self-learning systems, leading to suboptimal decisions and vulnerabilities. Given 6G's reliance on AI-driven functionalities and ultra-reliable low-latency communication, malicious data injection can disrupt real-time operations. This attack can erode trust in the network's intelligence, undermining its promise of enhanced connectivity and smarter applications.} Specifically, the intrusion poisoning attack consists of \textcolor{black}{applying} poisoning attacks in the intrusion detection datasets Venkatesan \textit{et al.} \cite{venkatesan2021poisoning}. An intrusion detection system (IDS) is a network monitoring system that detects anomalous network behavior and alerts when anomalous activity is detected. An IDS is a computer program that analyzes a system or network for malicious activity or policy violations. However, \textcolor{black}{to} disturb the IDS systems, the attackers use intrusion poisoning attacks by implementing the poisoning attacks in the IDS dataset. The following datasets can be used for evaluating the performance of ML-based-IDS: Edge-IIoT dataset\cite{ferrag2022edge} and X-IIoTID dataset \cite{al2021x}.  

\subsubsection{Label flipping attack}
\textcolor{black}{Label flipping attacks involve maliciously altering the data labels to mislead training processes. In the context of 6G, which promises enhanced machine learning and AI integrations, a label-flipping attack could compromise the integrity of AI-driven network functions. Consequently, this could degrade network performance, compromise user experience, and potentially introduce vulnerabilities to other integrated systems.}
Therefore, this attack is conducted by flipping the labels of specific data samples from one class (the source class) to another (the target class), as shown in Figure \ref{fig:fig-flip}. Jebreel \textit{et al.} \cite{jebreel2022defending} introduced a new security system against the label-flipping attack in federated learning that \textcolor{black}{dynamically extracts} the target and source class potential gradients of local peer updates, performs an application of a clustering method on these gradients, and then evaluates the resulting groups to extract potential bad updates \textcolor{black}{before} the aggregation of the model. The following three data sets are used in the experiments: MNIST, CIFAR10, and IMDB. The results of three models CNN, ResNet18, and BiLSTM show that the proposed defense system can offer a lower attack success rate, higher source class accuracy, and lower test error. The following steps define the generation of label-flipping attacks in FL:

\begin{itemize}

\item Step 1: The malicious clients poison their local training data by flipping the labels of training examples from a source class ${Class}_{source}$ to a target class ${Class}_{target}$ without changing the features of the input data.

\item Step 2: The honest and malicious clients train their local models using the same hyper-parameters, model architecture, optimization algorithm, and loss function sent by the server.

\item Step 3: The honest and malicious clients train their local models (i.e., bad updates from malicious clients and good updates from honest clients) to the server. 
\end{itemize}

\subsubsection{Poisonous label attack} 
Compared to the poisonous attack under the white box requirement, the poisonous label attack is the blue box, which can enhance the misclassification error under a more constrained and workable condition of the poisonous label attack. \textcolor{black}{A Poisonous label attack on 6G could manipulate training data, skewing the model's understanding and behavior. As 6G heavily relies on AI-driven systems for optimization and decision-making, such attacks could compromise network performance and security. Malicious actors could exploit this vulnerability, causing communication disruptions or unauthorized data access.} The poisonous label attack involves its capability, the attacker’s knowledge, and the attacker’s goal. The attackers’ goals are identified by three components: error specificity, attack specificity, and security violation. Liu \textit{et al.} \cite{liu2021poisonous} propose the poisonous label attack, which injects fake images with the poisonous label in the training dataset than directly modifying the label of the images. The knowledge budget for the poisonous label attack is defined as:

\begin{equation}
{Budget}_{know}=\frac{N(w_{know})}{N(w)}+\frac{N({Dataset}_{know})}{N({Dataset}_{train})}
\end{equation}

Where $N(w)$ is the number of parameters of the victim model, $N({Dataset}_{train})$ is the number of samples in the training dataset, $N({Dataset}_{know})$ is the number of samples known by the malicious clients, $N(w_{know})$ is the number of parameters known by the malicious clients.

\subsubsection{Clean-label poisoning attack}
This is a class of poisoning attacks where the attacker has no control over the labeling procedure. In this particular threat model, the poison patterns are generated by inserting undetectable changes that will cause the model to misbehave in reaction to particular target inputs. \textcolor{black}{Clean-label poisoning attacks introduce subtly modified, malicious samples into the training data of machine learning models. In the context of 6G, which heavily relies on AI for network optimization and decision-making, such attacks could degrade network performance, compromise security protocols, and disrupt intelligent service provisions. The ubiquity and complexity of 6G networks amplify the potential impact and challenges of defending against these sophisticated threats.} Aghakhani \textit{et al.} \cite{aghakhani2020bullseye} propose a scalable clean-label poisoning attack, which enhances the current state-of-the-art attack success rate by 26.75\% in transfer learning while improving the speed of attack by a factor of 12. \textcolor{black}{This attack is based on injecting poisoned data into the training dataset so that the model learns to classify the target input incorrectly.} Based on the method that is based only on first-order information, Zheng \textit{et al.} \cite{zheng2021first} proposed a clean-label data poisoning attack based on first-order information. \textcolor{black}{The proposed attack is evaluated with the CIFAR-10 dataset on multiple network architectures: ResNet, VGG, and ConvNet. The attack uses a Convex Polytope (CP) to solve an optimization problem involving minimizing the distance between the target feature vector and the poisoned samples set. The coefficients of the poisoned samples set are specified by $V^i_j$. The attack aims to find a set of poisoned samples that can mislead the model when it encounters specific inputs.} The CP adopted solves the following optimization problem:

\begin{equation}
\frac{1}{2n}\sum^n_{i=1}{\frac{{\left\|{\alpha }^i\left(x_t\right)-\sum^m_{j=1}{V^i_j{\alpha }^i(x^j_p)}\right\|}^2}{{\left\|{\alpha }^i\left(x_t\right)\right\|}^2}}
\end{equation}

where $x^j_p$ is a poison samples set, ${\alpha }^i$ is the target feature vector, and $V^i_j$ specifies the $\ j$-th poison's coefficient. \textcolor{black}{The objective function is a convex combination of squared Euclidean distances, where a positive coefficient weights each distance. The denominator of each term ensures that the objective function is normalized by the length of the target feature vector $\alpha^i(x_t)$, which makes it invariant to the magnitude of the feature vector. The minimization problem seeks to find the value of $x_t$ that minimizes the objective function. Because the objective function is convex, any local minimum is also a global minimum. Therefore, any optimization algorithm that converges to a local minimum will also converge to the global minimum.}

\subsubsection{Data poisoning attack}
Data poisoning attacks against machine learning models have emerged as an influential adversarial machine learning research area. This form of attack occurs during the training phase of machine learning models. \textcolor{black}{Data poisoning attacks in 6G can introduce maliciously crafted data into the network's training datasets, undermining the AI-driven functions vital for 6G optimizations. As 6G relies heavily on machine learning for various processes, tainted data can lead to misinformed network decisions and potential service disruptions.} The attacker attempts to infect the training data by inserting well-formed samples to impose a harmful model on the learner. Zhang \textit{et al.} \cite{zhang2021data} proposed a data poisoning attack scheme called IMF. The IMF scheme adopts the bi-level optimization problem to formulate the poisoning attack. Specifically, the upper-level problem estimates the strength of the actual fabricated poisoning sample, while the lower-level problem implements the model parameters based on these poisoning samples. The iterative learning algorithm is implemented by the IMF scheme to solve the bi-level optimization problem. The performance evaluation on two datasets, namely, the Stanford Dogs dataset and the Osteoporotic Fracture dataset, shows that the MFI scheme has the potential to reshape interpretations of the target samples successfully. However, Takiddin \textit{et al.} \cite{takiddin2020robust} present a sequential ensemble detector against data poisoning attacks, which is based on feed-forward neural networks, gated recurrent units (GRUs), and a deep auto-encoder with attention (AEA).


\subsubsection{Generative poisoning attack}
This type consists of deploying a generative adversarial network (GAN) model from the attacker's side to impersonate other participants' training dataset patterns, which can initiate the poisoning attack successfully under an assumption of a more feasible threat. \textcolor{black}{A GAN-based poisoning attack in a 6G environment involves adversarial manipulation of training datasets, misleading network algorithms into making incorrect decisions.} Zhang \textit{et al.} \cite{zhang2020poisongan} propose a novel poisoning attack, PoisonGAN, which can be applied in federated learning settings. Specifically, the PoisonGAN attack is based on generative adversarial networks as presented in Figure \ref{fig:gan}. Chen \textit{et al.} \cite{chen2021deeppoison} propose a novel adversarial network, DeepPoison, based on one generator and two discriminators. More precisely, the generator automatically extracts hidden features from the target class and integrates them into benign training patterns. The one discriminator monitors the poisoning perturbation ratio. The second discriminator operates as a target model to witness the poisoning impacts. The idea of GAN is to build an adversarial game between a generator ${ Gen}$ and a discriminator ${Disc}$, where ${ Disc}$ is trained on the actual samples and the generated samples simultaneously, causing the generator coupled ${Gen}$ to produce samples that are close to the true ones. The objective functions used by the GAN model are shown as follows:

\begin{equation}
{{Learn\_algo}}_{{G}}\left({{Model_g}}\right){=}{{E}}_{{z}{\sim }{p}\left({{z}}_{{noise}}\right)}\left[{log}\left({Disc}\left({Gen}\left({z}\right)\right)\right)\right] 
\end{equation}

\begin{multline}
{{ Learn}{ \_}{ algo}}_{{ D}}\left({{{ Model}}_{{ d}}{ ,\ }{ Model}}_{{ g}}\right){ =}{{ E}}_{{ z}{ \sim }{ p}\left({{ x}}_{{ real}}\right)}\\\left[{ log}\left({ Disc}\left({ x}\right)\right)\right]{ +}{{ E}}_{{ z}{ \sim }{ p}\left({{ z}}_{{ noise}}\right)}\left[{ log}\left({ 1}{ -}{ Disc}\left({ Gen}\left({ z}\right)\right)\right)\right]
\end{multline}

\begin{figure*}
\centering
\includegraphics[width=0.9\textwidth]{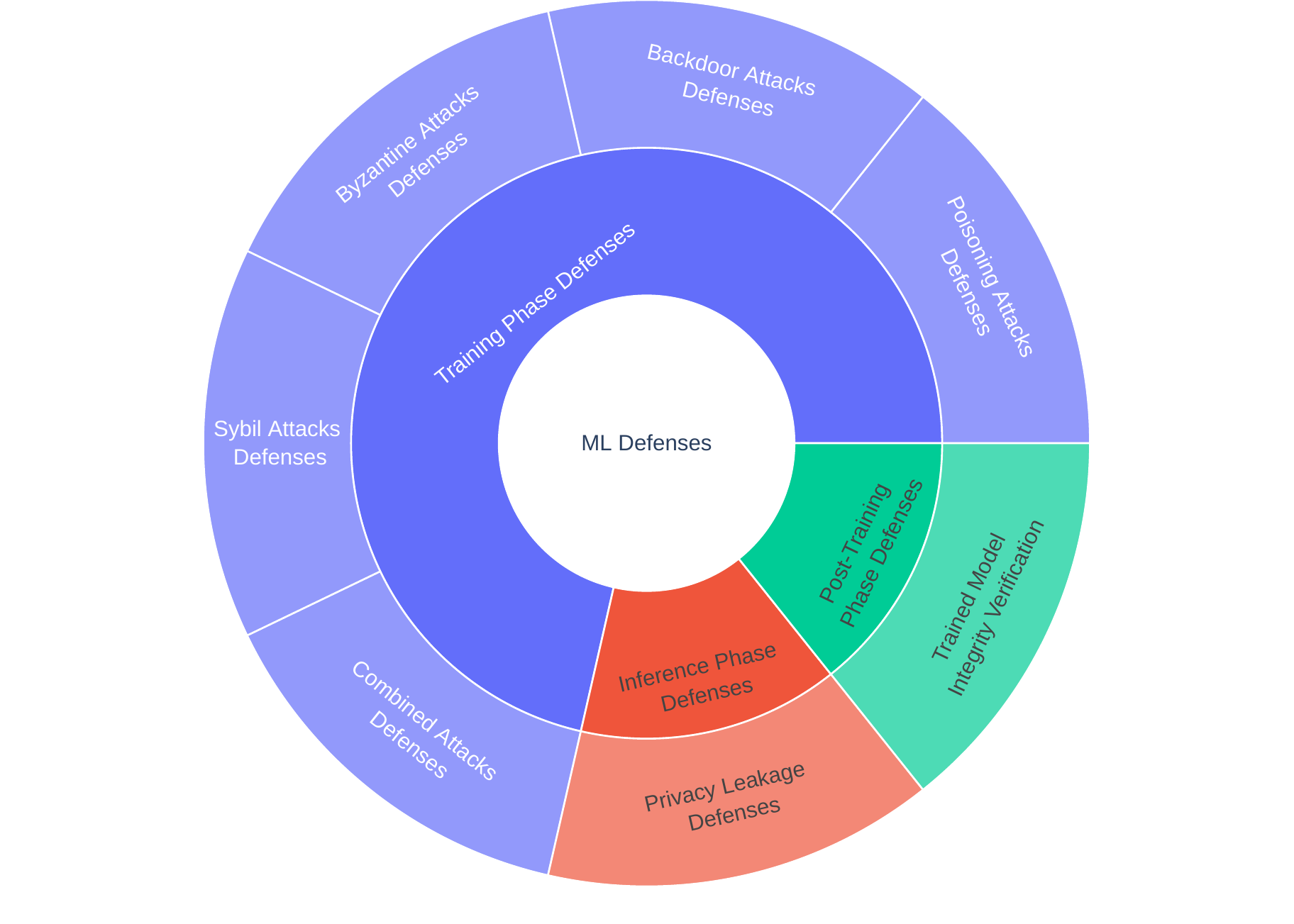}
\caption{Defense methods against machine learning vulnerabilities}
\label{fig:fig8}
\end{figure*}

\subsubsection{Over-the-air spectrum poisoning attack}
\textcolor{black}{Over-the-air spectrum poisoning attacks in 6G can maliciously introduce interference in the radio frequency spectrum, leading to degraded network performance. By exploiting the intelligent and adaptive nature of 6G's signal processing, adversaries can inject, amplify, or manipulate spectrum data. This disrupts the AI-driven operations in 6G networks, potentially leading to unreliable communication, misallocated resources, or complete service outages.} This poisoning attack focuses on the spectrum sensing period and handling the transmitter's input data in the test and training stage. An adversary performs these attacks primarily to confuse the sender into making poor transmission decisions (i.e., an evasion attack) or to tamper with the sender's recycling process (i.e., a causal attack). \textcolor{black}{To carry out an over-the-air spectrum poisoning attack, an attacker must know the transmitter's sensing parameters and the communication protocol used. Once the attacker has this information, he can craft adversarial examples that exploit the vulnerabilities of the transmitter's decision-making process.} Sagduyu \textit{et al.} \cite{sagduyu2019adversarial} present over-the-air spectrum sensing poisoning attacks based on the application of adversarial machine learning. Benchmarking shows that over-the-air spectrum sensing poisoning attacks are superior to conventional jamming attacks and substantially decrease the transmitter throughput. \textcolor{black}{Therefore, over-the-air spectrum poisoning attacks can be more effective than traditional jamming attacks, which flood the communication channel with noise to disrupt communication. Additionally, these attacks can significantly reduce the throughput of the transmitter, which can have significant consequences in applications such as wireless networks and IoT devices.}

\begin{table*}[h!]
\centering
\setlength{\tabcolsep}{2.5pt}
\renewcommand{\arraystretch}{1}
\caption{Defense methods against machine learning vulnerabilities.}
\centering
\label{tab:tab4}
\scriptsize
\begin{tabular}{|p{0.66in}|p{0.2in}|p{0.66in}|p{0.66in}|p{0.66in}|p{0.66in}|p{0.66in}|p{0.8in}|p{1.7in}|}
\hline
\textbf{Defense framework} & \textbf{Year} & \textbf{Threat model} & \textbf{Mitigation solution} & \textbf{Learning mode} & \textbf{Targeted ML} & \textbf{Classifiers} & \textbf{Datasets} & \textbf{Pros (+)  Open Issues (-)} \\ \hline
Blanchard \textit{et al.} \cite{blanchard2017machine} & 2017 & Byzantine attacks & Aggregation rules & Centralized learning & Spam filtering and image classification & A multilayer perceptron  & MNIST dataset  & + The proposed approach works against attacks involving up to 33\% of adversely affected parties \newline - Federated poisoning attack is not considered  \\ \hline
Fung \textit{et al.} \cite{fung2018mitigating} & 2018 & Sybil attacks & Contribution similarity & Federated learning & Attack classification, Image classification, and Face recognition &  SqueezeNet model & MNIST, VGGFace2, KDDCup, and Amazon datasets & +  Uses diversity of client updates to identify Sybil's poisoning.  \newline - Distributed backdoor attack is not considered \\ \hline
Qu \textit{et al.} \cite{qu2020blockchained} & 2020 & Poisoning attacks & Bio-inspired & Federated Learning & Image classification & - CNN models & CIFAR-10 dataset & + Fast convergence \newline + Defense against poisoning attacks \newline - Distributed backdoor attack is not considered \\ \hline
Wang \textit{et al.} \cite{wang2020deep} & 2021 & Data poisoning attacks & Experience-based learning & Centralized learning & Image classification & - Deep Q-Network \newline- LSTM model \newline- DT model & Beijing PM 2.5 dataset & + Robustness under data poisoning attacks \newline+ Training speed enhancement \newline - Generative poisoning attack is not considered    \\ \hline
Jangseung \textit{et al.} \cite{wolf2021jangseung} & 2021 & Poisoning attacks & Preprocessor-based & Centralized learning & Image classification & SVM model & MNIST and VCI Wisconsin breast cancer datasets& + The protected model outperformed the unprotected model in all best-case scenarios \newline - Federated poisoning attack is not considered    \\ \hline
Li \textit{et al.} \cite{li2021detection} & 2021 & Label-flipping attacks & Dimensionality-reduction and clustering & Federated learning & Image classification & DNN model & CIFAR-10 and Fashion-MNIST datasets & + Detect and mitigate data-poisoning attacks in Federated learning \newline - The authentication using cryptography is not applied\\ \hline
Chan \textit{et al.} \cite{chan2021transfer} & 2021 & Label-flipping attacks & Knowledge Transfer & Centralized learning &  Security-related applications & SVM model & Phishing Website Detection, Spam Assassin, and Letter Recognition datasets &  + Can significantly reduce the attack success rate \newline - Distributed backdoor attack is not considered
 \\ \hline
Stokes \textit{et al.} \cite{stokes2021preventing} & 2021 & Data poisoning attacks & Cryptography-based authentication & Centralized learning &  Machine learning software & N/A & Text-based datasets & + Authenticate both the trained model and the evaluation set \newline - Federated poisoning attack is not considered  \\ \hline
Shi \textit{et al.} \cite{shi2021mitigation} & 2021 &  Untargeted attack and targeted attack & Historical distance detection & Federated learning & Image Classification & CNN model & MNIST dataset & + It can standardize the scenario of federated learning \newline - Generative poisoning attack is not considered 
 \\ \hline
Andreina \textit{et al.} \cite{andreina2021baffle} & 2021 & Backdoor attacks & Feedback-based & Federated learning & Image classification &  ResNet18 CNN model & CIFAR-10 and FEMNIST datasets & + Achieve a false-positive rate below 5\% with a detection accuracy of 100\% \newline - The authentication using cryptography is not applied \\ \hline
Gu \textit{et al.} \cite{gu2021detecting} & 2021 &  Byzantine attacks & Architectural Style & Federated learning & Image classification & Conditional Variational Autoencoder & MNIST and FEMINIST datasets & + Can withstand Byzantine attacks and targeted model poisoning attacks \newline - Generative poisoning attack is not considered \\ \hline
Liu \textit{et al.} \cite{liu2021blockchain} & 2021 &  Combined attacks & Asynchronous convergence &  Federated learning & Image classification & CNN model & MNIST and CIFAR-10 datasets & + Achieve accuracy rates of 98.96\% with horizontal mode and 95.84\% vertical FL mode \newline - The authentication using cryptography is not applied \\ \hline
Qiu \textit{et al.} \cite{qiu2021mt} & 2021 & Combined attacks & Moving target defense & Centralized learning & Sign recognition and Image classification & DNN model & GTSRB and Imagenet datasets & + Resistance against Trojaning attacks \newline - Federated poisoning attack is not considered    \\ \hline
Shayan \textit{et al.} \cite{shayan2020biscotti} & 2021 & Sybil attacks & Verifiable random functions & Federated learning & Image classification & A logistic regression and softmax classifiers & Credit Card fraud and MNIST datasets & + Ability to withstand poisoning attacks and node churn \newline - The authentication using cryptography is not applied
 \\ \hline
Xie \textit{et al.} \cite{xie2022defending} & 2022 & Local poisoning attacks & Bio-inspired & Federated learning &  Image and text classification & - CNN models \newline - LSTM models & IMDB, CIFAR-10, MNIST, and SST-5 datasets &  + Lower success attack rate  \newline + Prevents attacks by extracting the benign client's features \newline - Distributed backdoor attack is not considered \\ \hline
Chen \textit{et al.} \cite{chen2022dynamic} & 2022 & Poisoning attacks & Reputation-awareness & Federated learning & Image classification & CNN model & MNIST, Fashion-MNIST, and CIFAR-10 datasets & + Achieve a 30\% reduction in learning time and is strong in resistance to representative poisoning attacks \newline - The authentication using cryptography is not applied
 \\ \hline
 Hou \textit{et al.} \cite{hou2021mitigating} & 2022 & Backdoor attacks & Federated Filters & Federated learning & Image classification & CNN model & MNIST and CIFAR10 datasets & + Backdoor recognition with the accuracy up to 99\%  \newline - The authentication using cryptography is not applied\\ \hline            
\textcolor{black}{Wang \textit{et al.}\cite{wang2023adaptive}} & \textcolor{black}{2023} & \textcolor{black}{Backdoor attacks} & \textcolor{black}{Adaptive clustering} & \textcolor{black}{Federated learning} & \textcolor{black}{Image classification} &  \textcolor{black}{CNN model} & \textcolor{black}{CIFAR-10, MNIST, and FEMNIST datasets} &  \textcolor{black}{+ Defend against backdoor attacks  \newline - Increased computational overhead and longer training times} \\ \hline
\textcolor{black}{Jiang \textit{et al.} \cite{jiang2023data}} & \textcolor{black}{2023} & \textcolor{black}{Label Flipping Attacks} & \textcolor{black}{Data quality detection mechanism} & \textcolor{black}{Federated learning} & \textcolor{black}{Image classification} & \textcolor{black}{CNN model} & \textcolor{black}{CIFAR-10 and Fashion-MNIST datasets}& \textcolor{black}{ + Defend against label flipping attacks with the lightweight generator  \newline - Introduce an additional layer of complexity and potential vulnerability} \\ \hline
\end{tabular}
\end{table*}

\subsection{Sybil attacks}
\textcolor{black}{Sybil attacks involve the creation of multiple fake identities to manipulate the training process of a federated learning system. Different types of Sybil attacks are presented, including Sybil-based poisoning attacks and Sybil-based collusion attacks \cite{fung2018mitigating}.} A Sybil attack involves using a single node inside the targeted network (or system) \textcolor{black}{to generate and operate multiple active false entities simultaneously}. The primary objective of these attacks is to gain the majority of influence within the targeted system to facilitate its manipulation.

\subsubsection{Sybil-based poisoning attack}
\textcolor{black}{A Sybil-based poisoning attack in 6G networks involves malicious actors creating multiple fake identities to influence the network's decision-making processes. By overwhelming the network with deceptive information from these fake nodes, the attack can degrade the quality of service, data integrity, and trustworthiness. This can lead to compromised network performance, erroneous decisions, and reduced security in the 6G ecosystem.} Sybils conduct federated learning poisoning attacks by delivering status updates that lead the distributed model to a poisoned common target.  Based on contribution similarity, Fung \textit{et al.} \cite{fung2018mitigating} propose a new defense system against Sybil-based poisoning attacks. The proposed system adapts the learning rate of clients based on contribution similarity.

\subsubsection{Sybil-based collusion attack} \textcolor{black}{In 6G networks prioritizing decentralized, AI-driven infrastructures and massive device connectivity, Sybil-based collusion attacks can introduce numerous malicious identities, misleading the system's decision-making processes. By artificially amplifying specific metrics or information, attackers can disrupt network services, degrade trustworthiness, and compromise AI-driven functionalities.} Xiao \textit{et al.} \cite{xiao2022sca} propose the Sybil-based collusion attack, in which the attacker employs the label flipping scheme to train the poison data locally and collide with other poisoned patterns. Meanwhile, The malicious adversary will virtually implement several Sybil nodes in the network so that the server chooses the collusion pattern to aggregate with a higher probability, constructing a global poisoning pattern. CIFAR-10 and Fashion-MNIST are benchmark datasets to evaluate attack performance in federated learning settings with convolutional Neural Networks (CNN). The experiment analysis shows that the proposed attack can achieve a more substantial attack effect.

\subsection{Byzantine attacks}
These attacks consist of Byzantine failures in federated learning, which can exploit the parameters of the local model on the affected devices during the learning phase. \textcolor{black}{Specifically, Byzantine attacks in federated learning involve malicious participants that deliberately manipulate the local model to corrupt the training process. These attacks can severely compromise the integrity and accuracy of the final global model. The participants performing Byzantine attacks are known as Byzantine faulty or malicious. These participants can exhibit arbitrary behavior and send incorrect or intentionally corrupted updates to the central server, leading to the poisoning of the global model. The attacker's knowledge of Byzantine attacks can be classified into No, Partial, and Complete. \textcolor{black}{Without knowledge}, the attacker has no specific knowledge of the distributed machine learning model, its architecture, or the data it's working with. An attacker with partial knowledge has some information about the distributed machine learning model or its architecture but not complete details. In the scenario of complete knowledge, the attacker can access all information related to the distributed machine learning system, including its architecture, model parameters, training data, and even updates from other nodes.}

\subsubsection{Federated learning-Byzantine attack} \textcolor{black}{Federated learning in 6G networks decentralizes training across multiple devices, boosting privacy and efficiency. However, a Byzantine attack, where malicious nodes provide faulty data or computations, can degrade the model's accuracy and compromise the collaborative learning process. Without robust defense mechanisms, 6G networks might suffer reduced performance, security vulnerabilities, and data integrity issues.} Different from the existing data poisoning attacks that affect the data collection integrity of the training datasets, Fang \textit{et al.} \cite{fang2020local} propose Byzantine attacks against federated learning which affect the integrity of the learning phase in the training process. The Byzantine attacks are formulated as optimization problems. The performance evaluation on four real-world datasets (i.e., Breast Cancer Wisconsin (Diagnostic), CHMNIST, Fashion-MNIST, and MNIST datasets) shows that the attacks against Byzantine-robust federated learning methods can substantially increase the error rates.

\subsection{Inference attacks}
In this type of attack, malicious users tend to exploit the already-trained models.

\subsubsection{\textcolor{black}{Model Inversion attack}}

\textcolor{black}{\textcolor{black}{Model inversion attacks in 6G could exploit the advanced machine learning frameworks, potentially revealing user data processed by these algorithms. As 6G networks rely heavily on AI for optimizing and personalizing services, such attacks could undermine user privacy. Without robust security measures, compromised data confidentiality might overshadow the envisioned benefits of 6G.} Model inversion attacks are a type of inference attack where an adversary attempts to infer sensitive information about the training data, such as inputs or parameters, by analyzing the output of the trained model. Fredrikson \textit{et al.} \cite{fredrikson2015model} demonstrated the feasibility of such attacks on machine learning models trained on sensitive data, such as medical records, by using optimization techniques to reconstruct input data that results in a given model output.}

\subsubsection{Membership inference attack}
\textcolor{black}{Membership inference attacks target machine learning models to determine whether a specific data point was part of the training set. In the context of 6G, which promises enhanced machine learning integration and ultra-reliable low-latency communications, such attacks could compromise user privacy by deducing user-specific data used during network optimizations.} A membership inference attack (MIA) is this kind of security vulnerability exposure where an adversary seeks to investigate whether an intended sample has been utilized for training the target ML model or not \textcolor{black}{based on} the model's behavior and output. Shokri \textit{et al.} \cite{shokri2017membership} conducted a blue-box MIA attack against ML with a binary NN classification task-based formalization with a shadow learning technique to identify members of the training set from non-members. Tang \textit{et al.}  \cite{tang2022mitigating} classified these attacks into two categories, namely, 1) direct attacks, which often make a single query to interrogate the targeted sample \textcolor{black}{directly}, and 2) indirect attacks, which often makes numerous queries to interrogate vicinity of the targeted sample to deduce membership.

\subsection{\textcolor{black}{Highlights the impacts of vulnerabilities in Peer-to-Peer IoT learning}}
\textcolor{black}{In this section, we have reviewed a variety of attacks that can significantly disrupt operations and compromise data in peer-to-peer IoT learning. Backdoor and poisoning attacks introduce malicious behavior into models or influence learning processes, while adversarial examples and evasion attacks can cause widespread model malfunctions or incorrect classifications. Combined attacks utilize multiple methods, making detection difficult \cite{mitev2023physical}. Sybil and byzantine attacks can disproportionately impact model learning or introduce unpredictability into the system. Dropping attacks disrupt the efficiency of the learning process. Inference, model extraction, and inversion attacks \textcolor{black}{threaten} data privacy by inferring sensitive information, copying proprietary models, or reconstructing original inputs from outputs. These vulnerabilities underscore the importance of strong security measures to protect data integrity, user privacy, and model robustness and accuracy.}

\begin{table*}[h!]
\centering
{\color{black}
\caption{\textcolor{black}{Impact of ML Attacks on 6G Features.}}
\label{tab:tab6gnet2}
\begin{tabular}{|p{2in}|p{1in}|p{1in}|p{1in}|p{1in}|}
\hline
ML Attack Type & URLLC & AI-Centric Networking & D2D Communication & Ubiquitous Sensing \\
        \hline
        RNN Backdoor  & $\checkmark$ & $\checkmark$ & $\checkmark$ & $\checkmark$ \\\hline
        Imperceptible Backdoor Pattern Attack & & $\checkmark$ & $\bigstar$ & \\ \hline
        Distributed Backdoor Attack & $\bigstar$ & & $\checkmark$ & \\ \hline
        Pixel-space Backdoor Attack & & $\checkmark$ & & $\checkmark$ \\ \hline
        Clean-label Backdoor Attack & & $\bigstar$ & & $\checkmark$ \\ \hline
        White-box Adversarial Attack & $\bigstar$ & $\checkmark$ & & \\ \hline
        Blue-box Adversarial Attack & $\bigstar$ & $\checkmark$ & & \\ \hline
        Grey-box Adversarial Attack & $\bigstar$ & $\checkmark$ & & \\ \hline
        Synergetic Attack & $\checkmark$ & $\checkmark$ & $\checkmark$ & $\checkmark$ \\ \hline
        Federated Poisoning Attack & & $\bigstar$ & $\checkmark$ & \\ \hline
        Intrusion Poisoning Attack & $\checkmark$ & & $\checkmark$ & $\bigstar$ \\ \hline
        Label Flipping Attack & & $\checkmark$ & & $\checkmark$ \\ \hline
        Poisonous Label Attack & & & & $\checkmark$ \\ \hline
        Clean-label Poisoning Attack & & $\bigstar$ & & $\checkmark$ \\ \hline
        Data Poisoning Attack & & & $\checkmark$ & $\checkmark$ \\ \hline
        Generative Poisoning Attack & & $\checkmark$ & & $\bigstar$ \\ \hline
        Sybil-based Poisoning Attack & & & $\checkmark$ & \\ \hline
        Sybil-based Collusion Attack & & & $\checkmark$ & \\ \hline
        Federated learning-Byzantine Attack & $\checkmark$ & $\bigstar$ & $\checkmark$ & \\ \hline
        Model Inversion Attack & & $\checkmark$ & & $\bigstar$ \\ \hline
        Membership Inference Attack & & $\checkmark$ & & $\checkmark$ \\ \hline
\end{tabular}\\
"$\checkmark$" indicates that the attack significantly impacts or exploits the specific feature of 6G.\\
"$\bigstar$" indicates a partial impact on the specific 6G feature.\\
An empty cell indicates that the attack isn't directly related to the 6G feature.}
\end{table*}

\subsection{\textcolor{black}{Impact of ML Attacks on 6G Features}}

\textcolor{black}{As the next frontier in wireless communications, 6G is set to bring a myriad of advancements, particularly in areas such as Ultra-Reliable and Low-Latency Communication (URLLC), AI-Centric Networking, Device-to-Device (D2D) Communication, and Ubiquitous Sensing \& Data Collection. However, with these advancements come potential vulnerabilities, especially in the context of machine learning attacks.}

\textcolor{black}{Table \ref{tab:tab6gnet2} gives an in-depth look into how various ML attack types may exploit or be impacted by the key features of 6G. Notably, RNN Backdoor Attacks appear to exploit all the highlighted 6G features, underlining the multifaceted threats posed by this attack type. Moreover, the interplay between AI-Centric Networking and various adversarial attacks, including White-box, Blue-box, and Grey-box, emphasizes the potential risks of deeply embedding AI into 6G's core functions.}

\textcolor{black}{Device-to-Device communication, allowing for direct device interaction without centralized routing, presents another vulnerable point, susceptible to Distributed Backdoor Attacks, Federated Poisoning Attacks, and others. This shows the importance of establishing robust security mechanisms for direct communication protocols in 6G networks.}

\textcolor{black}{Lastly, new vulnerabilities may arise with 6G's potential to expand sensing technologies and become more context-aware through Ubiquitous Sensing. This is evident in its potential susceptibility to Pixel-space Backdoor Attacks, Label Flipping Attacks, and Data Poisoning Attacks. As 6G develops, understanding and mitigating these vulnerabilities becomes paramount to ensure a secure, reliable network.}

\section{Defense Methods Against \textcolor{black}{Edge Learning} Vulnerabilities}\label{sec:5}
The previous section highlighted the potential risks facing future 6G-IoT systems, as AI will be a key player in ensuring these systems' proper functioning, optimizing, and safeguarding. However, to take full advantage of AI's benefits while avoiding its associated security risks, the cybersecurity research community has made significant efforts to make this possible. In this section, we will attempt to comprehensively shed light on the importance of such efforts and provide a classification of defensive mechanisms against ML attacks, as presented in Figure \ref{fig:fig8}. Table \ref{tab:tab4} lists state-of-the-art methods proposed for securing machine learning systems. Table \ref{tab:tab5} summarizes defense methods against federated machine learning vulnerabilities.

\begin{figure*}
\centering
\includegraphics[width=0.9\textwidth]{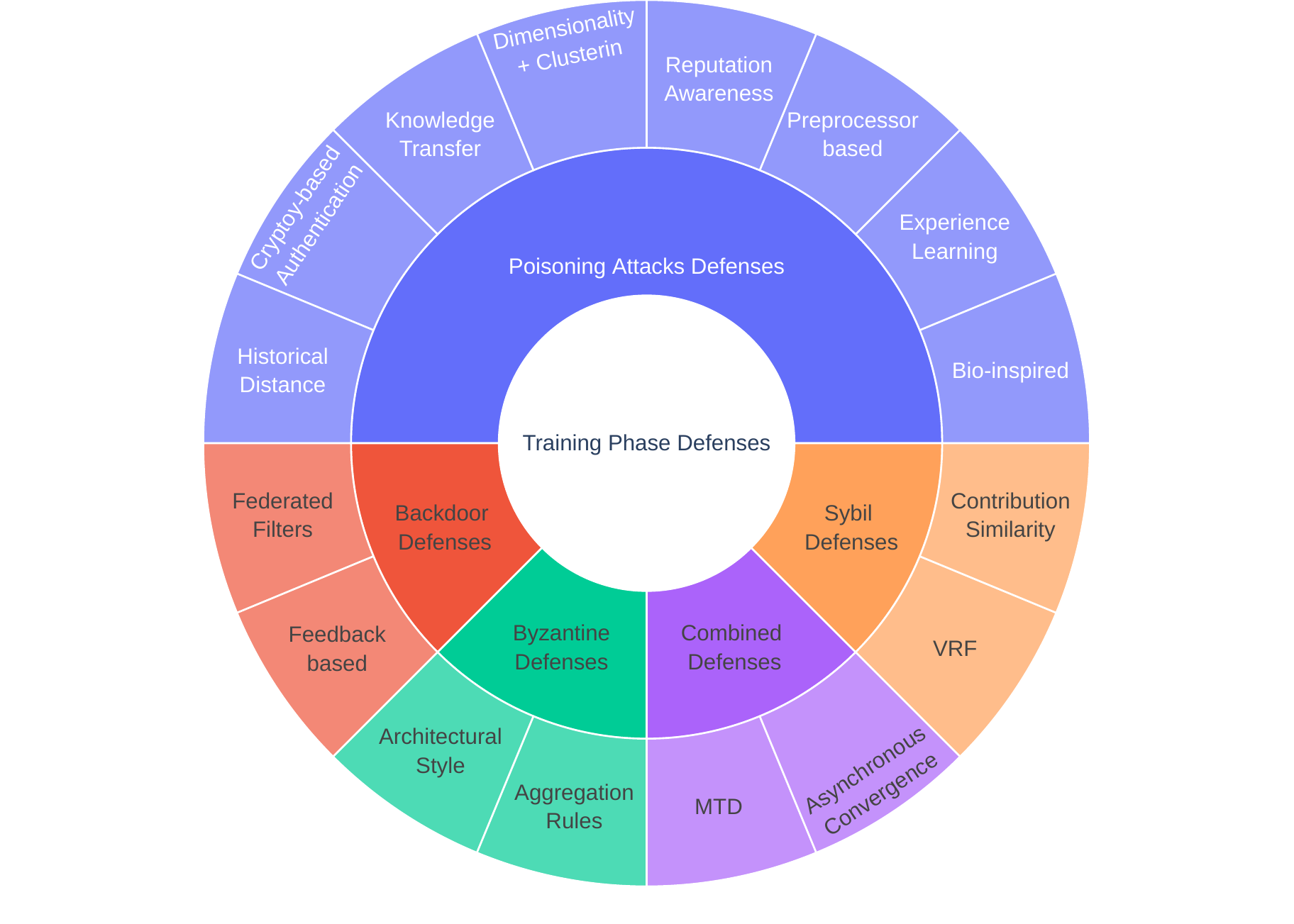}
\caption{Training phase defense methods against machine learning vulnerabilities}
\label{fig:fig9}
\end{figure*}

\begin{table*}[h!]
\centering
\setlength{\tabcolsep}{2.5pt}
\renewcommand{\arraystretch}{1}
\caption{Summary of defense methods against edge learning vulnerabilities.}
\centering
\label{tab:tab5}
\scriptsize
\begin{tabular}{|p{1in}|p{1in}|p{1in}|p{2.5in}|p{0.7in}|}
\hline
\textbf{Defense methods}                                                                         & \textbf{Defense mechanisms}                                                                        & \textbf{Defense strategy}                 & \multicolumn{1}{c|}{\textbf{Methodology}}                                                                                                                          & \textbf{Reference}                                                                                  \\ \hline
\multirow{17}{*}{\begin{tabular}[c]{@{}l@{}}Training Phase \\ Defense Methods\end{tabular}}      & \multirow{9}{*}{\begin{tabular}[c]{@{}l@{}}Poisoning Attacks \\ Defense Mechanisms\end{tabular}}   & \multirow{2}{*}{Bio-inspired}             & The idea is inspired by the   \textit{biological immune system} and deployed in the multi-party learning   scenario                                    & Xie   \textit{et al.} \cite{xie2022defending}                  \\ \cline{4-5} 
                                                                                                 &                                                                                                    &                                           & The authors propose a decentralized   cognitive computing model (D2C)                                                                                                  & Qu   \textit{et al.} \cite{qu2020blockchained}                 \\ \cline{3-5} 
                                                                                                 &                                                                                                    & Experience-based Learning                 & A Deep Q-network (DQN)-based feature   selection approach is proposed to mitigate poisoning attacks                                                                     & Wang \textit{et al.}   \cite{wang2020deep}                     \\ \cline{3-5} 
                                                                                                 &                                                                                                    & Preprocessor-based                        & A preprocessor for minimizing the   effects of poisoning attacks without affecting model performance                                                                    & Jangseung \textit{et al.}   \cite{wolf2021jangseung}           \\ \cline{3-5} 
                                                                                                 &                                                                                                    & Reputation-Awareness                      & Provides a dynamic reputation   measurement system among clients participating in learning                                                                              & Chen \textit{et al.}   \cite{chen2022dynamic}                  \\ \cline{3-5} 
                                                                                                 &                                                                                                    & Dimensionality-reduction   and Clustering & A Kernel Principal Component Analysis   (KPCA) and K-mean-based defense strategy against FL-targeted data-poisoning   attacks                                           & Li \textit{et al.}   \cite{li2021detection}                    \\ \cline{3-5} 
                                                                                                 &                                                                                                    & Knowledge   Transfer                      & A transfer learning-based defense   against label-flipping attacks                                                                                                      & Chan \textit{et al.}   \cite{chan2021transfer}                 \\ \cline{3-5} 
                                                                                                 &                                                                                                    & Cryptography-based   Authentication       & A cryptography-based authentication   technique, called VAMP, to protect the integrity of training data                                                                 & Stokes \textit{et al.}   \cite{stokes2021preventing}           \\ \cline{3-5} 
                                                                                                 &                                                                                                    & Historical   Distance Detection           & A defense strategy based on the   analysis of the statistical relationship of the Euclidean distance between   clients' models                                          & Shi   \textit{et al.} \cite{shi2021mitigation}                 \\ \cline{2-5} 
                                                                                                 & \multirow{2}{*}{\begin{tabular}[c]{@{}l@{}}Backdoor Attacks \\ Defense Mechanisms\end{tabular}}    & Federated Filters                         & The federated backdoor filter defense   is proposed in order to identify backdoor inputs                                                                                & Hou   \textit{et al.} \cite{hou2021mitigating}                 \\ \cline{3-5} 
                                                                                                 &                                                                                                    & Feedback-based                            & A feedback-based FL backdoor detection   technique called BaFFLE is proposed, which is based on several clients' data   for both training and model tampering discovery & Andreina \textit{et al.}   \cite{andreina2021baffle}           \\ \cline{2-5} 
                                                                                                 & \multirow{2}{*}{\begin{tabular}[c]{@{}l@{}}Byzantine   Attacks \\ Defense Mechanisms\end{tabular}} & Architectural   Style                     & An unsupervised Conditional Variational   Autoencoder (CVAE)-based training anomaly detection system                                                                    & Gu \textit{et al.}   \cite{gu2021detecting}                    \\ \cline{3-5} 
                                                                                                 &                                                                                                    & Aggregation   Rules                       & Chooses and eliminates the most distant   vectors with respect to the calculated average from the aggregated gradient   updates                                         & Blanchard \textit{et al.}   \cite{blanchard2017machine}        \\ \cline{2-5} 
                                                                                                 & \multirow{2}{*}{\begin{tabular}[c]{@{}l@{}}Sybil   Attacks \\ Defense Mechanisms\end{tabular}}     & Verifiable   Random Functions             & A secure blockchain-based framework for   fully decentralized multi-party privacy-preserving ML                                                                         & Shayan \textit{et al.}   \cite{shayan2020biscotti}             \\ \cline{3-5} 
                                                                                                 &                                                                                                    & Contribution   Similarity                 & A defensive discriminant technique to   distinguish sybils from legitimate users                                                                                        & Fung \textit{et al.}   \cite{fung2018mitigating}               \\ \cline{2-5} 
                                                                                                 & \multirow{2}{*}{\begin{tabular}[c]{@{}l@{}}Combined   Attacks \\ Defense Mechanisms\end{tabular}}  & Moving   Target Defense                   & an MTD-based defense strategy against   Trojan attacks on DNN networks                                                                                                  & Qiu \textit{et al.} \cite{qiu2021mt}                           \\ \cline{3-5} 
                                                                                                 &                                                                                                    & Asynchronous   Convergence                & A security architecture dedicated to a   secure FL-based learning workflow                                                                                              & Liu \textit{et al.}   \cite{liu2021blockchain}                 \\ \hline
\multirow{2}{*}{\begin{tabular}[c]{@{}l@{}}Post-Training  \\ Phase Defense Methods\end{tabular}} & \multirow{2}{*}{\begin{tabular}[c]{@{}l@{}}Trained   Model \\ Integrity Verification\end{tabular}} & Cryptographically-protected   Provenance  & A security scheme based on Blockchain   and computing fuzzy hash values                                                                                                 & Unal \textit{et al.}   \cite{unal2021integration}              \\ \cline{3-5} 
                                                                                                 &                                                                                                    & Optimization-based                        & A Bayesian Compromise Detection (BCD)   algorithm to solve an optimization problem                                                                                      & Kuttichira   \textit{et al.} \cite{kuttichira2022verification} \\ \hline
\multirow{2}{*}{\begin{tabular}[c]{@{}l@{}}Inference Phase \\ Defense Methods\end{tabular}}      & \multirow{2}{*}{\begin{tabular}[c]{@{}l@{}}Privacy Leakage \\ Defense Mechanisms\end{tabular}}     & Gradients Shielding                       & Secures exchanged aggregation with   homomorphic encryption and differential privacy                                                                                    & Hao   \textit{et al.} \cite{hao2019efficient}                  \\ \cline{3-5} 
                                                                                                 &                                                                                                    & Self-Distillation                         & Generates equivalent member/non-member   feedback to alleviate blue-box membership inference threats                                                                   & Tang   \textit{et al.} \cite{tang2022mitigating}               \\ \hline
\end{tabular}

\end{table*}

\subsection{Training Phase Defense Methods}
We start with the most vulnerable stage of ML, namely the training phase. We classify the deference mechanisms in the training phase against five threat categories: poisoning,  backdoor,  byzantine, Sybil, and combined attacks, as shown in  Figure \ref{fig:fig9}.

\subsubsection{Poisoning Attacks Defense Mechanisms}
Based on a selection of relevant recent literature, we classify the works that belong to this class into eight subclasses, namely bio-inspired, experience-based learning, preprocessor-based, reputation-awareness, dimensionality-reduction and clustering, knowledge transfer, cryptography-based authentication, and historical distance detection, as presented below:

\begin{itemize}
    \item \textbf{\textit{Bio-inspired:}} These defensive mechanisms are motivated by biological functions or structures. For example, Xie \textit{et al.} \cite{xie2022defending} propose an AI-targeted local poisoning attacks defense mechanism, which is inspired by the \textit{biological immune system} and implemented in a multiple-party training environment. The proposed system supports antigen identification, immune reaction, and immunological recall with an adversarial pipeline, with no restrictions concerning malicious clients' numbers or \textcolor{black}{the period} they participate in the process. Also, it can make adaptive judgments regarding the aggregation weights of different local models. Detailed experimental outcomes across image and text datasets, including MNIST, CIFAR-10, IMDB, and SST-5, with different neural networks, including CNN and LSTM,  and under different poisoning attacks, such as local label flipping attacks. The results proved the merit of the proposed system concerning model efficiency and resilience against poisoning attacks. Specifically, the accuracy reaches 95\% even with 75\% of the participating clients being compromised. Another Bio-inspired technique used for defending against poisoning attacks is the cognitive computing paradigm, which attempts to reproduce human cognitive processes within a computerized model. Qu \textit{et al.} \cite{qu2020blockchained} propose a \textit{decentralized cognitive computing} model (D2C) using a Blockchained FL framework for IIoT, specifically, for Industry 4.0 systems. The authors introduced blockchain FL into cognitive computing learning to enhance its privacy-preserving abilities in a fully decentralized manner. In addition, the authors introduced a modified Markov decision process (MDP) for performance optimization purposes. Evaluation of the CIFAR-10 dataset regarding three different aspects,  global model accuracy,  model convergence, and resistance to poisoning attacks, indicated the feasibility of the proposed system.
    
    \item \textbf{\textit{Experience-based Learning:}} such defensive mechanisms incorporate practical, real-life experiences into the systems learning workflow. Wang \textit{et al.} \cite{wang2020deep} propose a Deep Q-network (DQN)-based feature selection approach for cleaning multi-source data and mitigating poisoning attacks, which \textcolor{black}{RL influences} in terms of an experience-based learning paradigm. Specifically, the problem is formulated as a contest in dynamic states among agents and the environment. To circumvent the problem of high computational complexity, the authors provided SS, a spatial search algorithm for quicker learning of the DQN agent. Performed simulations on Beijing PM 2.5, Colon Data, and MNIST datasets proved that the proposed approaches can successfully mitigate data poisoning attacks.
    
    \item \textbf{\textit{Preprocessor-based:}} These defensive mechanisms check the authenticity of the data before using it for training. Jangseung \textit{et al.} \cite{wolf2021jangseung} present Jangseung, a preprocessor for minimizing the effects of poisoning attacks without affecting model performance. The system is specifically developed to protect SVMs models from adversarial perturbations through anomaly detection engines. The approach involves \textcolor{black}{identifying} whether a given data point is an outlier based on historical training data, and newly entered data elements would be flagged and rejected as either faulty or malicious. The performance evaluation step using the MNIST and VCI Wisconsin breast cancer datasets involved training two replicate models under the same adversarial points, except one is protected by the proposed system. The results demonstrated that the protected model outperformed the unprotected model in all best-case scenarios, with 96.2\% versus 53.2\% with the MNIST dataset, as well as 88.18\% versus 75.51\% with the VCI breast cancer Wisconsin dataset.
    
    \item \textbf{\textit{Reputation-Awareness:}} such defensive mechanisms check the genuineness level of the clients before allowing them to participate in the model training. Chen \textit{et al.} \cite{chen2022dynamic} propose RAPFDL, a dynamic asynchronous anti-poisoning FL framework. The main concept is to provide a dynamic reputation measurement system among clients participating in learning. Individual clients can collect points by committing to update their local model parameters and exchange the earned points. As a result, customers are incentivized to download more updates to earn more reward points.\textcolor{black}{ The correlation coefficient between the accuracy of the final model and the individual contributions of the parties measures the level of fairness}. Clients use a differentially private GAN (DPGAN) to generate artificial private data samples, and all of the exchange updates are encrypted using the improved additive homomorphic encryption. Then, the updates are stored in the blockchain as unmodifiable records, which ensures both audibility and transparency. It is demonstrated by the experimental results on MNIST, Fashion-MNIST, and CIFAR-10 datasets that RAPFDL can achieve a 30\% reduction in learning time and is strong in resistance to representative poisoning attacks.

    \item \textbf{\textit{Dimensionality-reduction and Clustering:}} These defensive mechanisms facilitate data verification. Li \textit{et al.} \cite{li2021detection} present a Kernel Principal Component Analysis (KPCA) and K-mean-based defense strategy against FL-targeted data-poisoning attacks, and precisely, label-flipping attacks. The main reason for this combination is that KPCA, an algorithm for dimensionality reduction \textcolor{black}{can} handle non-linear data efficiently while coping well with linear data (e.g., skewed data). The KPCA algorithm is consequently effective in locating and filtering out malicious updates, as these updates contain unique features, while k-mean clustering is used to reduce noise.
    \item \textbf{\textit{Knowledge Transfer:}} In these defensive mechanisms, the knowledge already acquired is used to sustain new training. Chan \textit{et al.} \cite{chan2021transfer} propose a transfer learning-based defense against label-flipping attacks. \textcolor{black}{At the same time}, different studies aim to exclude malicious data from the training process. The authors considered the problem of fully using contaminated samples in training in cases where the first approach cannot be achieved. The problem is expressed as transfer learning where the footprint of malicious samples is downplayed by retrieving only information similar to non-malicious samples in the affected datasets. In addition, a weight initialization technique is proposed which allocates a sample weight according to the appropriateness of its predicted cost provided by a classifier against poisoning attacks.

    \item \textbf{\textit{Cryptography-based Authentication:}} These defensive mechanisms are designed to detect any possible manipulation of the previously verified data, which protects the data from tampering. Stokes \textit{et al.} \cite{stokes2021preventing} present a cryptography-based authentication technique, called VAMP, to protect the integrity of training data. To provide the metadata for the media objects, the manifest is used. The main idea behind this concept involves protecting the data sets for training and validation and safeguarding any existing software package utilized to train and evaluate the model. To \textcolor{black}{protect}, the data sets manifests for the training and validation datasets such as metadata and data bindings are created to be either published on the VAMP service or integrated directly into the respective datasets. Then, both the training and evaluation packages are then uploaded to the service. The proposed system uses SHA2 (256 or 512) to create manifests, while their serialization can be performed using JSON for textual datasets or CBOR for binary implemented datasets.
    
    \item \textbf{\textit{Historical Distance Detection:}} In such mechanisms, the objective is to inspect each client update before using it. Shi \textit{et al.} \cite{shi2021mitigation} propose a defense mechanism based on \textcolor{black}{analyzing} the statistical relationship of the Euclidean distance between clients' models. This defense mechanism is constructed based on the following statements. The distance between honest clients' models is not similar to the distance between harmful clients' models. \textcolor{black}{Most distances are similar because the number of malicious clients is much lower than that of honest clients}. Based on this, the primary solution proposed is to force the aggregation server to choose the possible benign clients, which are established on the minimal sum of their distance from each other. Experiment evaluations on MNIST with different data distribution techniques (IID and non-IID) validated the feasibility of the proposed system.
\end{itemize}

\subsubsection{Backdoor Attacks Defense Mechanisms}
We classify the works that belong to this class into two subclasses, namely federated filters and feedback-based mechanisms, as given below:
\begin{itemize}
    \item \textbf{\textit{Federated Filters:}} In these mechanisms, the implementation takes place on the aggregation server side and it can be used for client monitoring and security attack filter distribution. Hou \textit{et al.} \cite{hou2021mitigating} focused on proposing a defensive mechanism against backdoor attacks in IIoT-FL environments. The authors trained multiple backdoor filters and various combinations of XAI models and classifiers on the server side to guarantee the identification of the backdoor entries. In addition, the authors proposed a blur-label-flipping strategy to sanitize the backdoor locations, allowing data availability reclaiming. The results of experimental evaluations on MNIST and CIFAR10 datasets prove the system's validity with an accuracy of up to 99\% in recognizing backdoor samples.
    
    \item \textbf{\textit{Feedback-based:}} Such mechanisms are designed to monitor and provide authenticity feedback on participating clients continuously. Andreina \textit{et al.} \cite{andreina2021baffle} propose a feedback-based FL backdoor detection technique called BaFFLE, which uses several clients' data for training and model tampering discovery. The defense strategy consists of taking advantage of the existence of diverse datasets across clients by embedding a feedback loop within the FL workflow.  To address the various challenges of running a distributed protocol in a Byzantine environment, the authors used an off-the-shelf procedure to locally (clients-side) benchmark and compare the updated model's classification performance against the former one while excluding updates that demonstrate unexpected behavior. The proposed system is evaluated using the CIFAR-10 and FEMNIST datasets, which results show 100\% accuracy with less than 5\% false positive rate.
\end{itemize}

\subsubsection{Byzantine Attacks Defense Mechanisms}
We organize the works that belong to this class into two subclasses, namely architectural style and aggregation rules, as provided below:
\begin{itemize}
    \item \textbf{\textit{Architectural Style:}} Such mechanisms exploit the architectural learning style to detect and eliminate malicious updates. Gu \textit{et al.} \cite{gu2021detecting} proposed an unsupervised Conditional Variational Autoencoder (CVAE)-based training anomaly detection system, called Fedcvae, which aims to accurately discover and eradicate malicious model updates to negate their harmful consequences. The architecture of Fedcvae consists of an \textit{encoder-decoder} design. Using the encoder, the original variables are mapped into low-dimensional embeddings while the decoder rebuilds their initial variables. The performance evaluation under four different datasets, namely Vehicle, Synthetic, MNIST, and FEMINIST (under both IID and non-IID data distribution approaches) shows that the proposed system can withstand Byzantine attacks and targeted model poisoning attacks with up to 30\% of the participating clients being malicious.
    
    \item \textbf{\textit{Aggregation Rules:}} The main purpose of these mechanisms is to ensure safe learning, whether all clients are honest or there is a subset of malicious clients. Blanchard \textit{et al.} \cite{blanchard2017machine} investigated Byzantine failure robustness for distributed Stochastic Gradient Descent (SGD)-based frameworks. The research seeks to address the problem of Byzantine failure resilience of an SGD system with no limitation on the dimension or parameter space size. The proposed solution entails \textcolor{black}{formulating} a resilience property of the aggregation rule that grasps the core demands for ensuring convergence regardless of having a subset of Byzantine parties. This technique chooses and eliminates the most distant vectors \textcolor{black}{concerning} the calculated average from the aggregated gradient updates. The distance is computed using the Euclidean spacing among the gradient vectors. The approach works well against attacks involving up to 33\% of adversely affected parties.
\end{itemize}

\subsubsection{Sybil Attacks Defense Mechanisms}
We classify the works that belong to this class into two subclasses, including verifiable random functions and contribution similarity, as presented below:
\begin{itemize}
    \item \textbf{\textit{Verifiable Random Functions:}} In short, a verifiable random function (VRF) is given two inputs (a secret key and a seed) and returns two associated values (hash and proof). The proof allows everyone who is carrying the public key of the given peer to confirm that the given hash was indeed produced by a peer who has the private key. Shayan \textit{et al.} \cite{shayan2020biscotti} present Biscotti, a secure blockchain-based framework for fully decentralized multi-party privacy-preserving ML. The proposed system implements a robust protocol built upon the most recent block hash and VRFs for subgroup selection among peers carrying out individual steps (noise addition, updates validations, and secure aggregations). The proposed system is based on the Proof of Federation (PoF) protocol, a blockchain consensus protocol that employs FL defenses and renders them enforceable across decentralized P2P environments. As the selection of subgroups is performed by the VRF, the stake of the peer is treated as the individual's reputation acquired by contributing favorably to the shared model. This guarantees that an opponent \textcolor{black}{cannot} expand its leverage in the system by establishing too many peers and still not enhancing the model, which helps mitigate the effect of Sybils. The experimental results with 26 committee sizes provide immunity against an opponent who controls 30\% of the system's stakes.
    \item \textbf{\textit{Contribution Similarity:}} Fung \textit{et al.} \cite{fung2018mitigating} present a defensive discriminant technique to distinguish Sybils from legitimate users based on the diversity of their relative gradient updates. The proposed learning parameters \textcolor{black}{use} a per-client adaptive learning rate that relies on the contribution similarity among the clients. The identification of malicious users is based on a common goal. Therefore their model updates are susceptible to low variance, which is done by searching for similarities regarding indicative characteristics. Experimental evaluations on four datasets, namely MNIST, VGGFace2, KDDCup, and Amazon, with 5 attack scenarios indicated that the proposed system, in conjunction with other modules (such as Multi-Krum), can successfully mitigate a range of different attack types and even when Sybils submerge legitimate users.
\end{itemize}

\subsubsection{Defense Mechanisms Against Combined Attacks}
We organize the works \textcolor{black}{in} this class into two subclasses, including Moving Target Defense and Asynchronous Convergence.
\begin{itemize}
    \item \textbf{\textit{Moving Target Defense:}} MTD aims to shift the surface available for attacks actively. Qiu \textit{et al.} \cite{qiu2021mt} propose MT-MTD, an MTD-based defense strategy against Trojan attacks on DNN networks. The framework takes an attack-defense game approach, specifically, a signaling game. MT-MTD has four main steps. The first is a dimensional division of the training set by the defender and then passed to the attacker for training. The second is a random selection from the resulting collection of dimensional combinations. The third step is the weight adjustment. The last is the consensus process. 
    
    \item \textbf{\textit{Asynchronous Convergence:}} used when aggregation can be done asynchronously. Liu \textit{et al.} \cite{liu2021blockchain} propose a security architecture dedicated to a secure FL-based learning workflow. The architecture is composed of two parts. The first one, named FedBlock, is proposed to introduce decentralization in the learning process through the use of the blockchain, while the second one, named FedAC, permits the FL to carry out a global aggregation in an asynchronous manner while considering a staleness coefficient. The system is dedicated to securing the learning process against many threats, including Single Point Of Failure (SPOF), unconventional learning failures, and dedicated attacks. Experimental evaluation on a physical deployment, namely a Raspberry Pi (4b) under the MNIST dataset, shows that the system can reach accuracy scores as high as 98.96\% in horizontal FL and 95.84\% in vertical FL. The reported results of experimental evaluations on the CIFAR-10 and Fashion-MNIST datasets show that the proposed approach can be effective in identifying and differentiating honest updates from malicious ones.
\end{itemize}


\subsection{Post-Training Phase Defense Methods}
The defense mechanisms in this class tend to safeguard the model from tempering after the training phase. 
\subsubsection{Trained Model Integrity Verification}
We organize the works that belong to this class into two subclasses, namely cryptographically-protected Provenance and optimization-based techniques, as provided below:
\begin{itemize}
    \item \textbf{\textit{Cryptographically-protected Provenance:}} these defensive mechanisms protect the model trained using encryption. Unal \textit{et al.} \cite{unal2021integration} propose a security scheme based on Blockchain and computing \textit{fuzzy hash} values for protecting FL algorithms operating in IoT systems. Notably, the proposed scheme stores the model parameters in the blockchain, although this is true to some extent, the model parameters on the blockchain are not stored unchanged. In other words, the authors applied a one-way hash function upon these model parameters before their storage on the blockchain. Thus, the introduced scheme provides an efficient approach that does not affect FL's privacy concerns. In another approach under the same context, Stokes \textit{et al.} \cite{stokes2021preventing} propose a cryptography-based authentication, which provides a cryptography-based provenance module to protect trained models against alteration of their settings or underlying structure. The proposed authentication can protect against model poisoning attacks.
    \item \textbf{\textit{Optimization-based:}} in a work by Kuttichira \textit{et al.} \cite{kuttichira2022verification}, the authors propose a Bayesian Compromise Detection (BCD) algorithm that aims to address the potential security risk associated with malignant modification of stored models in the cloud. The main solution consists of solving an optimization problem that seeks to maximize the discrepancies in prediction outcomes by comparing the genuine and damaged models. Although this task seems easy at first glance, the \textcolor{black}{problem's difficulty} lies in three points, according to the authors. First, its non-convex nature; second, the large space dimensionality when searching through the training input distribution for the sensitive sample; and third, a blue-box view of the compromised model is all that cloud clients can have. The proposed solution consists of two parts. First, a Variational Autoencoder (VAE) is used to associate high-dimensional data with a low-dimensional nonlinear space, and second, Bayesian optimization (BO) is applied to determine the generally ideal sensitive sample, which can identify the model corruption with little overhead. Experiments on MNIST, Olivetti, and CIFAR-100 datasets illustrate the capability of the proposed approach, with results of up to 100\% detection rate.
\end{itemize}

\subsection{Inference Phase Defense Methods}
The defense mechanisms in this class are designed to prevent attacks against ML models during the inference phase. 
\subsubsection{Privacy Leakage Defense Mechanisms}
We classify the work belonging to this class into three subclasses, namely  privacy leakage, self-distillation, and overfitting control, as provided below:
\begin{itemize}
    \item \textbf{\textit{Gradients Shielding:}} The main objective of these mechanisms is to protect the model's gradient cryptographically. Hao \textit{et al.} \cite{hao2019efficient} implemented a privacy-preserving IDS for FL-based industrial environments named PEFL. The proposed system secures exchanged aggregation with homomorphic encryption and differential privacy. In essence, the homomorphically encrypted data of the private gradients is incorporated into the A-LWE (augmented learning with error). Benchmarking on the MNIST dataset exhibits good performance, along with computational and communication cost efficiency.
    \item \textbf{\textit{Self-Distillation:}} in a work by Tang \textit{et al.} \cite{tang2022mitigating}, the authors propose SELENA. This privacy-preserving model learning framework generates equivalent member/non-member feedback to alleviate blue-box membership inference threats. The proposed system consists of two major components. The first is the Split Adaptive Inference Ensemble (Split-AI), which allows the model to behave similarly to members and non-members samples. This is done by training sub-models through random subsets within the training set. The second is a self-distillation mechanism that transfers knowledge about the model created by Split-AI to deliver a protected end model. To do this, Split-AI is first interrogated with the exact training data to retrieve the related prediction sequences. Then, the protected end model is trained with these predictions as soft labels. Evaluations on three different datasets (Purchase100, Texas100, and CIFAR100), with ResNet-18 and a fully connected four-layer NN, and different MIA attacks, including direct single-query attacks, label-only attacks, and adaptive attacks, showed that the proposed system achieved a good tradeoff between practicality and privacy.
    \item \textbf{\textit{Overfitting Control:}} when given a data point on which models have been trained, they return a high aftereffect value on a class relative to the others, reflecting the underlying overfitting nature of ML models, which is taken to be one of the reasons for the effectiveness of MIAs. Hence, according to Salem \textit{et al.} \cite{salem2018ml} controlling overfitting is one way to mitigate such attacks. The authors propose using two approaches: classical and relatively new, namely \textit{dropout} and \textit{model stacking}. The first is a regularization technique that prevents complex co-matching on training data and is used explicitly for DL models. The second one is proposed to be used for other ML classifiers. The main concept is to organize several ML models to avoid overfitting hierarchically.
\end{itemize}


\section{Lessons learned, Open issues and Challenges}\label{sec:7}
Despite the significant efforts made by the scientific community to strengthen security in the cyber world in general, and for future networks such as 6G, in collaboration with emerging technologies such as IoT and AI in particular, there is however a long way to be traveled, if we are to achieve a fully secure cyber environment. This section presents the lessons learned, open issues, and challenges.

\subsection{Lessons learned}

Through comprehensive reviews and in-depth analysis, we were able to classify the datasets used by the scientific community for experimenting and evaluating ML techniques on cyber attacks into \textcolor{black}{nine main} categories, including, IoT-based multi-purpose security, attack classification, image classification, time series classification, human activity recognition, sentiment classification, location awareness, and text classification. From the attacks against machine learning systems, we found twenty-one attacks. According to the actual context of the attack in \textcolor{black}{Edge Learning}, we were able to classify them into eight categories, including backdoor attacks, adversarial examples, combined attacks, poisoning attacks, Sybil attacks, byzantine attacks, inference attacks, and drop attacks. Based on the deployment strategy of each security defense, we \textcolor{black}{classified} the defense methods against \textcolor{black}{Edge Learning} vulnerabilities into three categories, including training phase defense methods, post-training phase defense methods, and inference phase defense methods.

From the above analysis and reviews that we completed, we suggest the following steps for proposing defense methods against \textcolor{black}{Edge Learning} vulnerabilities in 6G-enabled IoT networks:

\begin{itemize}
    \item Definition of the infrastructure of the 6G-enabled IoT network and the emerging technologies adopted for each layer (e.g., Edge layer, Fog Layer, SDN Layer, Blockchain Layer, Digital twins layer...etc.).
    \item Definition of the edge learning model (i.e., centralized, federated, distributed learning). 
    \item Identification of the attacks against machine learning systems (e.g., backdoor attacks, adversarial examples, poisoning attacks, inference attacks, ... etc.)
    \item Systematize the threat models based on three dimensions: adversarial goal, attack strategies, and malicious client selection.
    \item Selection of the defense method against edge learning vulnerabilities (i.e., training phase defense methods, post-training phase defense methods, or inference phase defense methods).
    \item Selection of the datasets adopted for the 6G-enabled network (e.g., attack classification dataset, image classification dataset, human activity recognition dataset, sentiment classification dataset, and text classification dataset).
    \item Selection of the data distribution (i.e., IID and Non-IID).
    \item Experimenting and evaluating the defense method regarding accuracy, precision, recall, and attack rate.
    \item Study the system's performances with the application of the proposed defense method regarding scalability and interoperability.
\end{itemize}

\textcolor{black}{The use of network slicing, multi-access edge computing, cloud computing, virtualization, terahertz, and sub-terahertz communications, and artificial intelligence in the 5G/6G IoT testbeds demonstrates the complexity and diversity of the technologies being developed. These testbeds offer an opportunity to explore new possibilities in smart transportation, energy management, healthcare, and industrial automation.}

\textcolor{black}{In summary, the development of effective defense methods against \textcolor{black}{Edge Learning} vulnerabilities in 6G-enabled IoT networks requires a systematic approach that considers the network infrastructure, edge learning models, attack types, threat models, defense methods, datasets, data distribution, and system performance. With a comprehensive understanding of these factors, researchers can better develop and evaluate defense methods to improve the security and resilience of 6G-enabled IoT networks.}

\begin{table*}[]
\centering
\setlength{\tabcolsep}{2.5pt}
\renewcommand{\arraystretch}{1}
\caption{\color{black}Open Issues and Challenges in 6G-IoT \textcolor{black}{Edge Learning} vulnerabilities.}
\centering
\label{tab:tabcha}
\scriptsize
{\color{black}
\begin{tabular}{|p{1.5in}|p{2.2in}|p{2.4in}|}
\hline
\textbf{Challenge}                 & \textbf{Description}                                                                                                            & \textbf{Key Considerations}                                                                                                              \\ \hline
Reliable and Trustworthy Learning  & Ensuring the security of AI-based systems by building security into every aspect of AI, from data preparation to inference     & - Adoption of secure AI practices \newline - Development of secure hardware and software architectures \newline - Development of lightweight and efficient ML algorithms \\ \hline
Security Solutions Adaptability    & Developing security solutions that are highly adaptive to the dynamic and heterogeneous nature of wireless networks            & - Use of pre-trained models, unified datasets, and agreed-upon models between gNB and UEs.                                                 \\ \hline
Ethics by Design                   & Addressing the ethical dimensions of AI systems at the earliest stage of their design through the "Ethics by Design" approach  & - Adoption of ethical guidelines \newline - Consideration of the specific requirements and constraints of different applications and industries.     \\ \hline
Datasets Formation/Availability    & Creating dependable \textcolor{black}{Edge Learning} datasets as well as managing complexity, and ensuring high-quality data availability      & - Use of Explainable AI (XAI) techniques \newline - Introduction of a confidence level and validations, and layered security models                 \\ \hline
Learning Complexity                & Managing complexity in mobile communication systems and other complex environments through Explainable AI (XAI) techniques     & - Use of XAI to provide transparency, understanding, and trust in AI models                                                               \\ \hline
High-Quality Data Availability     & Ensuring the availability of high-quality data and using techniques that introduce a confidence level and validations for FL   & - Adoption of techniques that introduce a confidence level and validations                                                                \\ \hline
ML Vulnerabilities Elimination     & Balancing security, performance, profitability, simplicity and complexity, and backward compatibility with future security & - Striking a balance between security, performance, and profitability \newline - Implementing a layered security model                             \\ \hline
Defense Strategies Implementations & Ongoing research and development of security solutions to stay ahead of emerging threats, such as zero-day attacks            & - Use of ML-based security for detecting zero-day attacks                    \\ \hline
\textcolor{black}{Large language models at the Edge}            & \textcolor{black}{Deploying large language models at the edge for 6G-enabled IoT systems introduces several challenges}    &  \textcolor{black}{- Edge devices are often battery-powered and have strict energy consumption limits \newline -  Latency affect real-time responses and system efficiency.}   \\ \hline
\textcolor{black}{Real-time Decision Making}      & \textcolor{black}{Enabling 6G-IoT devices to make real-time decisions despite some ML models being computationally expensive} & \textcolor{black}{- Development of lightweight ML algorithms \newline - Optimizing computational resources for real-time processing } \\ \hline
\textcolor{black}{Model Generalization}     &     \textcolor{black}{Ensuring ML models generalize well across various 6G-IoT devices despite differences in data distributions} & \textcolor{black}{- Development of diverse training datasets \newline - Adoption of model validation techniques to assess generalization } \\ \hline
\textcolor{black}{Scalability}   & \textcolor{black}{Dealing with the increasing number of IoT devices and the data they generate} & \textcolor{black}{- Deployment of distributed ML algorithms \newline - Development of efficient data storage solutions} \\ \hline

\end{tabular}}
\end{table*}

\subsection{Open issues and Challenges}

Table \ref{tab:tabcha} overviews the open issues and challenges in 6G-IoT machine learning vulnerabilities. The challenges covered include reliable and trustworthy learning for 6G-IoT intelligence, security solutions adaptability, ethics by design, datasets formation/availability, learning complexity, high-quality data availability, ML vulnerabilities elimination, and defense strategies implementations.

\subsubsection{Reliable and Trustworthy Learning for 6G-IoT Intelligence}
AI capabilities have proven to be a key component of future technologies. However, there is more involved in ensuring security. Here we present some challenges when creating reliable and trustworthy learning environments.

\begin{itemize}

\item \textcolor{black}{Natively Secured AI: Adversary-aware AI, or natively secured AI, is an upcoming class for AI-based systems that focuses on building security into the data preparation, learning, storage, and inference stages. This paradigm is necessary because cyberattacks can compromise the integrity, confidentiality, and availability of AI models and their associated data, leading to adverse consequences. Similar to how cyber attacks forced developers to write secure code and libraries for programming languages, the increasing adoption of AI-based systems will necessitate the adoption of secure AI practices. This will involve not only the development of secure AI algorithms but also the design and implementation of secure hardware and software architectures to support these algorithms. Given the complexity and uncertainty associated with AI techniques, the challenges associated with implementing natively secured AI are considerable. Additionally, the limitations of IoT devices, such as their limited processing and storage resources, may pose a challenge to running complex AI models. To address this challenge, lightweight and efficient ML algorithms tailored for these devices must be developed, similar to the situation encountered in the case of lightweight IoT-based authentication protocols. In summary, natively secured AI is an upcoming class for AI-based systems that focuses on building security into every aspect of AI, from data preparation to inference. While \textcolor{black}{significant challenges} are associated with implementing this paradigm, ensuring the security of AI-based systems in the face of increasingly sophisticated cyber attacks is necessary.}

\item \textcolor{black}{Trustworthy AI: Trustworthy AI is a key component of natively secure AI, assuring that AI-based systems operate reliably, transparently, and ethically. Specifically, Trustworthy AI refers to AI designed to be transparent, explainable, and accountable. Trustworthy AI systems focus on minimizing the risks associated with AI algorithms and their decision-making processes by ensuring that they operate within specific ethical constraints. To achieve trustworthy AI, secure AI algorithms must be developed that incorporate ethical and legal standards, including human rights, data protection, and privacy. This is essential to ensure that AI-based systems are transparent and explainable and that users can understand how the system works, makes decisions and uses data.}

\item Security Solutions Adaptability: The dynamic nature of settings shifts across wireless networks requires the proposed security solutions to be highly adaptive. Because many operators collaborate, this diversity can pose a significant heterogeneity problem if not handled effectively. Solutions to this problem have been briefly discussed, including pre-trained models, unified datasets, and agreed-upon models between Next Generation NodeB (gNB) and user equipment (UEs) \cite{shehzad2022artificial}. However, real-world implementations and further tests are different stories with varying perspectives, posing a future challenge for these networks. \textcolor{black}{Hence, the adaptability of security solutions is crucial for the security of wireless networks, and ongoing research and testing are necessary to address the challenges posed by these networks' dynamic and heterogeneous nature.}

\item Ethics by Design: While machine intelligence is a promising addition to future networks, making them fully automated, they do not accommodate ethical guidelines in the same way as humans \cite{siriwardhana2021ai}. The "Ethics by Design" approach addresses the ethical dimensions of AI systems at the earliest stage of their design \cite{d2018towards}. And as much as it is not an easy subject for a machine to learn, the concept itself is difficult to define, as even among humans and nations, the concept sometimes has wide variations in how it is described. \textcolor{black}{By adopting the Ethics by Design approach, AI designers and developers can help ensure that AI systems are developed responsibly and ethically.}

\end{itemize}

\subsubsection{Datasets Formation/Availability for \textcolor{black}{Edge Learning}}
In addition to introducing security into AI, other challenges may arise along the way, including the availability and the creation of dependable \textcolor{black}{Edge Learning} datasets. We highlight two important issues: managing complexity and ensuring reliable data resource availability.

\begin{itemize}
\item Learning Complexity: In a complex environment such as mobile communication systems, which is likely to be even more complex with the projected introduction of a range of technologies for future networks such as 6G, and given the limited hardware capabilities of IoT devices and cellular entities, particularly in terms of processing power, battery life, and storage, this is going to be a serious problem in terms of how well these entities can handle complexity in the interests of functionality. Explainable AI (XAI) can solve part of the problem, unlike the black-box concept, as it effectively introduces understanding, management, and trust in AI models, making it easier to reduce complexity without affecting accuracy.  XAI is an important tool for managing complexity in mobile communication systems and other environments. \textcolor{black}{By providing transparency, understanding, and trust in AI models, XAI can help stakeholders navigate complex environments while maintaining functionality and accuracy.}

\item High-Quality Data Availability: The FL paradigm solves the problem of compromising the privacy of private data in the learning phase. However, as we have seen in the previous sections, this paradigm is subject to different threats. One is related to data, which can be a significant danger for future wireless networks since distributed learning is supposed to be a key component of future networks. It can be tricky to trust models trained on data the operator has not seen or validated, such as when relying on UEs for the training task. Similarly, even when dealing with internal components such as gNBs and Intelligent Radios (IRs), it can be dangerous to trust data collected from the surrounding environment and used for training directly without the proper validations. Techniques that introduce a confidence level and validations must be applied before incorporating the models directly into the networks. \textcolor{black}{Overall, high-quality data availability and techniques that introduce a confidence level and validations are critical to ensuring the success of federated learning in wireless networks. By adopting these techniques, operators can trust the models generated by the learning process and use them to make informed decisions about the network's configuration and performance.}
\end{itemize}


\subsubsection{ML Vulnerabilities Elimination and Defense Strategies Implementations}
Developing a security solution that guarantees equal levels for conflicting aspects is challenging. Here we discuss the trade-offs that can arise when trying to eliminate ML vulnerabilities and develop security solutions.

\begin{itemize}

\item Security vs. Performance and/or Profit: The high data rates promised by 6G will enable various time-sensitive IoE applications, requiring real-time security techniques to be deployed, as conventional security schemes are subject to short-term and long-term failure under such circumstances. This results in different trade-offs, on the one hand, between security and performance, e.g., between privacy and accuracy in FL, since introducing a lot of noise to protect the data will result in reduced accuracy, and vice versa. On the other hand, the effort for real-time security techniques is costly and depends on what the telecom industry is willing to pay and their priorities. \textcolor{black}{The effort for developing and implementing real-time security techniques is costly, and the telecom industry may have to prioritize security, performance, and profitability. In some cases, the industry may decide to prioritize performance or profitability over security, which could compromise the security of the network and the data transmitted over it. Therefore, it is essential to balance security, performance, and profitability while deploying real-time security techniques for 6G networks. This will require careful consideration of different applications and industries' specific requirements and constraints and collaboration between telecom, academia, and government stakeholders.}

\item Simplicity vs. Complexity: Since future networks are expected to be more complex than previous generations, which is natural, this has persisted in the evolution of all preceding generations. A trivial conclusion that could be drawn directly is that security solutions need to be complex as well; however, this can be seen as a double-edged sword since, to keep the system stable, more secure, and have fewer vulnerabilities, simplicity, and transparency are keys \cite{nguyen2021security}. \textcolor{black}{In the case of 6G-enabled IoT networks, which are expected to be more complex, it is important to consider both simplicity and complexity in security solutions. One approach could be implementing a layered security model, where multiple layers of security are used to protect against different attacks. Each layer can be designed with a simple and transparent solution that is easy to understand and audit while also being integrated into a larger, more complex security architecture. ML-based security is a promising approach for detecting zero-day attacks, which are attacks that exploit vulnerabilities that are unknown to security experts. However, zero-day attacks targeting ML algorithms can be difficult to detect. This highlights the need for ongoing research and development in security solutions to stay ahead of emerging threats.}

\item Backwards Compatibility vs. Future Security: In addition to the increased complexity, maintaining backward compatibility is another feature that persists in all previous generations. While it has benefits, including reduced deployment costs compared to a new standalone deployment and ease of upgrading, the backward compatibility functionality potentially exposes old vulnerabilities. In addition, AI-based security must also consider some of the previous generation's parameters to function properly. This may be something to be carefully examined in the future. \textcolor{black}{Overall, balancing the need for backward compatibility with the need for future security is a complex challenge that requires careful consideration. It may be necessary to prioritize security over backward compatibility in certain cases, but in other cases, it may be possible to reach a trade-off between the two.}
\end{itemize}

\subsubsection{\textcolor{black}{Large language models at the Edge}}
\textcolor{black}{Large language models, like OpenAI's GPT-4, FalconLLM, and Bert,  are increasingly being used in a wide range of applications, from content generation to customer service and much more \cite{vaswani2017attention, devlin2018bert, raffel2020exploring}. However, deploying large language models at the edge for 6G-enabled IoT systems introduces several challenges. First, security and privacy issues are magnified in a 6G environment due to the sheer volume and variety of devices \cite{ramezanpour2023security}. The immense data flow, which includes sensitive data, presents a significant risk of cyber threats and vulnerabilities. Second, despite 6G's promise of higher data rates and lower latencies, managing the scale and complexity of these models can be a challenge due to the inherent resource constraints of IoT devices, potentially affecting model performance and efficiency. The high density of IoT devices in 6G networks could also strain network resources and introduce latency issues, impacting real-time responses and decision-making \cite{shen2023five}. Third, maintaining consistency of the model's deployment across a diverse range of IoT devices is difficult, and software updates become challenging due to the vast distribution of devices. Therefore, deploying large language models at the edge of 6G IoT environments necessitates careful attention to these challenges to ensure security, efficiency, and compliance.}

\subsubsection{\textcolor{black}{Universal Multi-prompt and Multi-model attacks}}

\textcolor{black}{Creating adversarial attacks for discrete forms of data such as text is substantially more challenging than for continuous data types like images \cite{li2020bert}. This is primarily due to the difficulties in generating adversarial samples through gradient-based techniques. Existing successful strategies for attacking text data typically involve heuristic replacement methods at either the character or word level \cite{zhang2020adversarial}. Recently, Zou et al. \cite{zou2023universal} presented a novel method for creating adversarial attacks on large language models (LLMs) to generate objectionable content. This method, unlike previous human-centric or inefficient automated techniques, generates adversarial suffixes through automated greedy and gradient-based search techniques, prompting LLMs to produce undesired responses. These generated prompts were highly transferable across several LLMs, including ChatGPT, Bard, Claude, and others, with a notable susceptibility in GPT-based models. The study's findings advance the understanding of adversarial attacks against LLMs and highlight the need to improve these systems' resistance to producing inappropriate content. The challenge is developing effective defenses against adversarial attacks that prompt large language models to generate objectionable content.
}

\section{Conclusions} \label{sec:8}
While 5G is being deployed and commercialized around the world, researchers are getting excited about what 6G can and should be. One of the most endorsed views is that AI should be a core component of 6G rather than just an add-on utility. Edge learning involves making the edge of the network intelligent, where models are trained at the edge using coordinated distributed learning paradigms such as FL, with data available across various edge devices. However, given the existing vulnerabilities in ML, its adoption without adequate security considerations can expose the network to various threats. In this paper, we have surveyed the state-of-the-art of existing vulnerabilities and defenses of federated machine learning for 6G-enabled IoTs. We have summarized the existing surveys on machine learning for 6G–IoT security and machine learning-associated threats in three different learning modes: centralized, federated, and distributed. Through extensive research and analysis that has been conducted, we have classified the threat models against machine learning into eight categories, including backdoor attacks, adversarial examples, combined attacks, poisoning attacks, Sybil attacks, byzantine attacks, inference attacks, and dropping attacks. In addition, we have analyzed the state-of-the-art defense methods against federated machine learning vulnerabilities. Finally, as new attacks and defense technologies are realized, new research and future overall prospects for 6G-enabled IoTs are discussed. There still exist several challenging research areas on new attacks and defense technologies, which should be further investigated in the near future.


\bibliographystyle{IEEEtran}
\bibliography{ref} 

\end{document}